\documentclass[11pt]{article}
\usepackage{graphicx,epsf,subfigure,pstricks,pst-node,psfrag,amsthm,amssymb,amsmath,epstopdf}
\usepackage{latexsym, multicol, multirow, epsfig, color, setspace, array, multirow, color, amsbsy, epsfig, verbatim, diagbox, soul, setspace}
\usepackage{natbib}
\usepackage[latin1]{inputenc}
\usepackage{tikz}
\usetikzlibrary{matrix}
\usetikzlibrary{shapes,arrows, fit}
\usetikzlibrary{positioning,calc}
\usepackage{float, caption}
\usepackage{graphicx, setspace, float, diagbox}
\usepackage{amsmath}
\usepackage{mathtools}
\usepackage{SASnRdisplay}
\usepackage{algorithm}
\usepackage{algpseudocode}
\usepackage{lineno}
\usepackage{float}
\usepackage{pdflscape}
\usepackage{subfloat}
\usepackage{pbox}

\def\ProvidesPackageRCS#1{}
\newread\infile
\def\preparetable#1#2{\bgroup \openin\infile=#1
	\let\\=\relax \gdef\usetable{}\preparetableA #2,,}
\def\preparetableA #1,{\if,#1,\egroup \closein\infile \else \read\infile to\tmp
	\xdef\usetable{\usetable \tmp & #1 \\}\expandafter\preparetableA\fi}
	
\newcommand{\phone}{\phantom{-}1}

\bibpunct{(}{)}{;}{ay}{,}{,}

\begin{document}

\title{A-ComVar: A Flexible Extension of Common Variance Designs}

\author{Shrabanti Chowdhury\\Department of Genetics and Genomic Sciences\\Icahn School of Medicine at Mount Sinai\\New York, NY 10029 \and Joshua Lukemire\\ Department of Biostatistics and Bioinformatics\\Emory University\\Atlanta, GA 30322  \and Abhyuday Mandal\\Department of Statistics\\University of Georgia\\Athens, GA 30606}
\date{\today}
\maketitle

\begin{abstract}
We consider nonregular fractions of factorial experiments for a class of linear models. These models have a common general mean and main effects, however they may have different $2$-factor interactions. Here we assume for simplicity that $3$-factor and higher order interactions are negligible. In the absence of {\it a priori} knowledge about which interactions are important, it is reasonable to prefer a design that results in equal variance for the estimates of all interaction effects to aid in model discrimination. Such designs are called common variance designs and can be quite challenging to identify without performing an exhaustive search of possible designs. In this work, we introduce an extension of common variance designs called approximate common variance, or A-ComVar designs. We develop a numerical approach to finding A-ComVar designs that is much more efficient than an exhaustive search. We present the types of A-ComVar designs that can be found for different number of factors, runs, and interactions. We further demonstrate the competitive performance of both common variance and A-ComVar designs {\color{black} using several comparisons to other popular designs in the literature}.
\end{abstract}


\section{Introduction}

Fractional factorial designs are widely used in many scientific investigations because they provide a systematic and statistically valid strategy  for studying how multiple factors impact a response variable through main effects and interactions. When several factors are to be tested, often the experimenter does not know which factors have important interactions. Instead, the experimenter will need to perform model selection after conducting the experiment to identify important interactions. Generally this process will involve fitting different models under consideration and examining statistical significance of the interaction terms. Some techniques have been developed concerning finding efficient fractional factorial plans for this purpose. There is a rich literature on identification and discrimination to find the model best describing the data \citep{srivastava1976some, srivastava1976series, srivastava1979main}.

{\color{black} Consider a design with $m$ factors of interest. Following \cite{ghosh2017cv}, consider the following class of $s$ candidate models for describing the relationship between  $p$ $(\leq m)$ of the $m$ factors and the $n\times1$ vector of observations $\boldsymbol{y}$,} 
\begin{align}
E\left(\boldsymbol{y}\right) &= \beta_0\boldsymbol{j}_n + \boldsymbol{X}_1 \boldsymbol{\beta}_1 + \boldsymbol{X}_{2}^{(i)} \boldsymbol{\beta}_{2i}, \mbox{  } i = 1, \ldots, s \\
Var(\boldsymbol{y}) &= \sigma^2 \boldsymbol{I}, \nonumber
\end{align}

\noindent where $n$ is the number of runs, $\beta_0$ is the general mean, $\boldsymbol{j}_n$ is a vector of ones, $\boldsymbol{\beta}_1$ is the vector of $p$ main effects that are common in all $s$ models. The other parameters, $\boldsymbol{\beta}_{2i}$, are specific for the $i$th model and hence $\boldsymbol{\beta}_{2i} \neq \boldsymbol{\beta}_{2i^{\prime}}$ for $i \neq i^{\prime},$ \mbox{ }$i = 1, \dots, s$. We call these parameters ``uncommon parameters." The design matrices $\boldsymbol{X}_1$ and $\boldsymbol{X}_{2}^{(i)}$ correspond to the main effects and $i^{\mbox{th}}$ set of $2$-factor interactions, respectively. {\color{black} Following \cite{ghosh2013common} and \cite{ghosh2017cv} we consider the situation of $p=m$ for generating A-ComVar designs using our proposed approach, described in Section 4. However, we consider models with $p\leq m$ cases in our examples comparing the performance of our proposed designs with some popular designs from the literature.} 

Under the above setup, model selection consists of identifying the correct $i$ from the $s$ candidate models. This process is complicated by the fact that the variance estimates for the uncommon parameters are generally not the same, which can pre-bias the experiment towards identifying certain interactions as significant over others, i.e. making some $i$ more likely to be selected than others regardless of the true underlying model. To address this issue, \cite{ghosh2013common} introduced the notion of common variance designs for a single uncommon parameter. These designs estimate the uncommon parameter in all models with equal variance, which is desirable in the absence of any {\it a priori} information about the true model. \cite{ghosh2017cv} generalized this concept of common variance to $k$ $(k\geq 1)$ uncommon parameters in each model in the class. Under the situation of $k > 1$, \cite{ghosh2017cv}  defined a common variance design to be the one satisfying $|\boldsymbol{X}^{(i) \prime} \boldsymbol{X}^{(i)}|$ to be a constant, for all $i$, $\boldsymbol{X}^{(i)} = \left(\boldsymbol{j}_n, \boldsymbol{X}_1, \boldsymbol{X}_2^{(i)} \right)$.  

The concept of variance-balancedness is not totally new. Different types of ``variance-balanced designs" estimating all or some of the treatment-contrasts with identical variance were developed by \cite{calvin1986new}, \cite{cheng1986method}, \cite{gupta1983equireplicate}, \cite{hedayat1989relation}, \cite{khatri1982note}, \cite{mukerjee1985resolvable}, among others.

While common variance designs have been identified for two and three level factorial experiments with a single $2$-factor interaction \citep{ghosh2013common, ghosh2017cv}, it remains to develop a method which can find them for general number of factors and interactions. To date, these designs have been found using exhaustive searches, which becomes prohibitively expensive as the number of factors and runs increases. This leads us to introduce approximate common variance (A-ComVar) designs, which relax the requirement that the variance of the uncommon parameters be exactly equal. We introduce an objective function that allows us to rank designs under consideration, and we develop a genetic algorithm for searching for these designs. Moreover, we investigate the performance of both common variance and A-ComVar designs for model selection using the adaptive lasso regression technique in simulation \citep{kane2019}. We find comparable performance of common variance and A-ComVar designs to Placket-Burman designs, which further demonstrates the usefulness of designs that prioritize having a similar variance for the uncommon parameters in the model.

The rest of the article is organized as follows. In Section 2 we present the current state of knowledge for both two-level and three-level common variance designs. For three-level designs we also present the exhaustive search result for $m=3$.  {\color{black} In Section 3 we introduce our numerical approach for finding A-ComVar designs. In Section 4 we conduct extensive studies to both (i) examine our numerical approach's ability to find A-ComVar designs as we increase the number of factors and number of interactions in the model and (ii) compare these A-ComVar designs to potential competitor designs from the literature. Finally, Section 5 contains some discussion of the results and some future directions for our work. The Appendix contains an illustration of our genetic algorithm, as well as Tables corresponding to all results.}

\section{Common Variance Designs}\label{sec:knowledge}



\subsection{Two Level Designs}
The term ``common variance" for the class of variance-balanced designs was first introduced in \cite{ghosh2013common}.  As a more stringent criteria, the authors also introduced the concept of optimum common variance (OPTCV), which is satisfied by designs having the smallest value of common variance in a class of common variance designs with $p \leq m$ factors and $n$ runs. Several characterizations of common variance and optimal common variance designs were presented that provide efficient ways for checking the common variance or OPTCV property of a given design. These characterizations were obtained in terms of the projection matrix, eigenvalues of the model matrix, balancedness, and orthogonal properties of the designs. In Corollary 1 of \cite{ghosh2013common}, they stated one sufficient condition of common variance designs in terms of equality of the vectors of eigen values of $\boldsymbol{X}^{(i)\prime} \boldsymbol{X}^{(i)}$, $\boldsymbol{X}^{(i)} = \left(\boldsymbol{j}_ n, \boldsymbol{X}_1, \boldsymbol{X}_2^{(i)} \right)$, for all $i$. We present one design in Table \ref{table:design1} from \cite{ghosh2013common} for $m=5$ and $n=12$, that has identical vectors of eigen values for all $i$. In Section 4.3 we compare the performance of this particular design with that of  Plackett-Burman design for model selection to demonstrate further usefulness of such designs.

In their work, \cite{ghosh2013common} presented several general series of designs with the common variance property. For example, they identified two fold-over designs with the common variance property with all $m$ factors of the design and $n=2m$ and $n=2m+2$ runs respectively:


\[
d_m^{(2m)}=
\left[ {\begin{array}{c}
	\phantom{-}2\boldsymbol{I}_m -  \boldsymbol{J}_m\\
	-2\boldsymbol{I}_m + \boldsymbol{J}_m \\
	\end{array} } \right].
\]   

\[
d_m^{(2m+2)}=
\left[ {\begin{array}{c}
     \phantom{-}\boldsymbol{j}_m^{\prime} \\
     -\boldsymbol{j}_m^{\prime} \\
	\phantom{-}2\boldsymbol{I}_m -  \boldsymbol{J}_m\\
	-2\boldsymbol{I}_m + \boldsymbol{J}_m \\
	\end{array} } \right].
\] 

\noindent As reported in \cite{ghosh2013common}, both of these designs are balanced arrays of full strength and orthogonal arrays of strength 1, for all $m$. Moreover, the design $\boldsymbol{d}^{(2m)}_m$ is OPTCV for $m=4$ and $\boldsymbol{d}^{(2m+2)}_m$ is OPTCV for $m=3$.

\subsection{Three Level Designs}
\cite{ghosh2017cv} presented common variance designs for $3^m$ fractional factorial experiments. Consider the following model for a $3^m$ factorial experiment, with one $2$-factor interaction effect in the model, i.e. $k=1$:
\[
E\left(\boldsymbol{y}\right) = \beta_0\boldsymbol{j}_n + \boldsymbol{X}_1 \boldsymbol{\beta}_1 + \boldsymbol{X}_{2}^{(i)} \beta_{2i}, \mbox{ } Var(\boldsymbol{y}) = \sigma^2.
\]
A design for such an experiment would have the common variance property iff $\frac{Var(\hat{\beta}_{2i})}{\sigma^2}$ is constant for all $i=1, \dots, 4\binom{m}{2}$, for the situation $p=m$.

\cite{ghosh2017cv} presented two general series of $3^m$ fractional factorial common variance designs $d_1$ and $d_2$ with $n$ runs. The design $d_1$ has a common variance value given by $\frac{Var\left(\hat{\beta}_2^{(i)}\right)}{\sigma^2} = \frac{2-m+m^2}{9}$, for $m \geq 2$ and $n=2m+2$ runs, while design $d_2$ has a common variance value given by $\frac{Var\left(\hat{\beta}_2^{(i)}\right)}{\sigma^2} = \frac{m}{9(m-2)}$, for $m \geq 3$ and $n=3m$. Also, the design $d_1$ is efficient common variance (ECV, as termed in \cite{ghosh2017cv}) design for $m=2$, and design $d_2$ is ECV for $m=3$.

\cite{ghosh2017cv} also presented several sufficient conditions for general fractional factorial designs to have the common variance property, including the special case for $3^m$ designs in terms of the projection matrix of the design and the columns of $2$-factor interaction. For example, a design is common variance if (i) $\boldsymbol{PX}_2^{(i_1)} = \boldsymbol{PX}_2^{(i_2)}$, for $i_1, i_2 \in \{1, \dots, s\}$, where $\boldsymbol{P}$ is the projection matrix defined as $\boldsymbol{I}_n - \boldsymbol{X}_1 \left(\boldsymbol{X}_1^{\prime} \boldsymbol{X}_1\right)^{-1} \boldsymbol{X}_1^{\prime}$, and $\boldsymbol{X}_1$ contains the columns corresponding to the general mean and main effects from the model matrix $\boldsymbol{X}^{(i)} = \left(\boldsymbol{j}_n, \boldsymbol{X}_1, \boldsymbol{X}_2^{(i)}\right)$, and $\boldsymbol{X}_2^{(i)}$ corresponds to the $i^{th}$ $2$-factor interaction. Another set of sufficient conditions for having common variance is, for, $i_1, i_2 \in \{1, \dots, s\}$, (i) $\left(\boldsymbol{X}_2^{(i_1)} \pm \boldsymbol{X}_2^{(i_2)}\right)$ belongs to the column space of $\boldsymbol{X}_1$ and (ii) $\boldsymbol{X}_2^{(i_2)} = \boldsymbol{F} \boldsymbol{X}_2^{(i_1)}$ holds, where the permutation matrix $\boldsymbol{F}$  obtained from the identity matrix satisfies $\boldsymbol{F}^{\prime} \boldsymbol{P} \boldsymbol{F} = \boldsymbol{P}$.
 
For $3^3$ fractional factorial experiment, \cite{Chowdhury2016} conducted a complete search of common variance designs for $n=8$ to $n=27$, since $n=8$ is the minimum number of runs needed to estimate all the parameters considering all 3 factors are present in the model (one general mean, 6 main effects, one $2$-factor interaction effect). The results of this search are presented in Table \ref{completesearchresults}. The complete search revealed that common variance designs only exist for $n=8,9,10,11$ for $3^3$ factorial experiments. For each of the runs multiple groups of common variance designs were obtained, having different common variance values, among which 32 designs for $n=11$; 48 designs for $n=10$; 8256 designs for $n=9$; and 9600 designs for $n=8$, are the efficient common variance designs giving the minimum value of common variance in the respective classes.

\begin{table}[ht]
		\centering
		\resizebox{\textwidth}{!}{
		\begin{tabular}{|ccccccc|}
				\hline
			         & Possible                  & Satisfying                                 & No. of    & No. of   & No. of CV & \\
			  $n$ & Designs                  & Rank Condition                         & Non-CV  & CV        & designs with & CV value\\
			         & $= \binom{27}{n}$  & ($|\mathbf{X}'\mathbf{X}| > 0$) & designs   & designs & this value & \\
				\hline
				\multirow{2}{*}{11} & \multirow{2}{*}{13,037,895} &\multirow{2}{*}{6,926,898} &\multirow{2}{*}{6,924,772} & \multirow{2}{*}{2,096} &32 &0.2151\\
				& & & & &2,064 &0.2222\\
				\hline
				\multirow{5}{*}{10} & \multirow{5}{*}{8,436,285} & \multirow{5}{*}{2,792,387} & \multirow{5}{*}{2,775,747} &\multirow{5}{*}{16,640} &48 &0.2564\\
				& & & & &48 &0.2667\\
				& & & & &16 &0.2837\\
				& & & & &16,512 &0.2963\\
				& & & & &16 &0.4000\\
				\hline
				\multirow{5}{*}{9} & \multirow{5}{*}{4,686,825} & \multirow{5}{*}{636,348} & \multirow{5}{*}{588,348} & \multirow{5}{*}{48,000} &8,256 &0.3333\\
				& & & & &32 &0.3810\\
				& & & & &13,056 &0.4167\\
				& & & & &26,640 &0.4444\\
				& & & & &16 &0.5000\\
				\hline
				\multirow{2}{*}{8} & \multirow{2}{*}{2,220,075} & \multirow{2}{*}{49,628} & \multirow{2}{*}{23,340} &\multirow{2}{*}{26,288} &9,600 &0.6667\\
				& & & & &16,688 &0.8889\\
				\hline
		\end{tabular}
		}
			\caption{Complete search results for finding common variance designs for $3^3$ factorial experiments.}
		\label{completesearchresults}
			\end{table}

\section{Identifying Common Variance Designs}

\subsection{Challenges in Numerically identifying Common Variance Designs}

\cite{ghosh2013common} and \cite{ghosh2017cv} presented some general series of designs satisfying the common variance property for two- and three-level factorial experiments obtained via exhaustive searches of the design space. Such searches become extremely computationally challenging as the number of factors increases. For example, for a $3^3$ factorial experiment with one $2$-factor interaction ($k=1$) the possible set of candidate designs with 8 runs is $\binom{27}{8}=2220075$, with 9 runs is $\binom{27}{9}=4686825$, with 10 runs is $\binom{27}{10}=8436285$, and so on. For a $3^4$ factorial experiment, the cardinality of this set increases to $\binom{81}{10}=1.878392\times 10^{12}$, even for the designs with the smallest possible number of runs. This rapid growth in the size of the search space makes exhaustive searches for common variance designs impossible for anything but small design problems.

In light of the difficulty in finding common variance designs, we introduce a class of approximate common variance (A-ComVar) designs. Instead of having exactly equal variance for the uncommon parameters for the $s$ models under consideration, A-ComVar designs try to {\color{black}maximize} the ratio of the minimum variance to the maximum variance. In doing so, they contain common variance designs as a sub-case where the minimum variance is exactly equal to the maximum variance. In relaxing the requirement that the variances be exactly equal, we are able to develop an objective function and algorithm for identifying these A-ComVar designs without performing an exhaustive search.

\subsection{Proposed Algorithm: Genetic Algorithm for Finding A-ComVar Designs}

In this section we propose to use a genetic algorithm to identify A-ComVar Designs. We start by defining an objective function that seeks to quantify our goal. {\color{black}Denote the variance of the interaction effect for the $i$th model as $\sigma_{\beta_{2i}}^2$ and let $\overline{\sigma_{\beta_{2}}^2} = \frac{1}{s} \sum_{i} \sigma_{\beta_{2i}}^2$. The objective function for designs that discriminate between models with a single interaction term ($k=1$) is:}



\begin{equation}\label{obj_func}
f(d; \phi) = \frac{1 / \overline{\sigma_{\beta_{2}}^2} }{1 + \phi \times \sum_{i=1}^{s} (\sigma_{\beta_{2i}}^2 - \overline{\sigma_{\beta_{2}}^2} )^2 },
\end{equation}

\noindent where $\sigma_{\beta_{2i}}^2$ is replaced by the determinant of the lower-right $k \times k$ sub-matrix of {\color{black}the inverse of the Fisher information matrix} for $k > 1$, which bears some similarity to the idea behind $D$-optimal design of experiments. The value of the objective function increases as the variance of the estimates decreases through the numerator, encouraging designs with small variances for the interaction terms. However, this value is also strongly penalized towards zero as the {\color{black} individual model variances move away from the average model variance}. {\color{black} The strength of this penalty is controlled by the tuning parameter $\phi$, which we recommend setting to a very large value. In our experiments we found $\phi = 10 \times 10^{13}$ to be adequate. The $\phi$ parameter is just to force differences in variance across models under consideration to ``cost" more than the potential variance improvement from a design under some subset of those models, and thus setting it to any suitable large value should suffice.} Taken together the numerator allows us to differentiate between designs with common variance to select the better one, and the denominator encourages common variance designs by penalizing differing variance under alternative models under consideration.

This maximization approach will prefer A-ComVar designs with exactly common variance. Of course, in many experimental situations a common variance design may not exist. For example in the exhaustive search, \cite{Chowdhury2016} found that common variance designs did not exist for $3^3$ experiments for 13 runs. This leads us to the principal advantage of our approach: when a common variance design does not exist we can still find designs with variance that is as close as possible to being equal. To assess the quality of an A-ComVar design, we define the A-ComVar ratio

\begin{equation}
r_{ACV} = \frac{\min\limits_{i}\{ var(\hat{\beta}_{2i})\}}{\max\limits_{i}\{ var(\hat{\beta}_{2i}) \} }.
\end{equation}

\noindent Clearly when a design has common variance, $r_{ACV} = 1$. When a design does not have common variance, $r_{ACV}$ gives us an idea of how far we are from common variance. For example, if $r_{ACV} = 0.5$ then we know that {\color{black} among the models} under consideration, the largest variance of interaction terms is twice that of the smallest. This knowledge can hopefully help inform model selection.

Any off-the-shelf optimization algorithm could be used to try to maximize this objective function. We have chosen to use a genetic algorithm, as is common in the design literature \citep{mandal2015algorithmic, lin2015using}. Genetic algorithms are optimization techniques mimicking Darwin's idea of natural selection and survival of the fittest. This search expects that a good candidate solution will provide good offspring and imitates the way that chromosomes crossover and mutate when reproducing. Here, each chromosome is a design, and the fitness of a chromosome is determined by the corresponding objective function value. At each iteration the worst chromosomes are replaced with offspring generated by combining the settings from two better chromosomes, along with some small probability of a mutation. In the context of our problem, a mutation corresponds to randomly changing the settings for one of the factors in one of the runs. The algorithm terminates when either the maximum number of iterations has been reached, or a design with common variance has been found. The steps in our genetic algorithm are outlined in Algorithm \ref{acvalg}, {\color{black} and the Appendix provides a detailed example of how our algorithm is used.}

The genetic algorithm requires the user to specify {\color{black} the mutation probability, the number of chromosomes to replace at each iteration, and the maximum number of iterations. Our experience with the algorithm suggests using a small mutation probability to encourage only one or two mutations each time a new chromosome is created. Similarly, we have found replacing two chromosomes at each iteration to work for our purposes, and so throughout the remainder of this paper we fix this tuning parameter at two. Finally, we generally use a maximum of 10,000 iterations, although the algorithm is quite fast and this number can easily be increased if needed. } Our algorithm is implemented in Julia version 1.0.2 and is available for download from the author's website.

\begin{algorithm}
	\caption{Pseudo-code for the genetic algorithm to find A-ComVar designs.}
	\label{acvalg}
	\begin{algorithmic}[1]
		\Function{A-ComVarDESIGN}{design problem, mutation prob., num replace, max iter., $\phi$} 
		\For{Each chromosome}
		\State initialize chromosome to random design
		\State Calculate fitness 
		\EndFor
		\While{termination criteria not met}
		\State Identify worst {\it num replace} chromosomes
		\State Use a crossover to generate {\it num replace} new chromosomes
		\State Mutate the {\it num replace} new chromosomes
		\State Replace the worst chromosomes with the {\it num replace} new chromosomes
		\State Calculate fitness for new chromosomes
		\EndWhile
		\EndFunction
	\end{algorithmic}
\end{algorithm}

    \section{Numerical Examples}

\subsection{Example 1 $-$ Designs with One $2$-Factor Interaction}

We conducted a series of experiments to investigate the ability of our approach to find A-ComVar designs and to gain a better understanding of when common variance designs can be found. We started by examining designs with a single $2$-factor interaction. We consider $2^{m_1}$ and $3^{m_2}$ experiments, with $m_1 = 4, \ldots, 9$ and $m_2 = 3, \ldots, 6$ and consider the situation where all factors are present in the model as main effects. For the $2^{m_1}$ experiments, we considered run sizes of $n_{m_1} = m_1 + 2, \ldots, m_1 + 11$, and for the $3^{m_2}$ experiments, we considered run sizes of $n_{m_2} = 2m_1 + 2, \ldots, 2m_1 + 11$. For each combination of settings, we ran our genetic algorithm {\color{black} 100} times and stored the $r_{ACV}$ results. The tuning parameters used were a mutation probability of $0.05$ and a maximum of {\color{black} 10,000} iterations.

\begin{figure}
	\centering
	\includegraphics[width=1\linewidth]{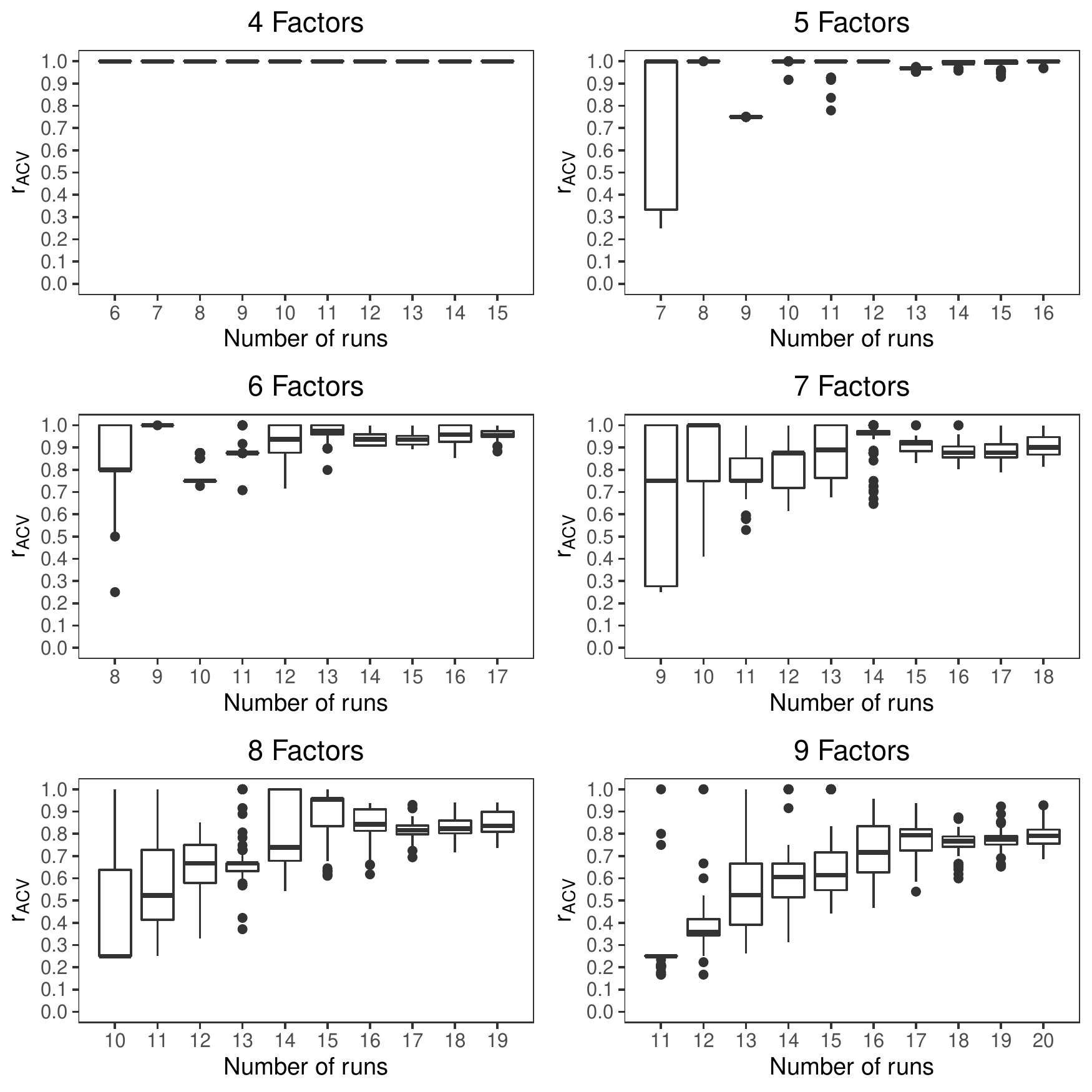}
	\caption{Ratios $r_{ACV}$ for the $2^{m_1}$ case across 1000 replicates for each experimental setting.}
	\label{fig:2levelresults}
\end{figure}

\begin{figure}
	\centering
	\includegraphics[width=1\linewidth]{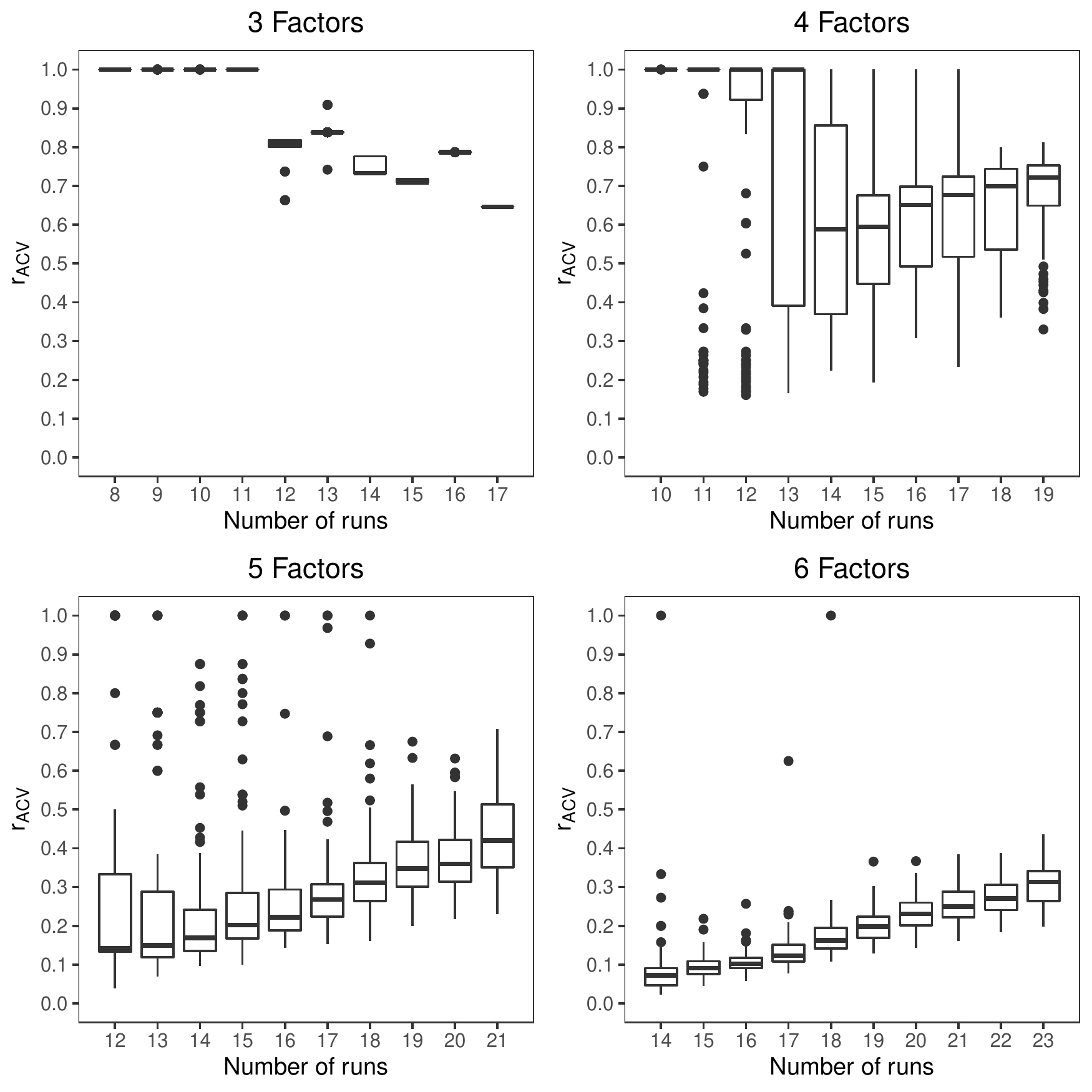}
	\caption{Ratios $r_{ACV}$ for the $3^{m_2}$ case across 1000 replicates for each experimental setting.}
	\label{fig:3levelresults}
\end{figure}

Figure \ref{fig:2levelresults} displays the results for the $2^{m_1}$ cases and Figure \ref{fig:3levelresults} displays the results for the $3^{m_2}$ cases. We first note that our results are consistent with the findings of  \cite{ghosh2017cv}, who used exhaustive searches to identify common variance designs. For example, \cite{ghosh2017cv} found that common variance designs exist for $3^3$ designs with 8 runs, which agrees with the boxplots in the first panel of Figure \ref{fig:3levelresults}. This supports our use of the genetic algorithm approach with the objective function described above. Furthermore, in cases where the common variance designs either do not exist or could not be found, our approach was able to find designs that attempt to get as close as possible to common variance. For example, it is known from exhaustive searches that no common variance design exists for a $3^3$ experiment with 12 runs. However, the proposed approach was able to find designs where the smallest variance was greater than $0.8$ times the largest variance, indicating that the design is quite close to having the common variance property.

\begin{figure}
	\centering
	\includegraphics[width=1\linewidth]{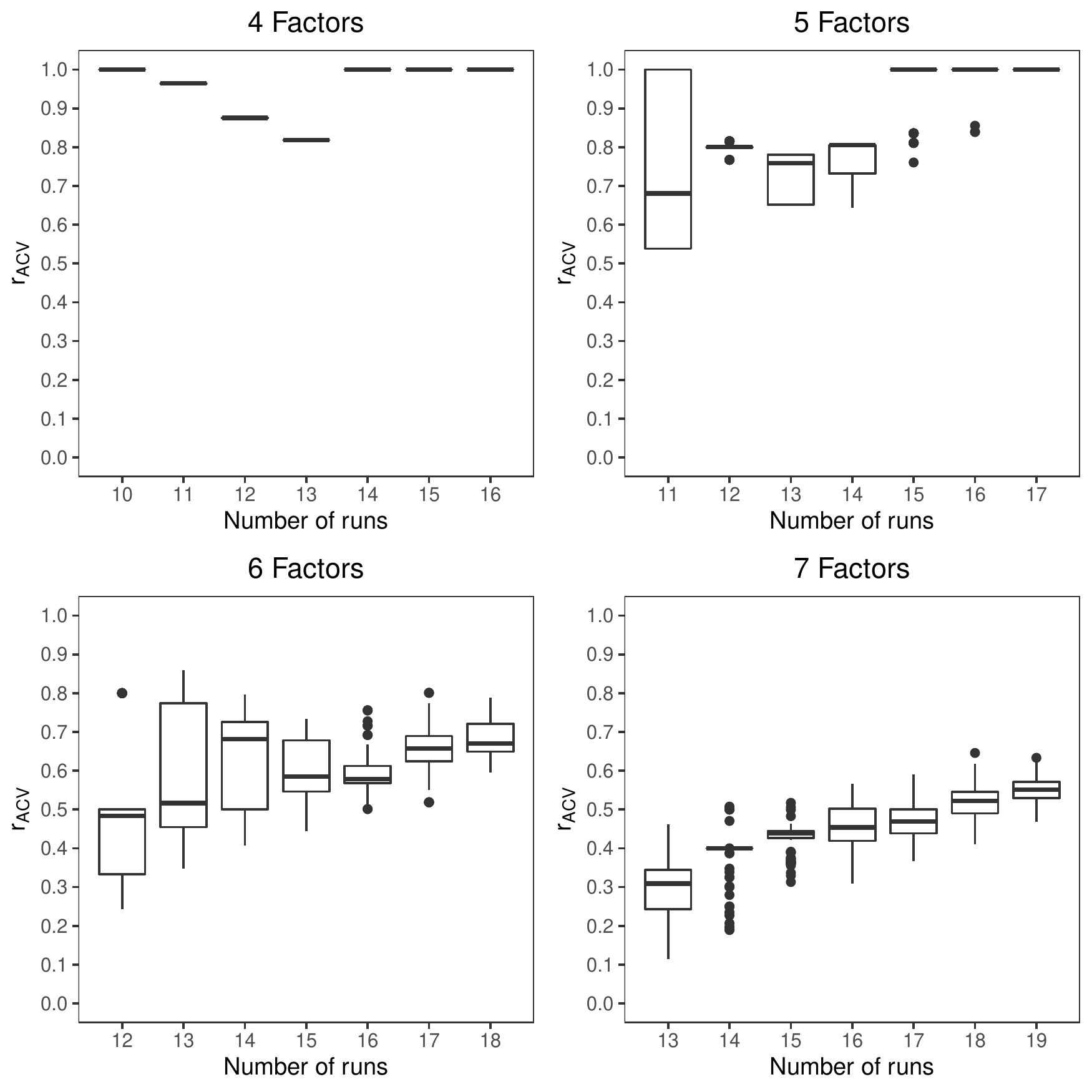}
	\caption{Ratios $r_{ACV}$ for the $2^{m_1}$ case across 1000 replicates for each experimental setting with $k=2$.}
	\label{fig:2levelresultsk2}
\end{figure}


\subsection{Example 2 $-$ Designs with Two $2$-Factor Interactions}
For designs with multiple $2$-factor interactions (i.e. $k > 1$), we generalize the objective function in (\ref{obj_func}) by replacing $var(\hat{\beta}_{2i})$ with the determinant of the block of the {\color{black} inverse of the Fisher information matrix} corresponding to the interactions terms. That is, we take the determinant of the bottom-right $k \times k$ sub-matrix of $var(\hat{\beta})$. 

To demonstrate the approach, we conducted another experiment with two $2$-factor interactions (i.e. $k=2$). We consider $2^{m_3}$ experiments, with $m_3 = 4, \ldots, 7$ and assume $p=m_3$. We considered run sizes of $n_{m_1} = m_3 + 6, \ldots, m_3 + 12$. For each combination of settings, we ran our genetic algorithm {\color{black} 100} times and stored the $r_{ACV}$ results. The tuning parameters used were a mutation probability of $0.05$ and a maximum of {\color{black} 10,000} iterations.

Figure \ref{fig:2levelresultsk2} shows the results. As before, we can see that in many cases the genetic algorithm is able to find common variance designs. In cases where common variance designs cannot be found, the approach is often able to identify a design resulting in relatively close to common variance. 


\subsection{Example 3 $-$ {\color{black} Common Variance Design and A-ComVar Designs for Model Selection }} \label{sim3pbsims}

{\color{black} We next perform a series of studies to demonstrate the advantages of pursuing A-ComVar designs. We do this by considering data generated from a variety of true models and testing whether a model selection procedure is able to identify the true model using observations collected using the designs under consideration. We used the adaptive lasso \citep{kane2019} to fit the model. We chose the adaptive lasso method of  \cite{kane2019} because they showed that this technique is suitable for identifying the correct model for designs with complex aliasing and that it outperforms other popular variable selection methods including the Dantzig Selector \citep{candes2007dantzig}, LARS \citep{yuan2007efficient}, and the Nonnegative Garotte estimator \citep{breiman1995better, yuan2009structured}. }

{\color{black} Our procedure is as follows. For a model with $p$ active main effects, let $F_1, \ldots, F_p$ denote the active factors, which are selected at random from the set of all factors of the designs at each replication. The corresponding effects, $\beta_1, \ldots, \beta_p$, as well as any interaction effects, are set to be either ``big" or ``small," where ``big" effects are drawn from a $U(1.5, 2.5)$ distribution and ``small" effects from a  $U(0.1, 0.3)$. Finally, the error standard deviation, $\sigma$, is chosen, completing the specification of the true underlying model. The total number of different models, effect sizes, and error standard deviations considered can be found in any of Tables \ref{table:sim1}$-$\ref{table:sim3}. The first column in each table corresponds to the true model under consideration, and the second column gives information about the strengths of the active effects (b $-$ ``big" and s $-$ ``small"). For example, row 25 of Table \ref{table:sim3} corresponds to a model with three active main effects ($F_1$, $F_2$, and $F_3$) as well as one active interaction ($F_1 F_3$). Here, the second column tells us that $F_1$ and $F_2$ have ``big" effects and $F_3$ and $F_1 F_3$ have ``small" effects.}

{\color{black} Next, for each design under consideration, a data set is generated from the true underlying model using the randomly selected factors of the design. A model is fit to this data set using the adaptive lasso, and we measure whether or not the true underlying model was identified. This process is then repeated 100 times for the same set of true active coefficients, and we store the percentage of the times the correct model was identified.}

{\color{black} For each model, design, and error standard deviation under consideration, this process of randomly selecting active factors in the model, generating observations from the design, and measuring how often the correct model is identified is repeated 50 times, resulting in 50 replicates per combination of settings. Here each replicate is a measurement of the percentage of times the data obtained using the design was able to correctly identify the true underlying model. Table \ref{table:sim3designs} displays a list of the model comparisons we made. In Tables \ref{table:sim1}$-$\ref{table:sim3} we report the average percentage of times (over 50 replications) the correct model was identified by the respective designs.}

\begin{table}
\centering
    \begin{tabular}{ccc} \hline
         Description & Design 1 & Design 2 \\ \hline
         \multirow{5}{*}{$2^5$ experiment with 12 runs} & Common Variance (Table \ref{table:design1}) & Plackett-Burman (Table \ref{table:design1}) \\
         & A-ComVar (Table \ref{table::acvdesigns}, $D^1$) & Plackett-Burman (Table \ref{table:design1})\\
         & A-ComVar (Table \ref{table::acvdesigns}, $D^1$) & Ghosh \& Tian (Table \ref{table::designexamples}, $D^6$)\\
         & A-ComVar (Table \ref{table::acvdesigns}, $D^1$) & Bayes Optimal (Table \ref{table::designexamples}, $D^4$)\\
         & A-ComVar (Table \ref{table::acvdesigns}, $D^1$) & Li \& Nachtsheim (Table \ref{table::designexamples}, $D^5$) \\ \hline
        $3^4$ experiment with 20 runs  & A-ComVar (Table \ref{table::acvdesigns}, $D^2$) & CCD (Table \ref{sim::3levelcompet}, $D^7$) \\ \hline
        $3^7$ experiment with 18 runs & A-ComVar (Table \ref{table::acvdesigns}, $D^3$) & OME (Table \ref{sim::3levelcompet}, $D^8$)\\ \hline
    \end{tabular}
\caption{Description of the designs used for Example 3. The tables listed in parenthesis are where the corresponding design is presented.}
\label{table:sim3designs}
\end{table}

{\color{black} The results with model $\times$ variance $\times$ design breakdown can be found in Tables \ref{table:sim1}$-$\ref{table:sim3} in the Appendix. Figures \ref{fig::simresults1}$-$\ref{fig::simresults3} present boxplots of the results for each standard deviation level, stratified by the number of interactions in the model (0, 1, or 2). From the figures we can see that while the common variance design outperforms Plackett-Burman design for all three types of models, a few general patterns are observed for A-ComVar designs. First, when the model contains only main effects and no interactions, the A-ComVar designs generally perform about as well as the competitor designs. This is important, as it suggests that there is not a strong disadvantage to seeking such designs in practice. Next, examining the boxplots for the models with one and two interactions, we can see that the A-ComVar designs generally outperform the other designs, especially for the models with two interactions. The exception to this pattern is the Ghosh and Tian design, which seems to outperform the A-ComVar design in several cases. This is likely because the \cite{ghosh2006optimum} design is optimal w.r.t all six standard optimality criteria, and thus it is hard to beat its performance. However, designs of this quality cannot always be obtained for arbitrary numbers of factors or runs, thus one advantage of our numerical approach is that it can be used for cases where such designs cannot be obtained via exhaustive search or by using theoretical results.}

\begin{figure}
\caption{Comparison of model selection performance of two-level common variance design $d_5^{(12)}$ design with 5 factors and 12 runs presented in Table 2 with PB design presented in Table 3. 
Here \protect\includegraphics[width=.2in,height = 0.02\textheight]{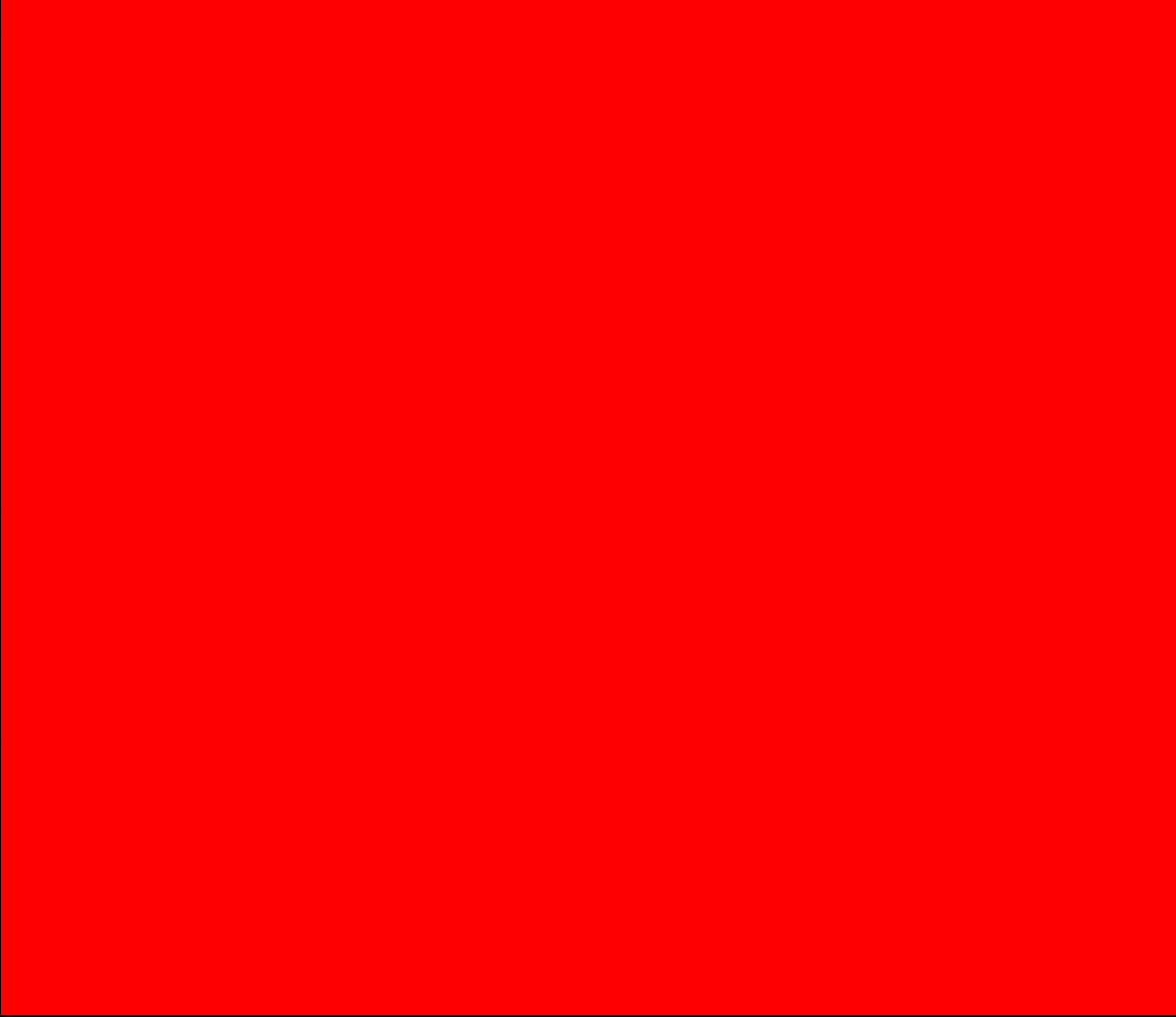} represents the common variance design $d_5^{(12)}$ and \protect\includegraphics[width=.2in,height = 0.02\textheight]{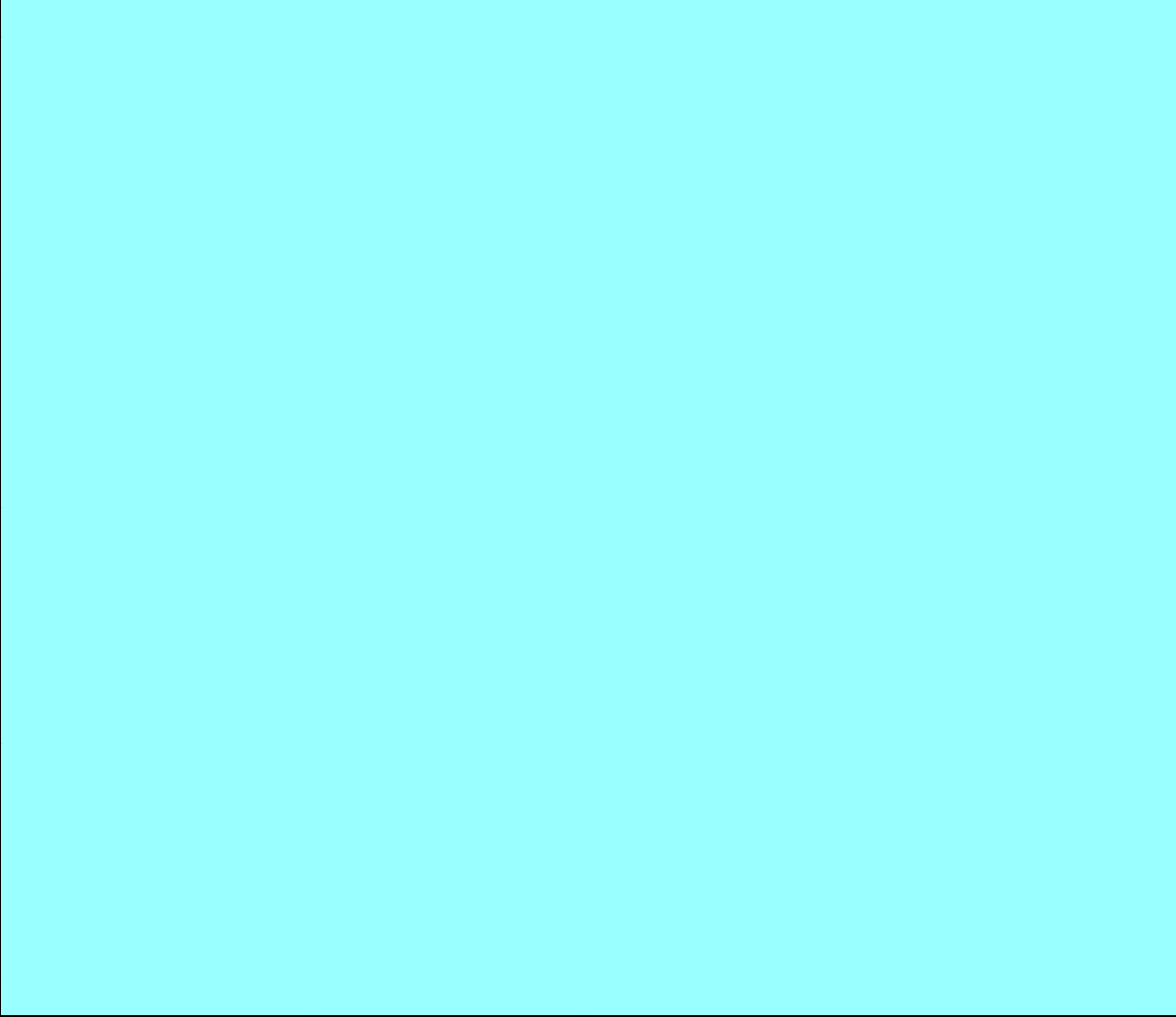} represents Plackett-Burman design.}
\label{fig::simresults1}
\centering
\begin{tabular}{ccc}
\multicolumn{3}{c}{Common Variance vs Plackett-Burman} \\
\includegraphics[scale=.22]{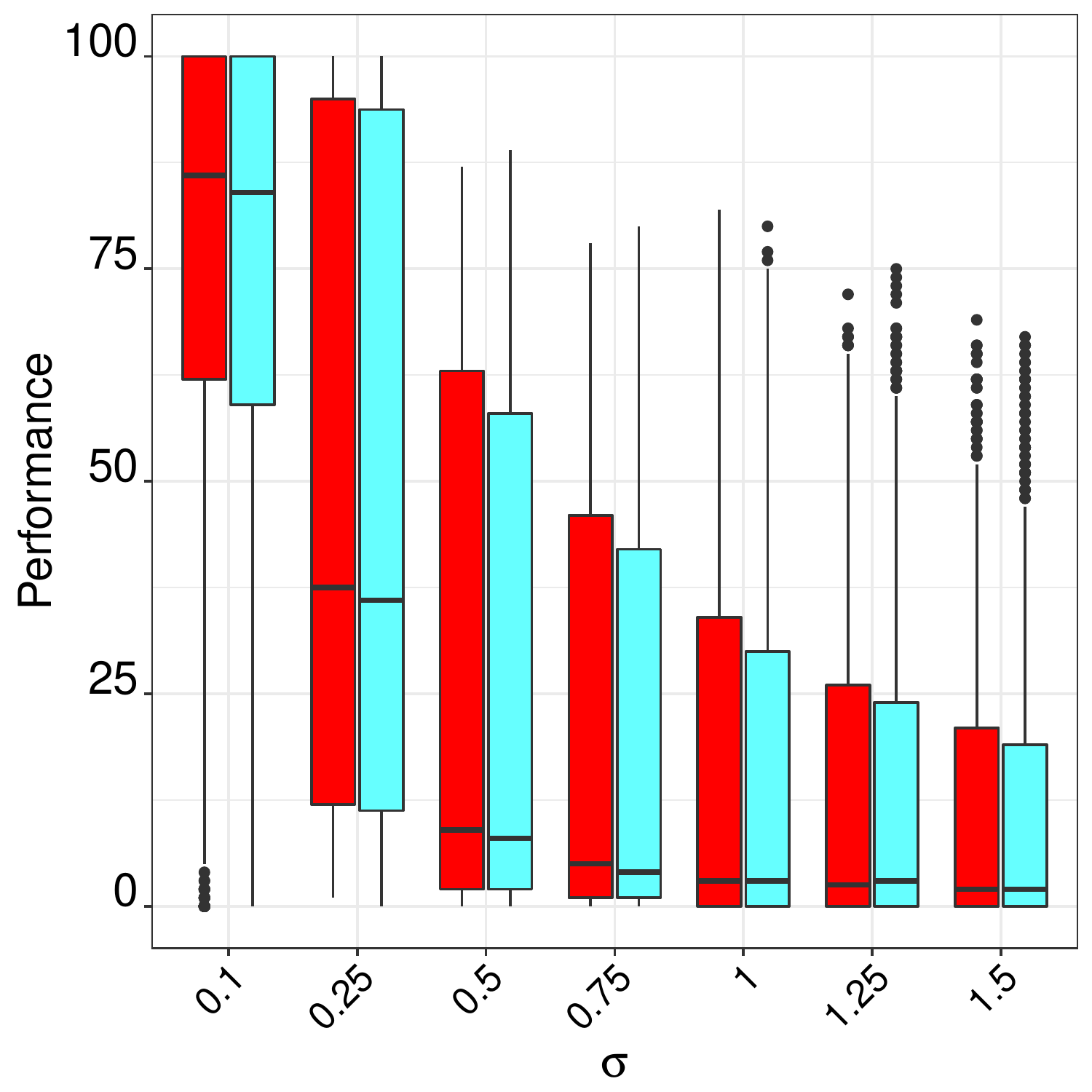} &
\includegraphics[scale=.22]{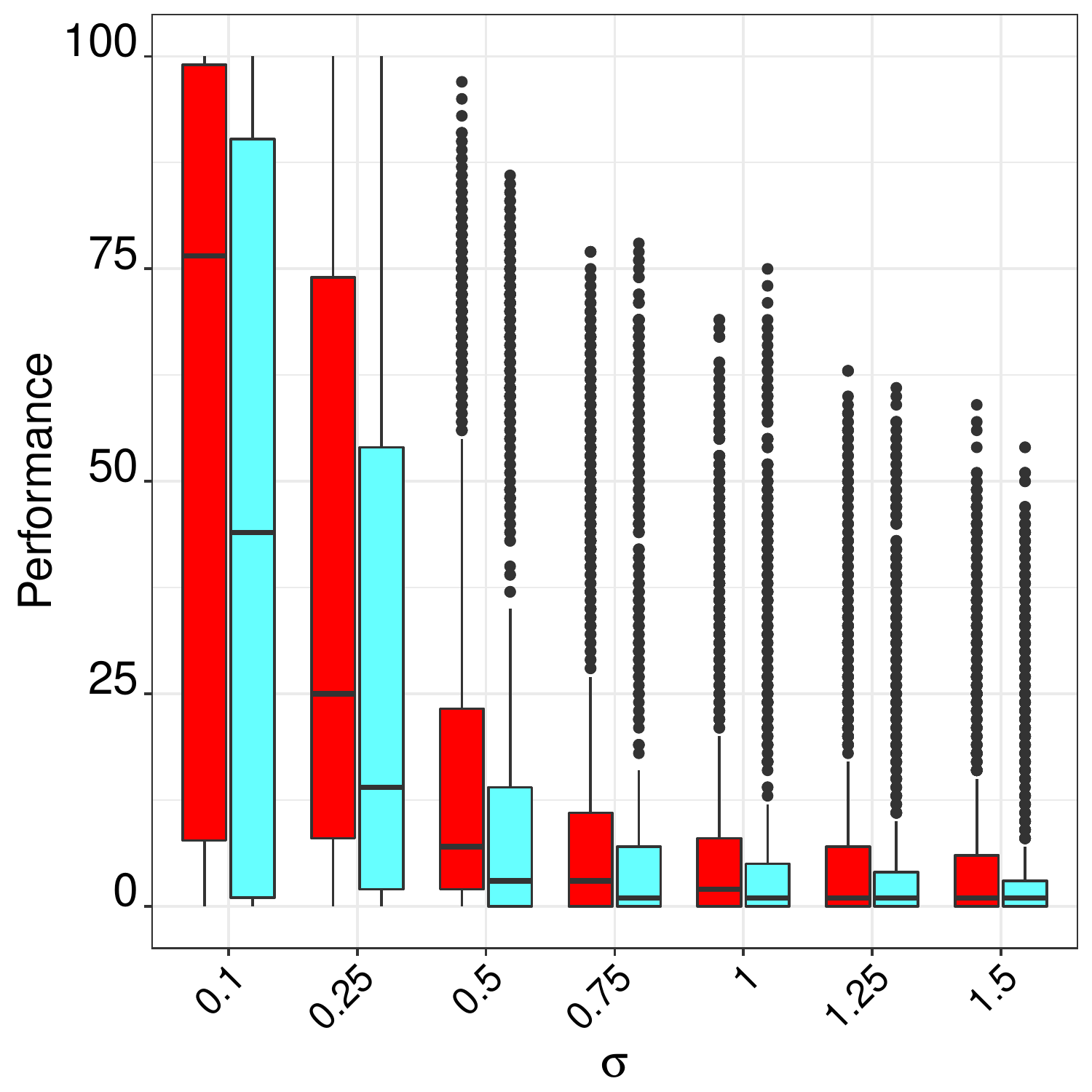} &
\includegraphics[scale=.22]{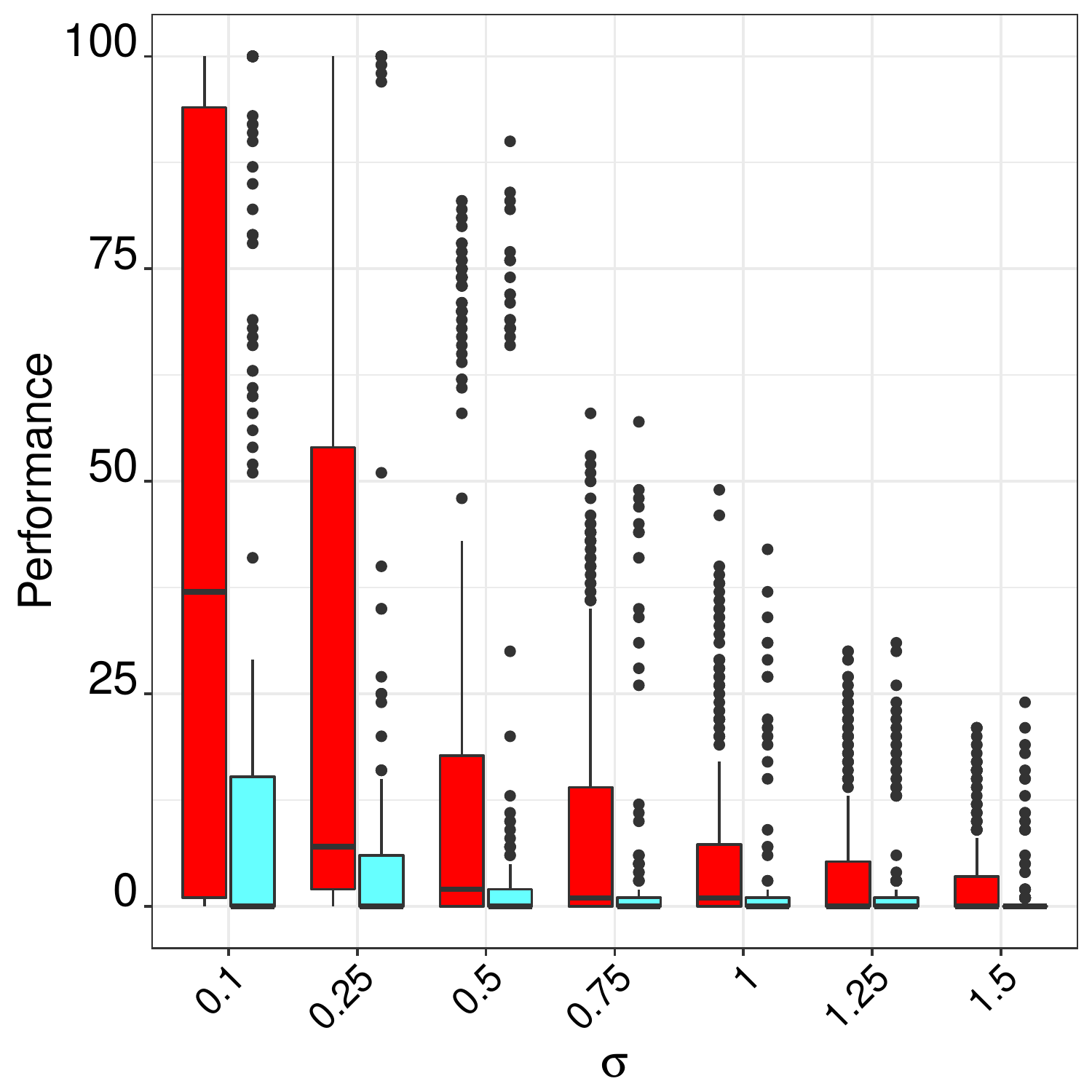}
\end{tabular}
\end{figure}

\begin{figure}
\caption{Comparison of model selection performance of two-level A-ComVar design with 5 factors and 12 runs with other designs. 
Here \protect\includegraphics[width=.2in,height = 0.02\textheight]{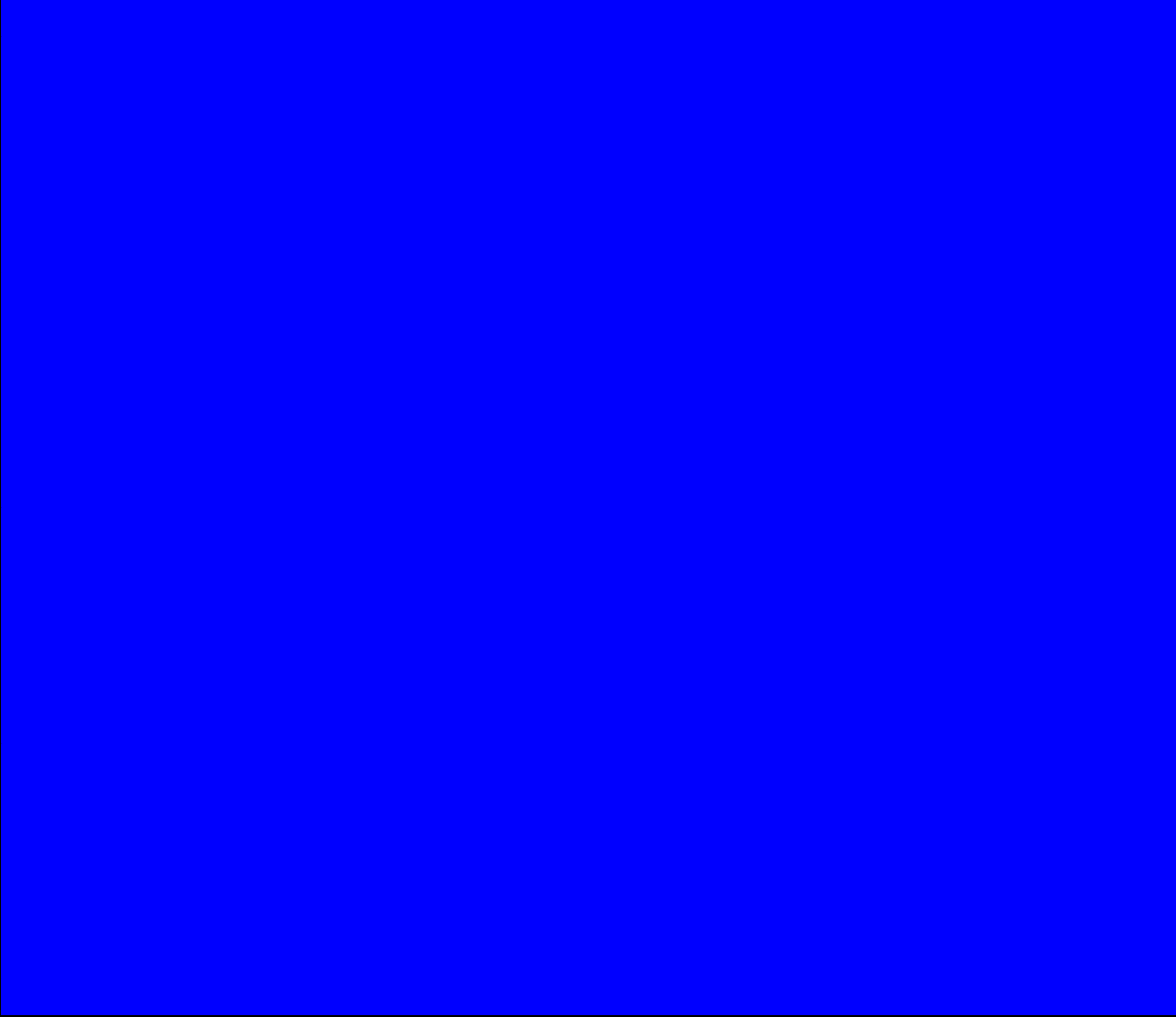}  represents our A-ComVar design with 2 levels, 
\protect\includegraphics[width=.2in,height = 0.02\textheight]{lightblue.pdf}  represents Placket-Burman design,
\protect\includegraphics[width=.2in,height = 0.02\textheight]{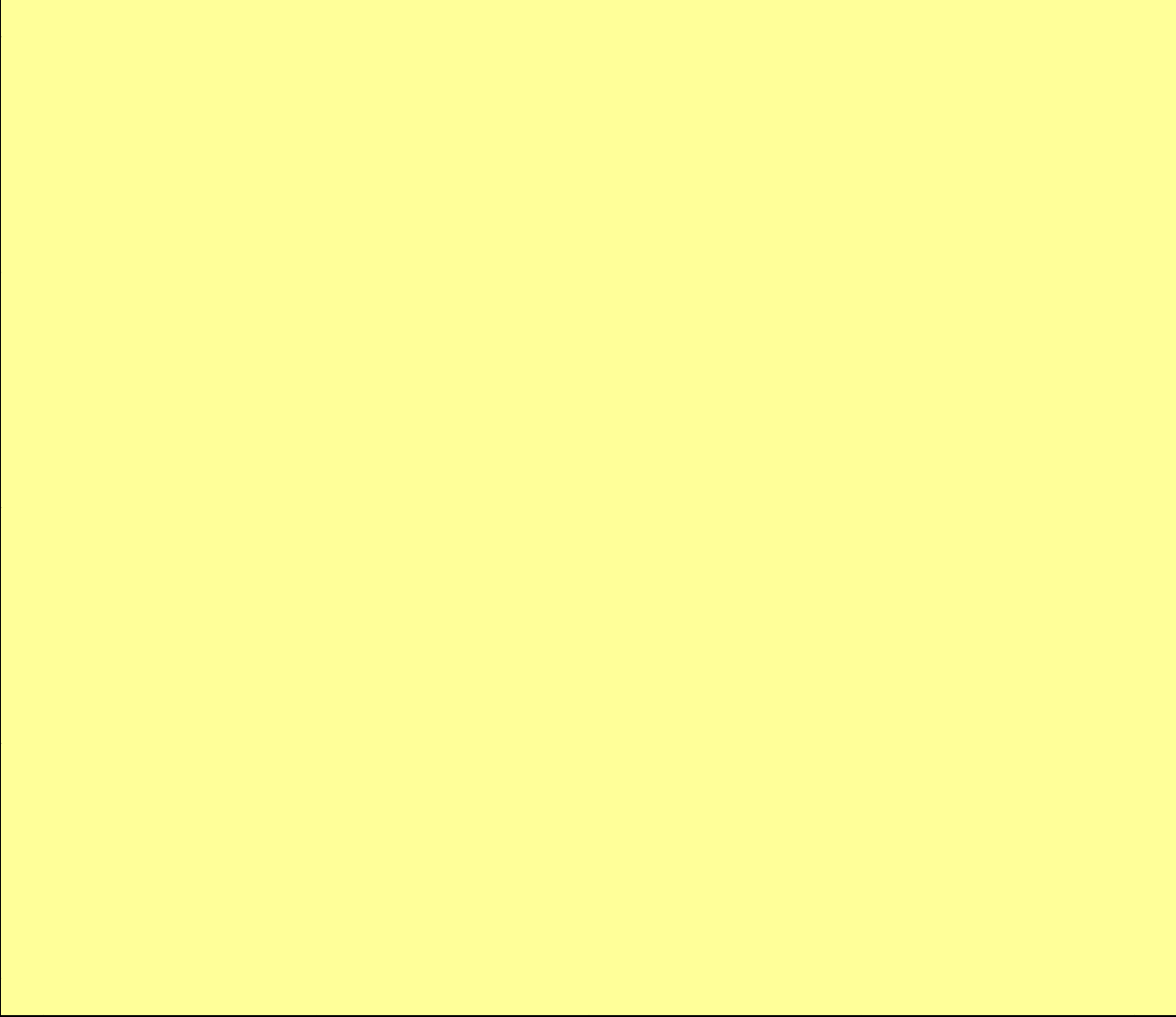}  represents a two-level design with 5 factors and 12 runs from \cite{ghosh2006optimum}, \protect\includegraphics[width=.2in,height = 0.02\textheight]{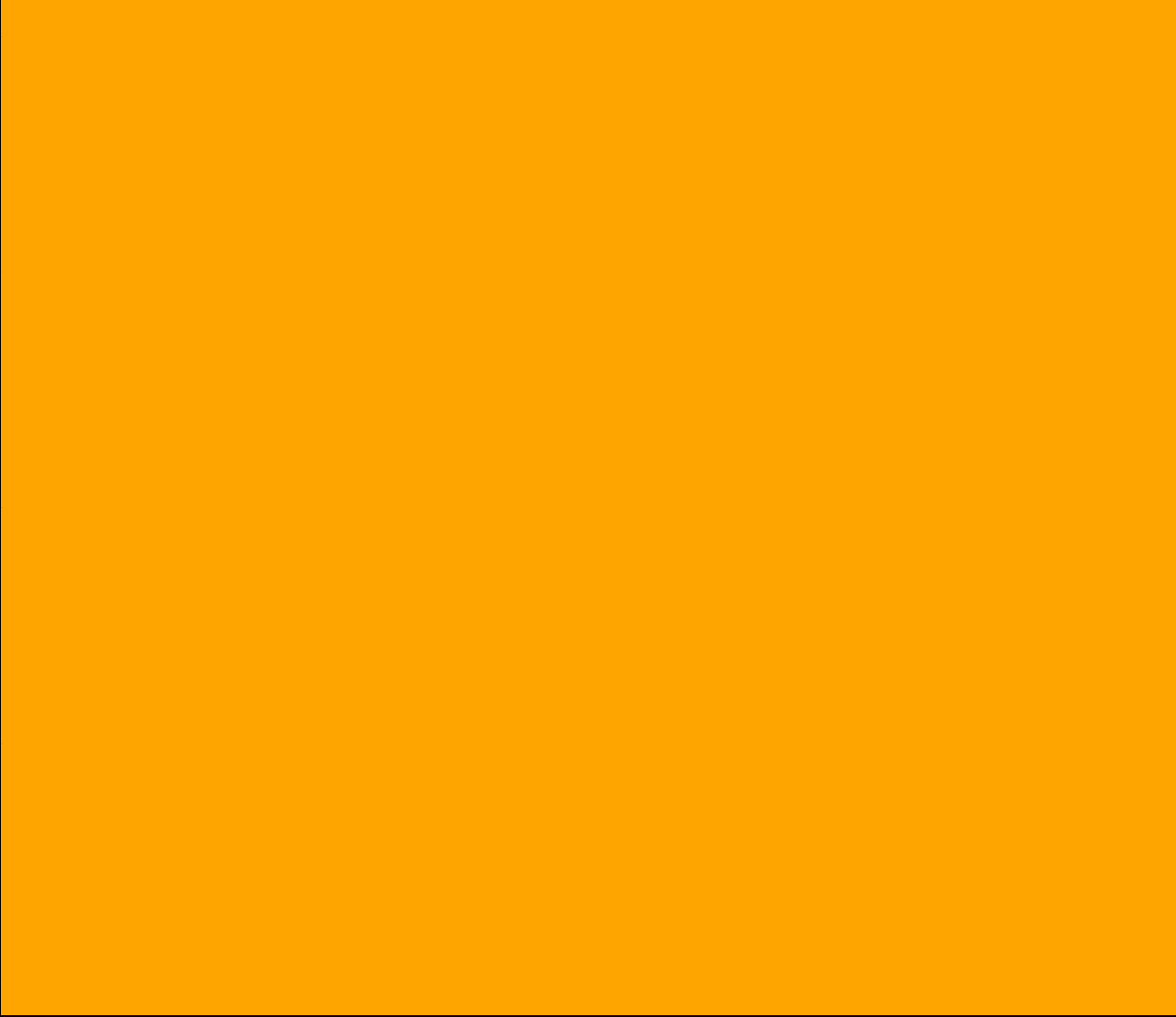}  represents a two-level Bayes optimal design with 5 factors and 12 runs from \cite{bingham2007incorporating},
\protect\includegraphics[width=.2in,height = 0.02\textheight]{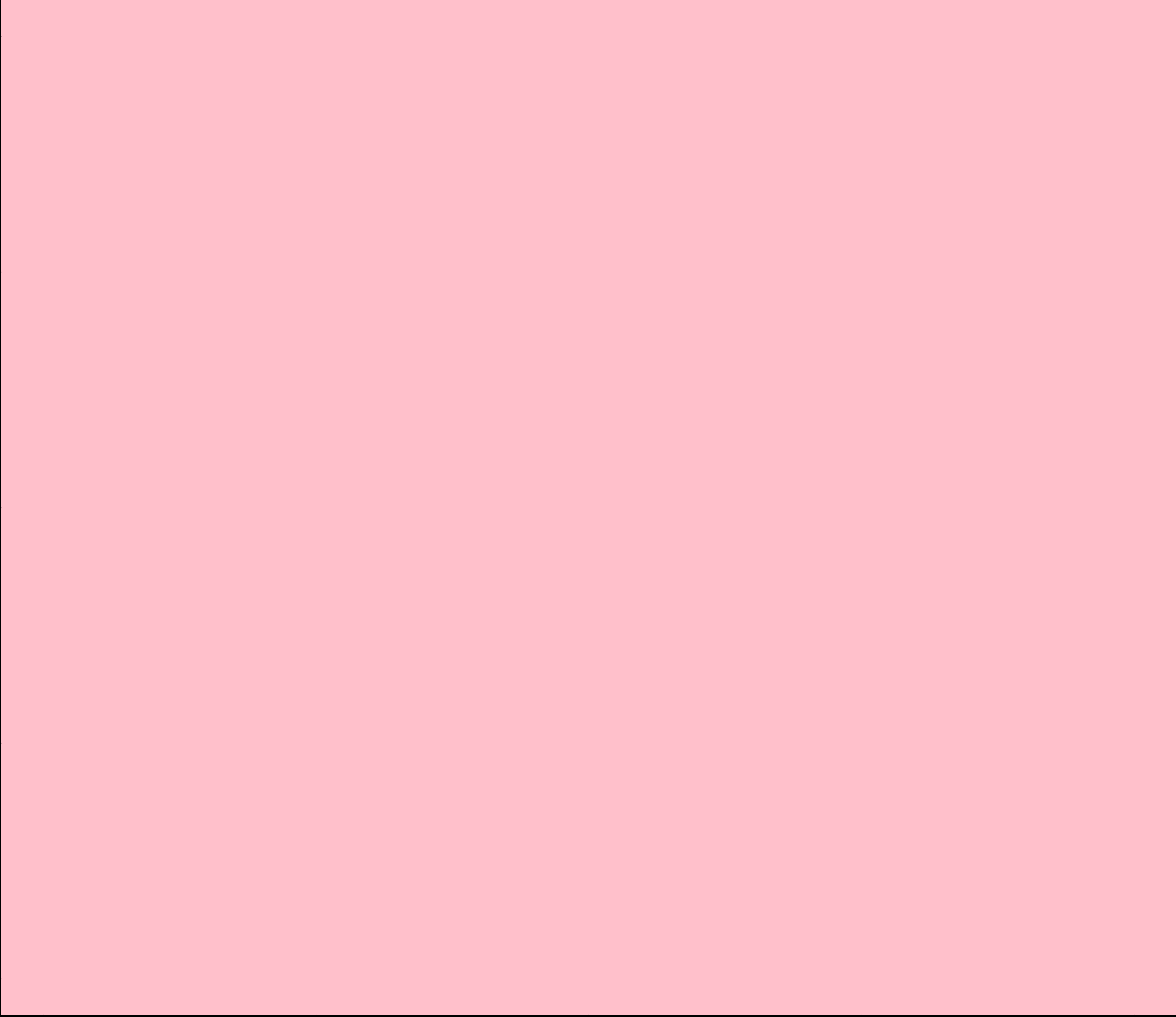}  represents a two-level design with 5 factors and 12 runs from \cite{li2000model}.}
\label{fig::simresults2}
\centering
\begin{tabular}{ccc}
\multicolumn{3}{c}{A-ComVar vs Placket-Burman} \\
\includegraphics[scale=.22]{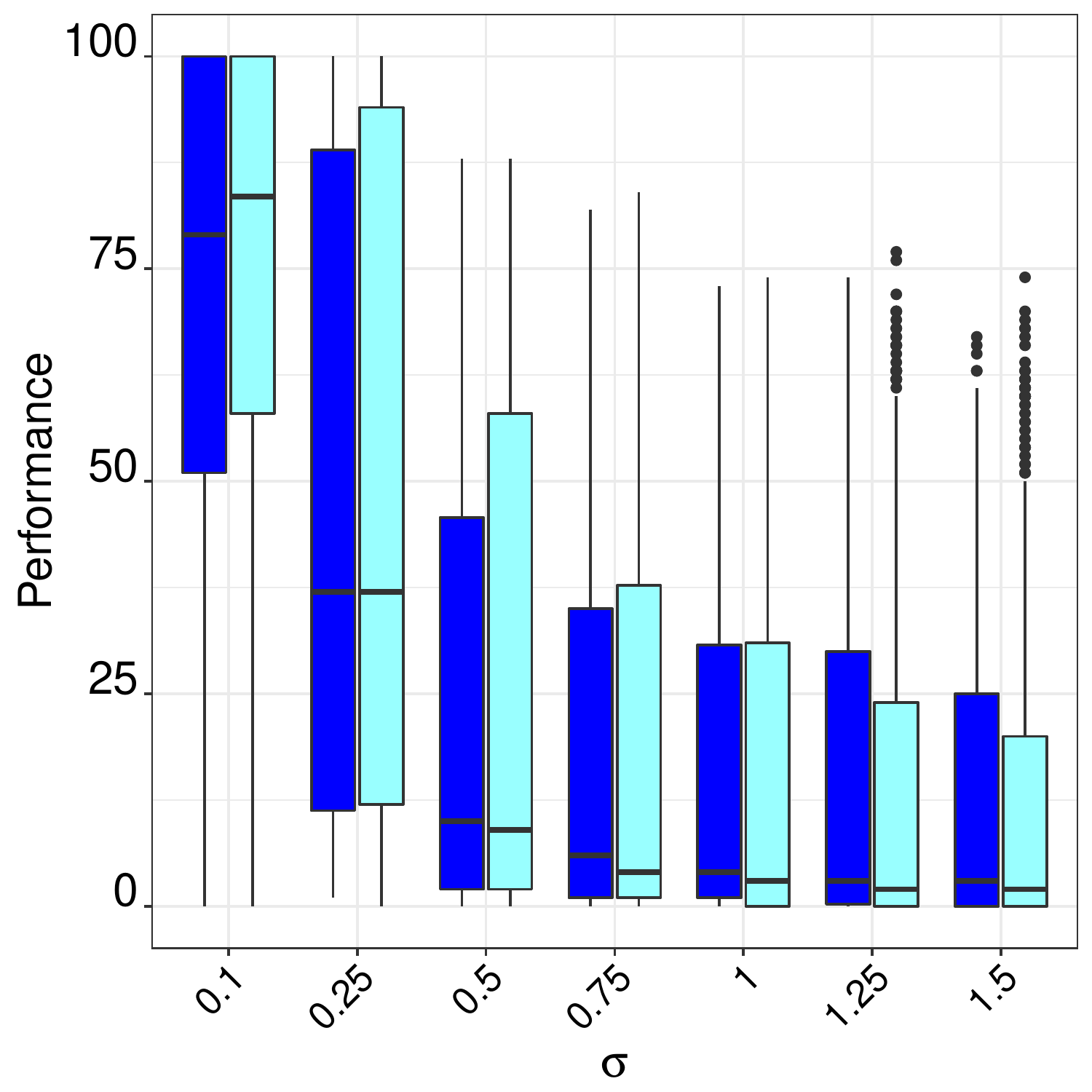} &
\includegraphics[scale=.22]{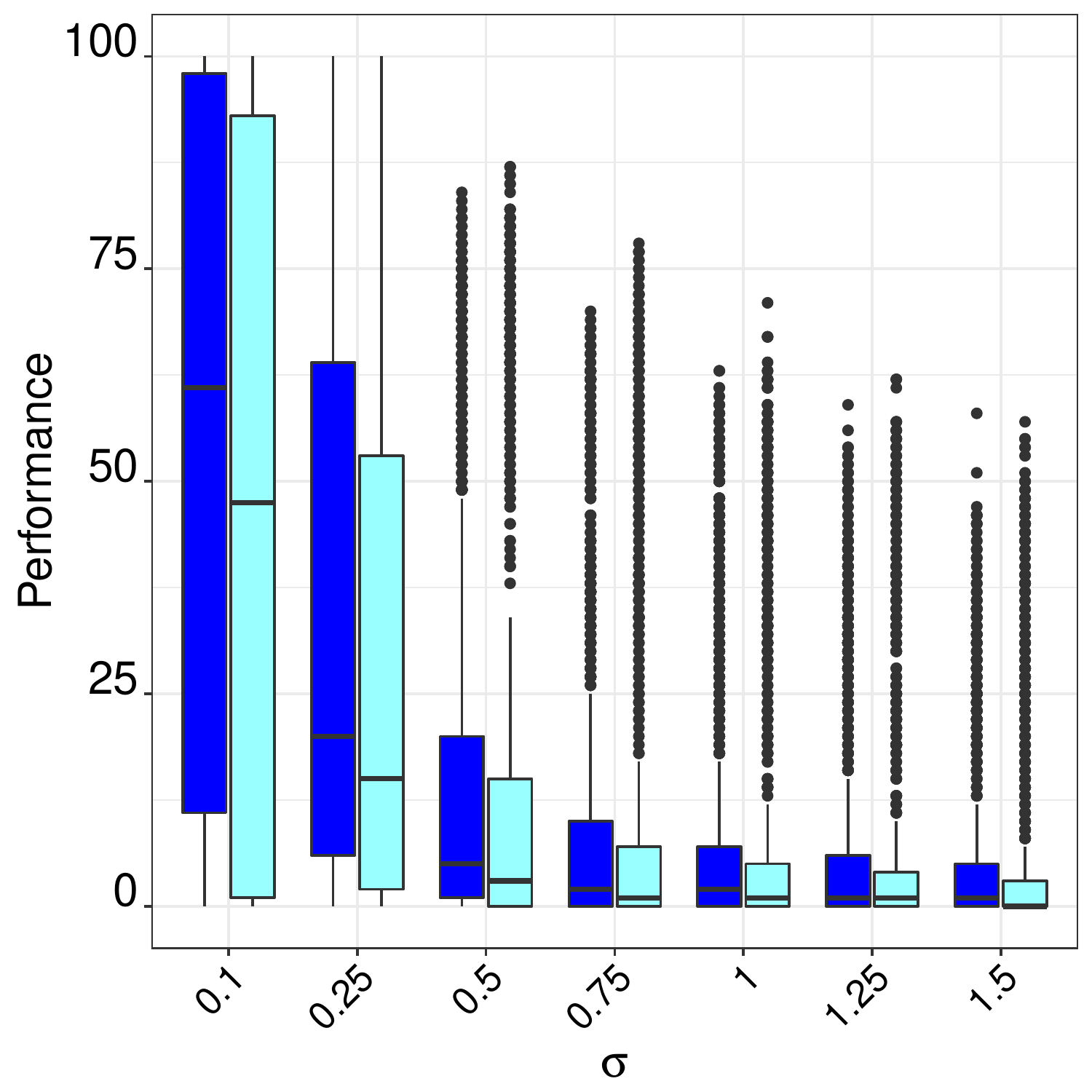} &
\includegraphics[scale=.22]{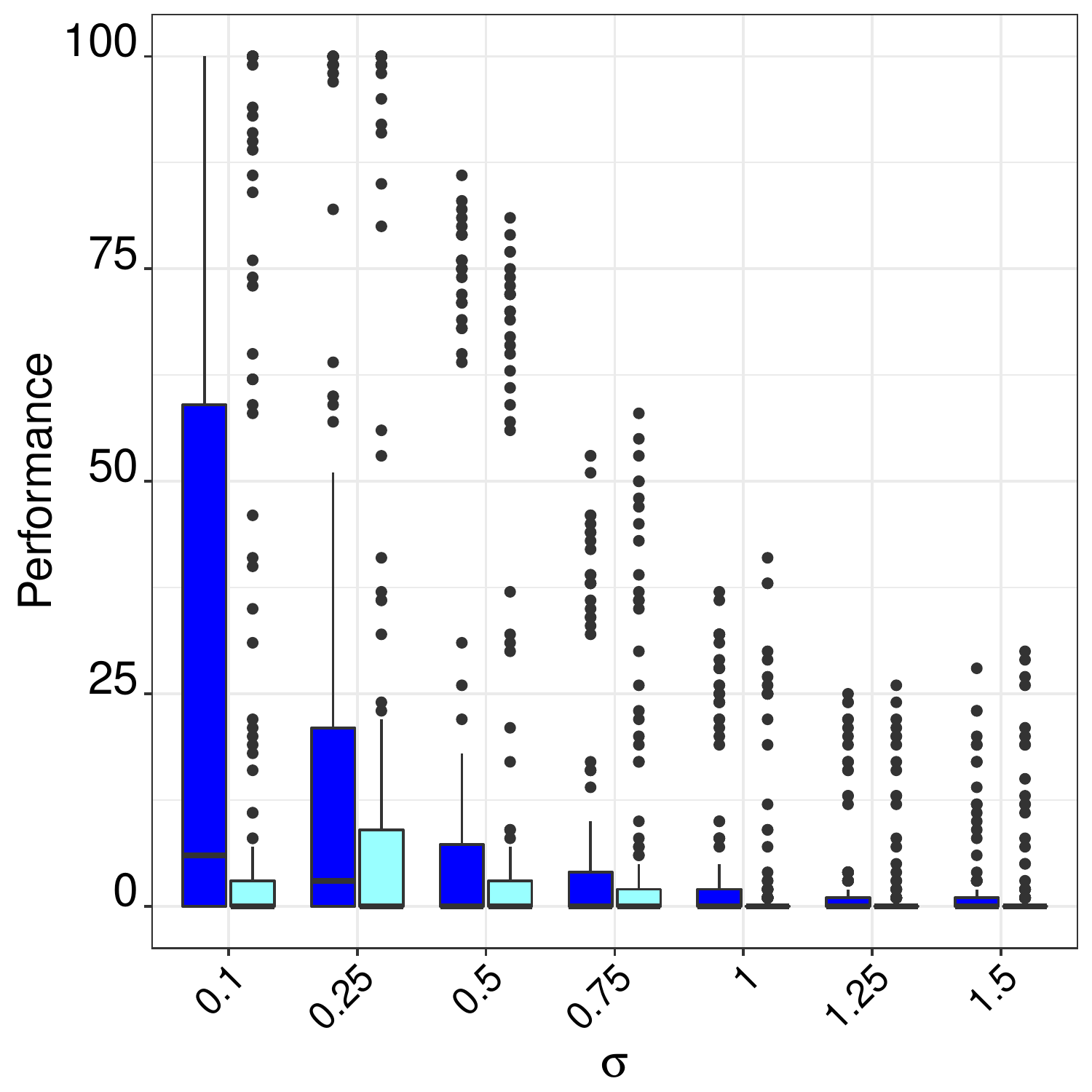} \\

\multicolumn{3}{c}{A-ComVar vs Ghosh and Tian (2007)} \\
\includegraphics[scale=.22]{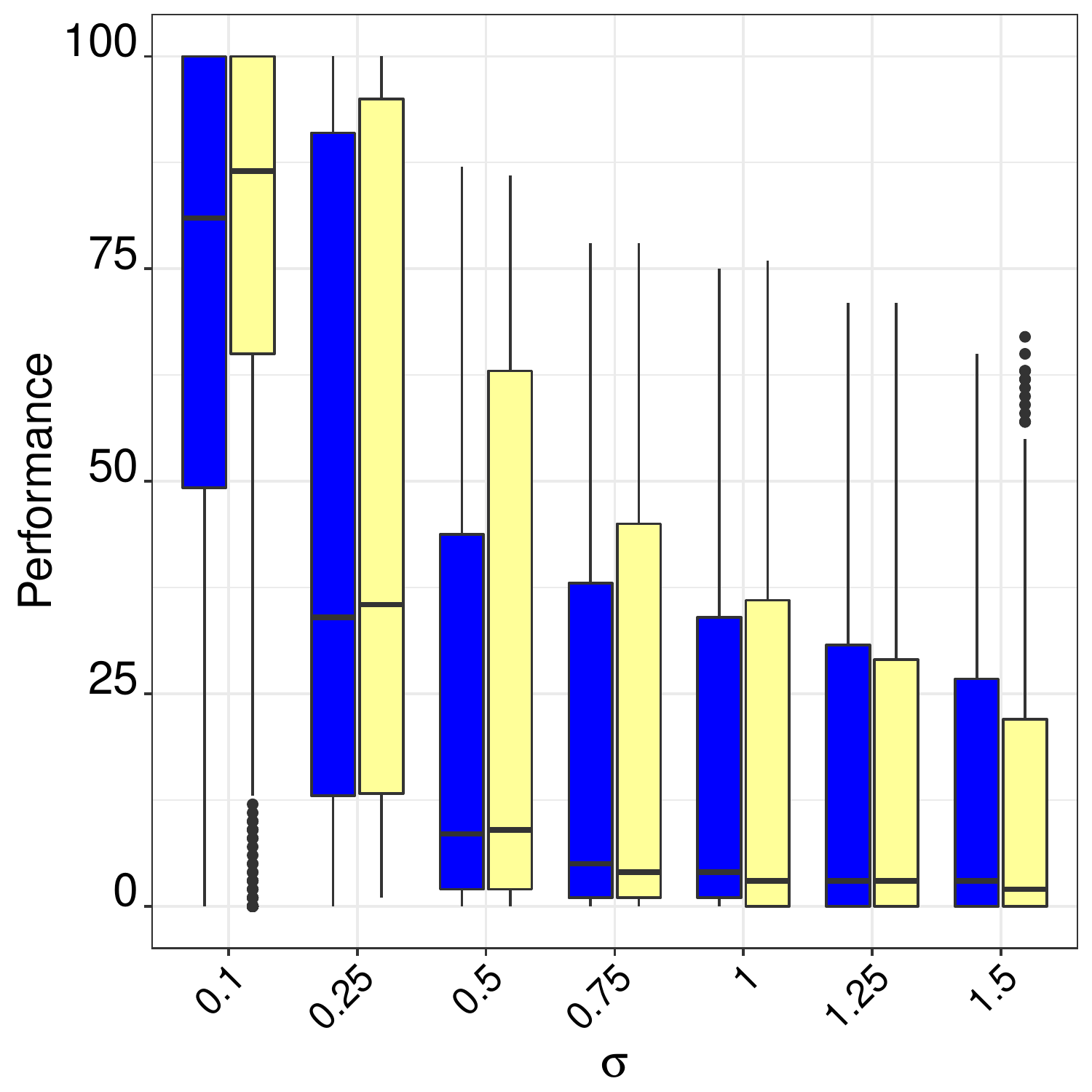} & 
\includegraphics[scale=.22]{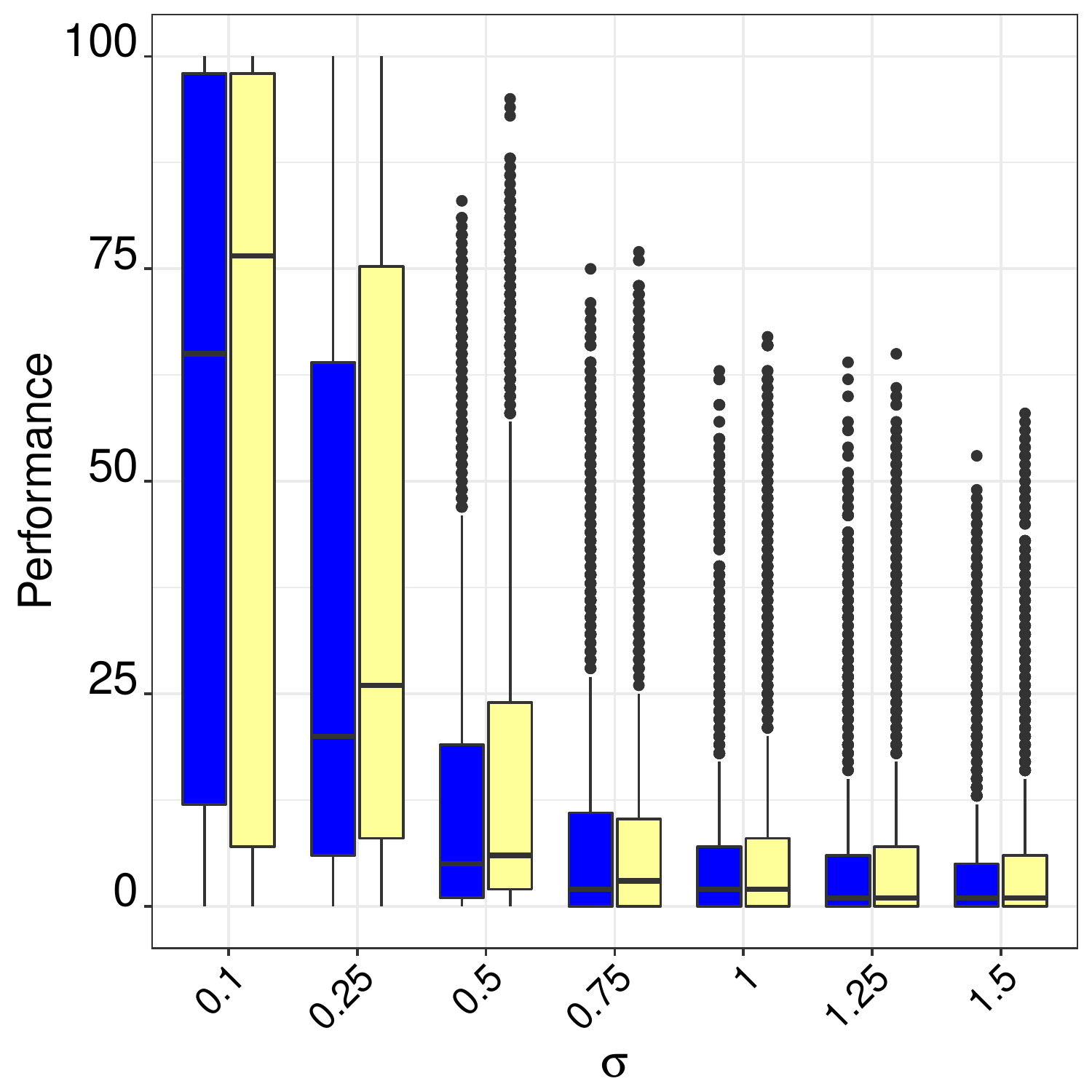} & 
\includegraphics[scale=.22]{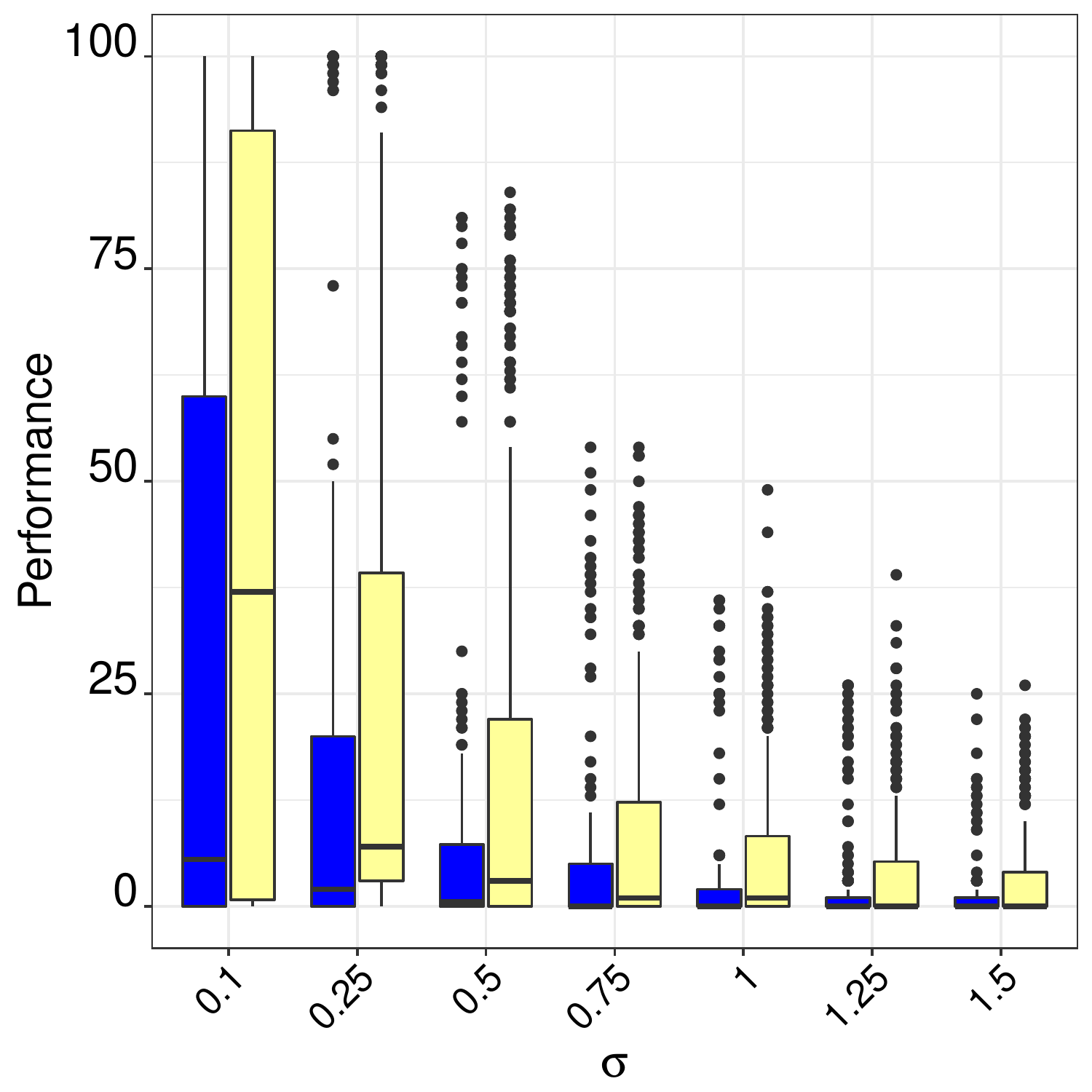} \\

\multicolumn{3}{c}{A-ComVar vs Bayes Optimal \citep{bingham2007incorporating}} \\
\includegraphics[scale=.22]{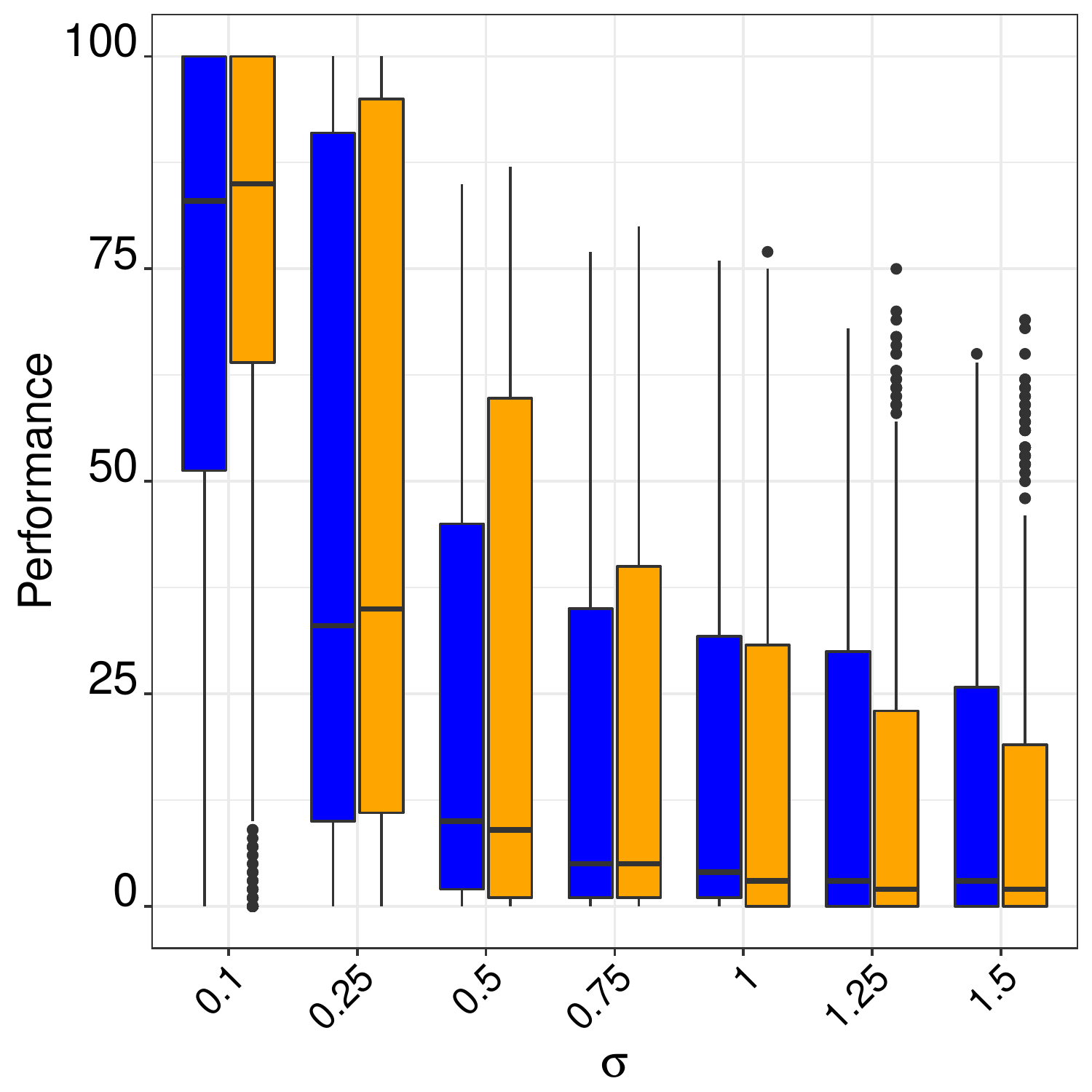} &
\includegraphics[scale=.22]{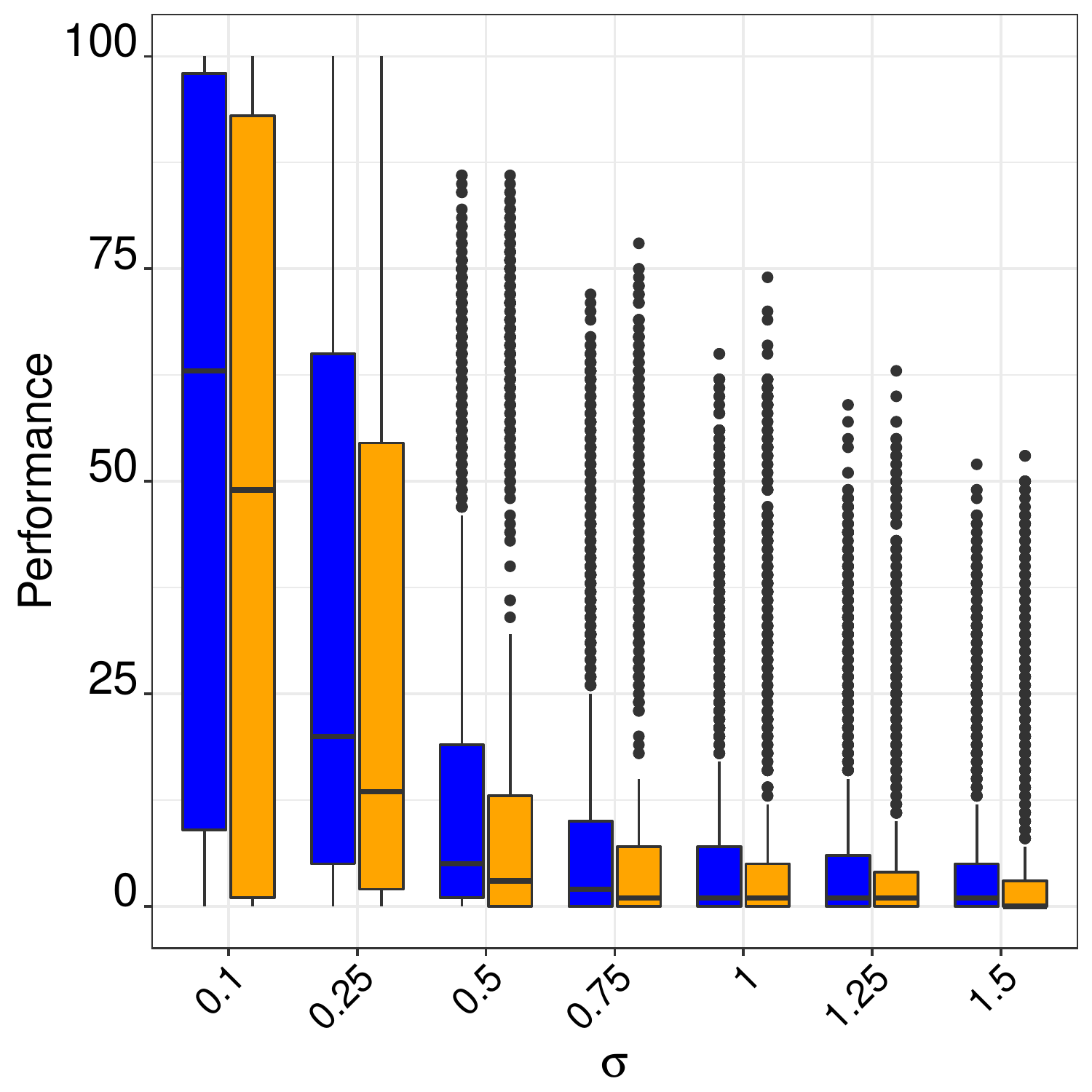} &
\includegraphics[scale=.22]{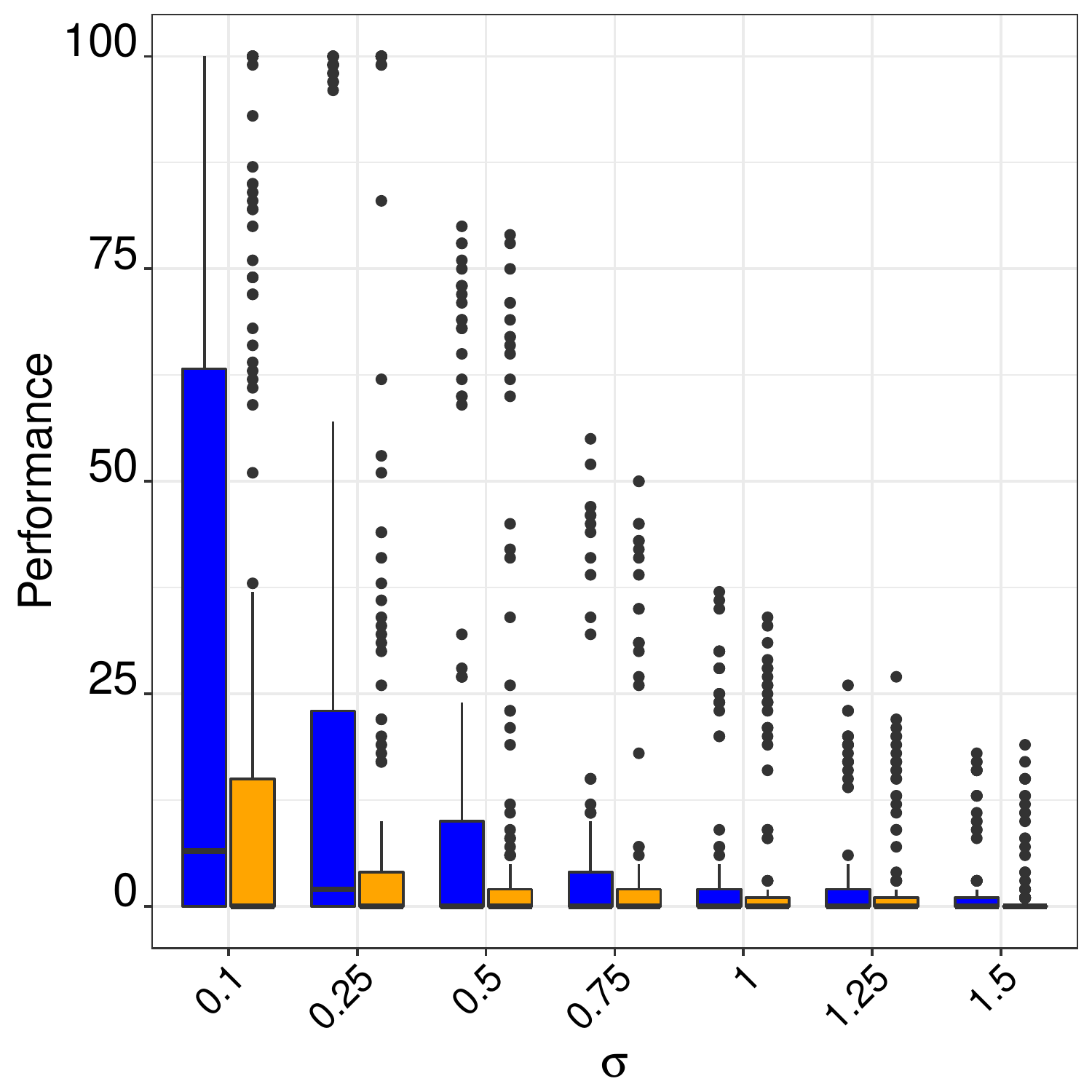} \\

\multicolumn{3}{c}{A-ComVar vs Li and Nachtsheim (2000)} \\
\includegraphics[scale=.22]{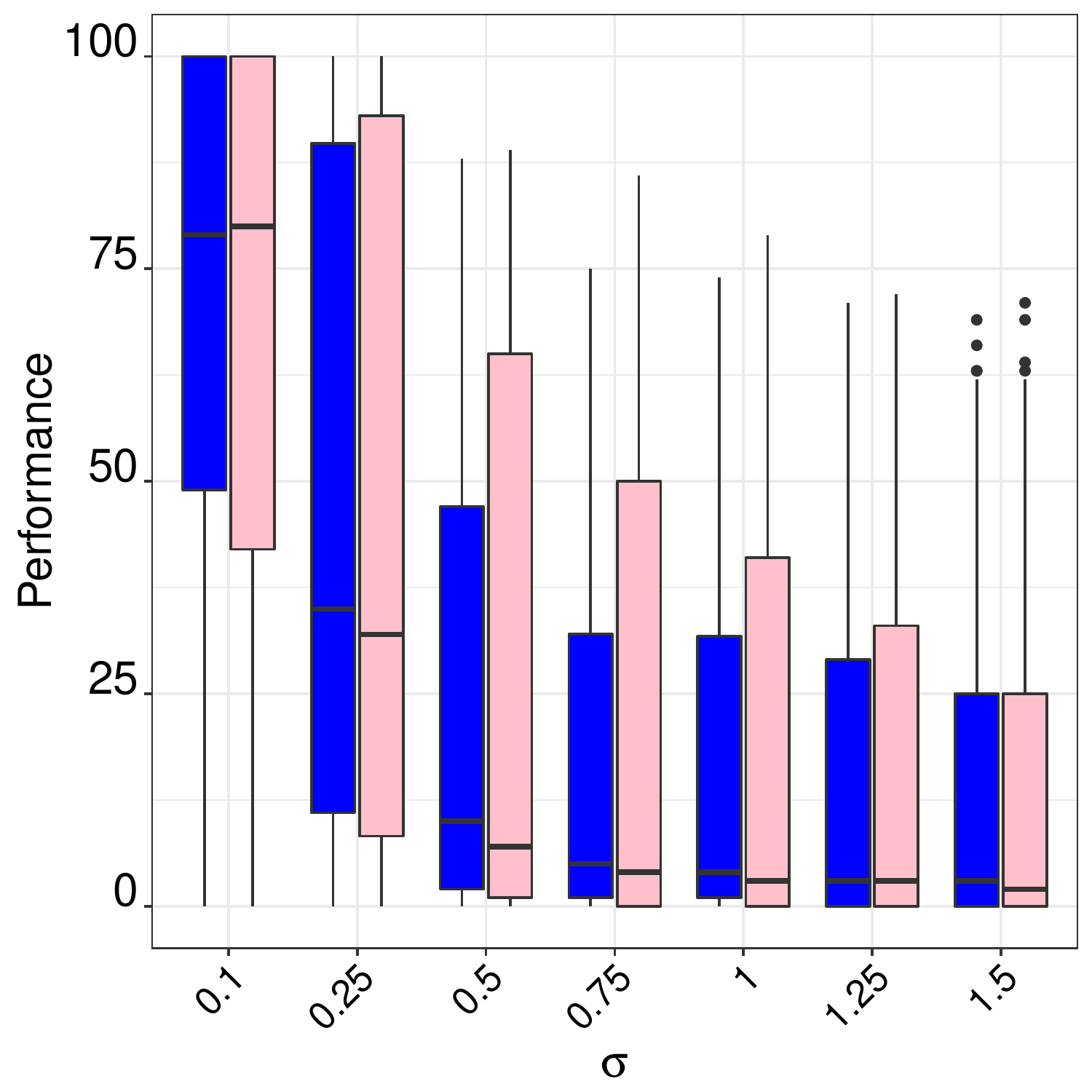} &
\includegraphics[scale=.22]{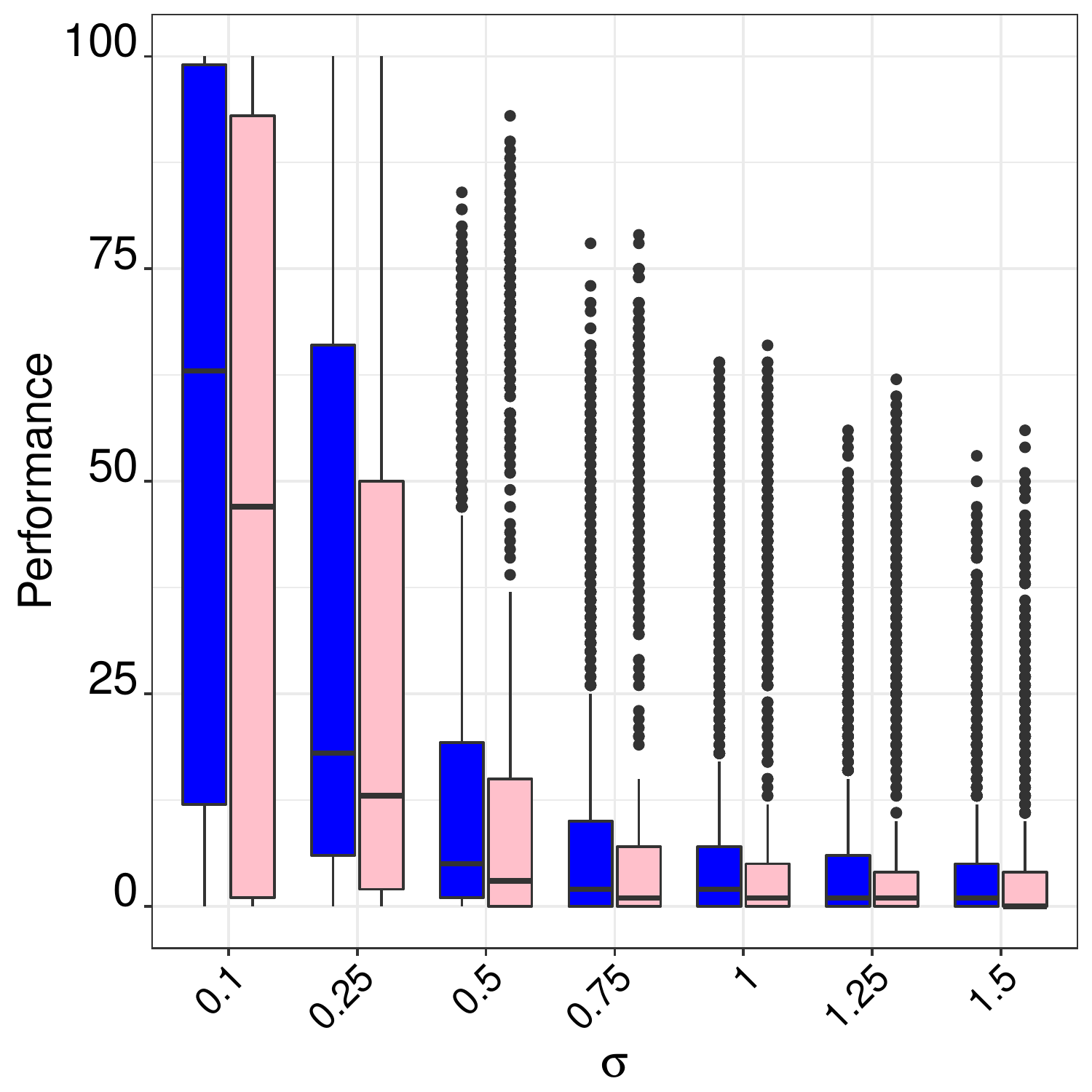} & 
\includegraphics[scale=.22]{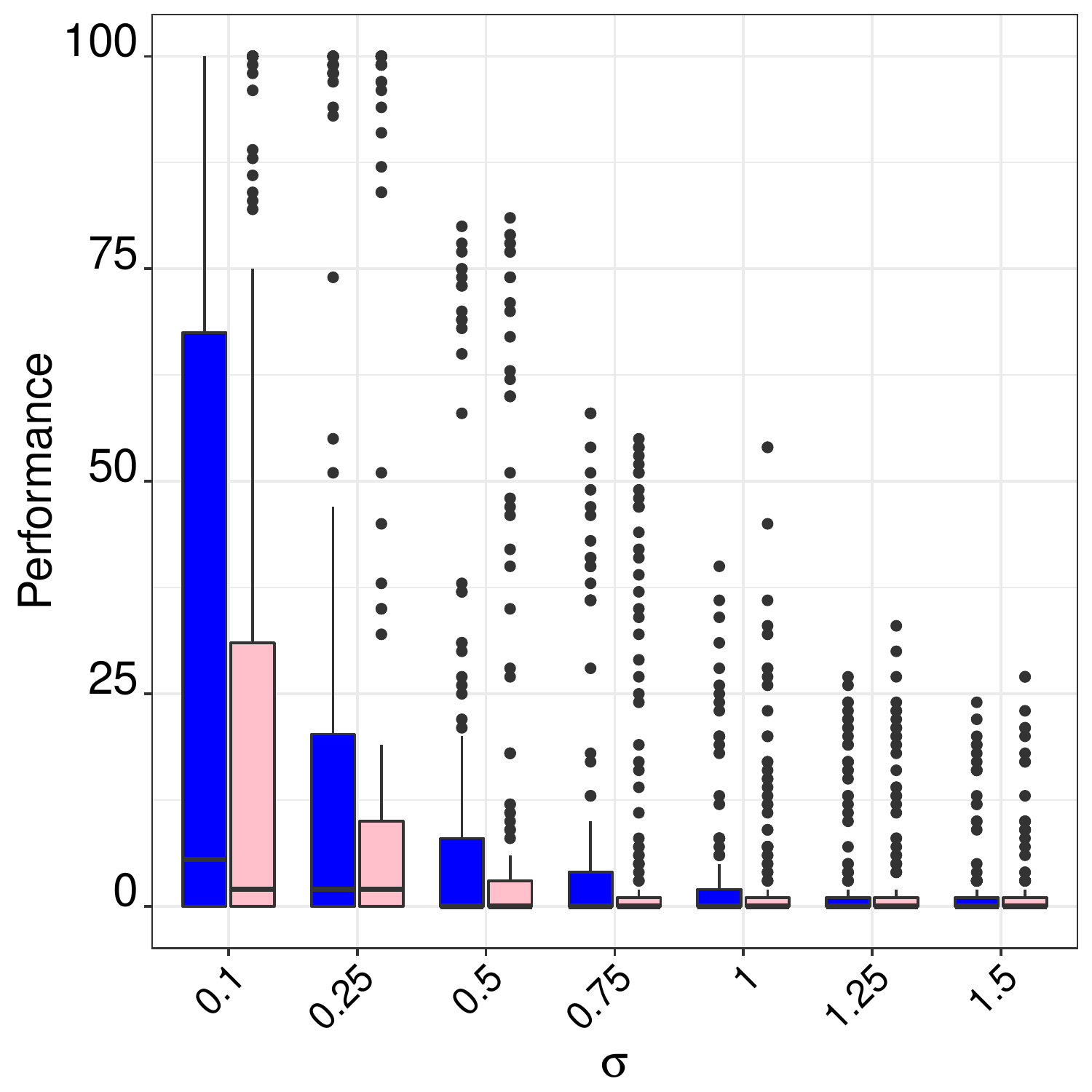} \\
\end{tabular}
\end{figure}

\begin{figure}
\caption{Comparison of model selection performance of three-level A-ComVar designs with central composite and orthogonal main effect designs.
Here \protect\includegraphics[width=.2in,height = 0.02\textheight]{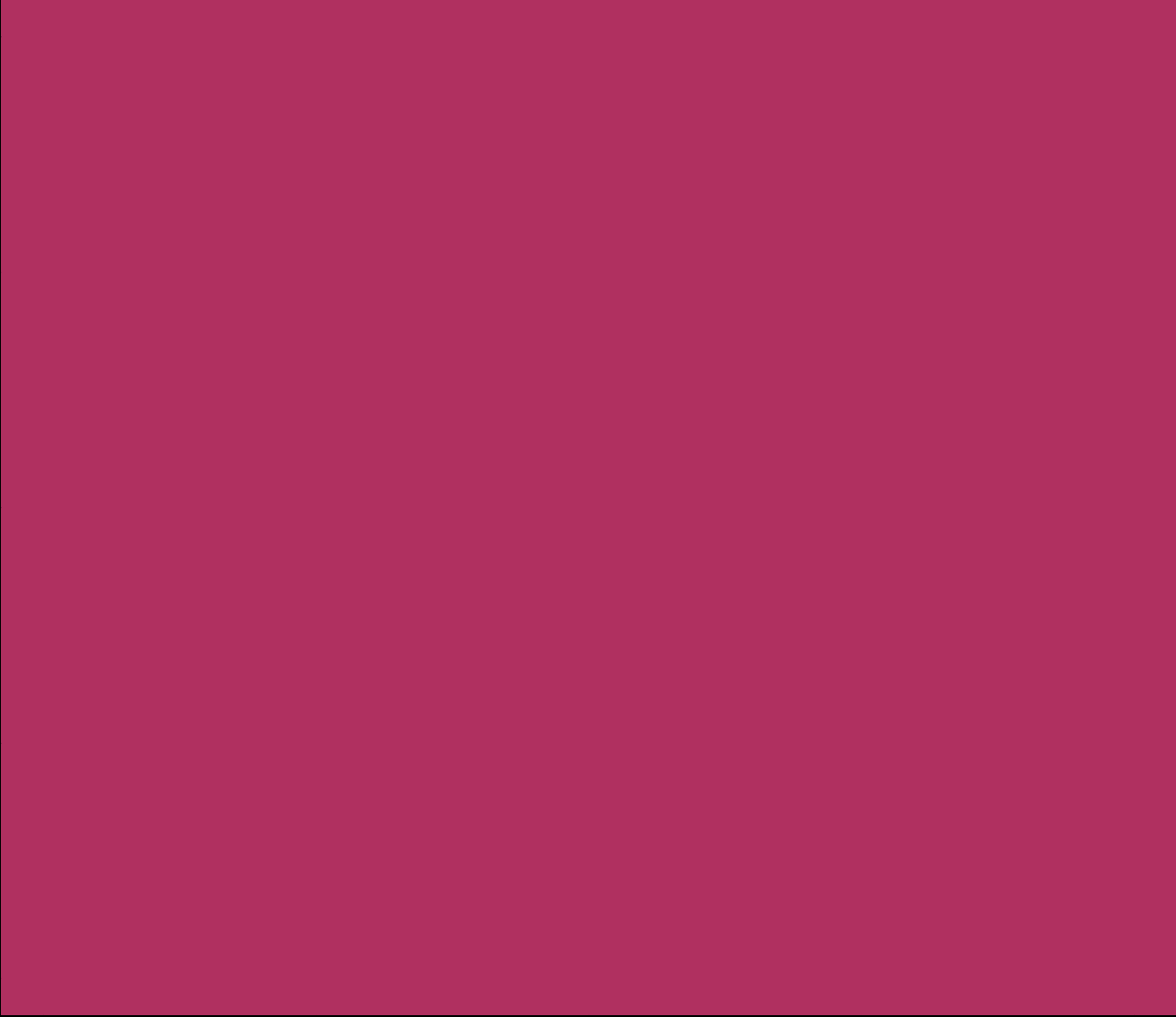}  represents our three-level A-ComVar design with 4 factors and 20 runs, 
\protect\includegraphics[width=.2in,height = 0.02\textheight]{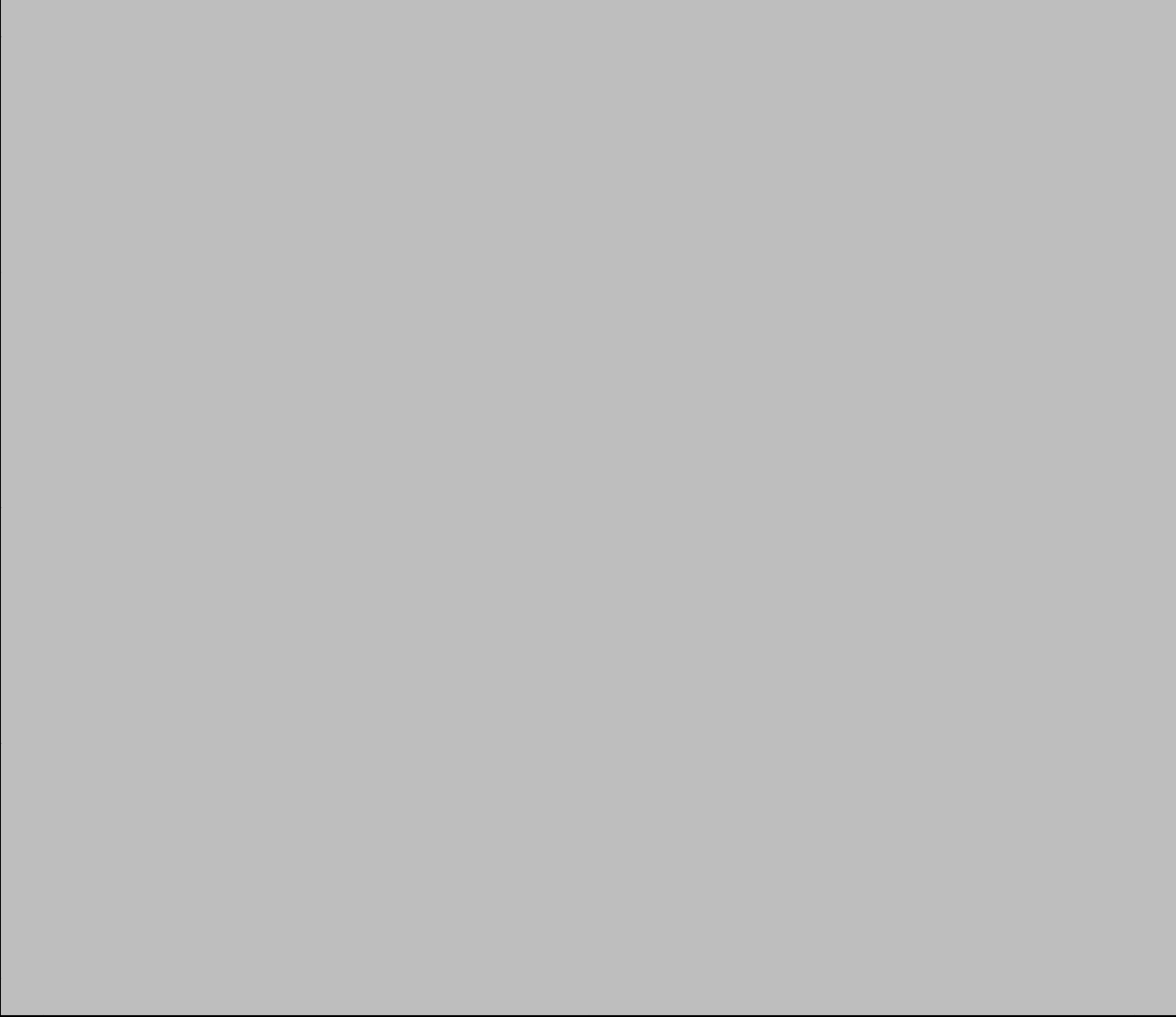} represents a central composite design with 3 factors and 20 runs,
\protect\includegraphics[width=.2in,height = 0.02\textheight]{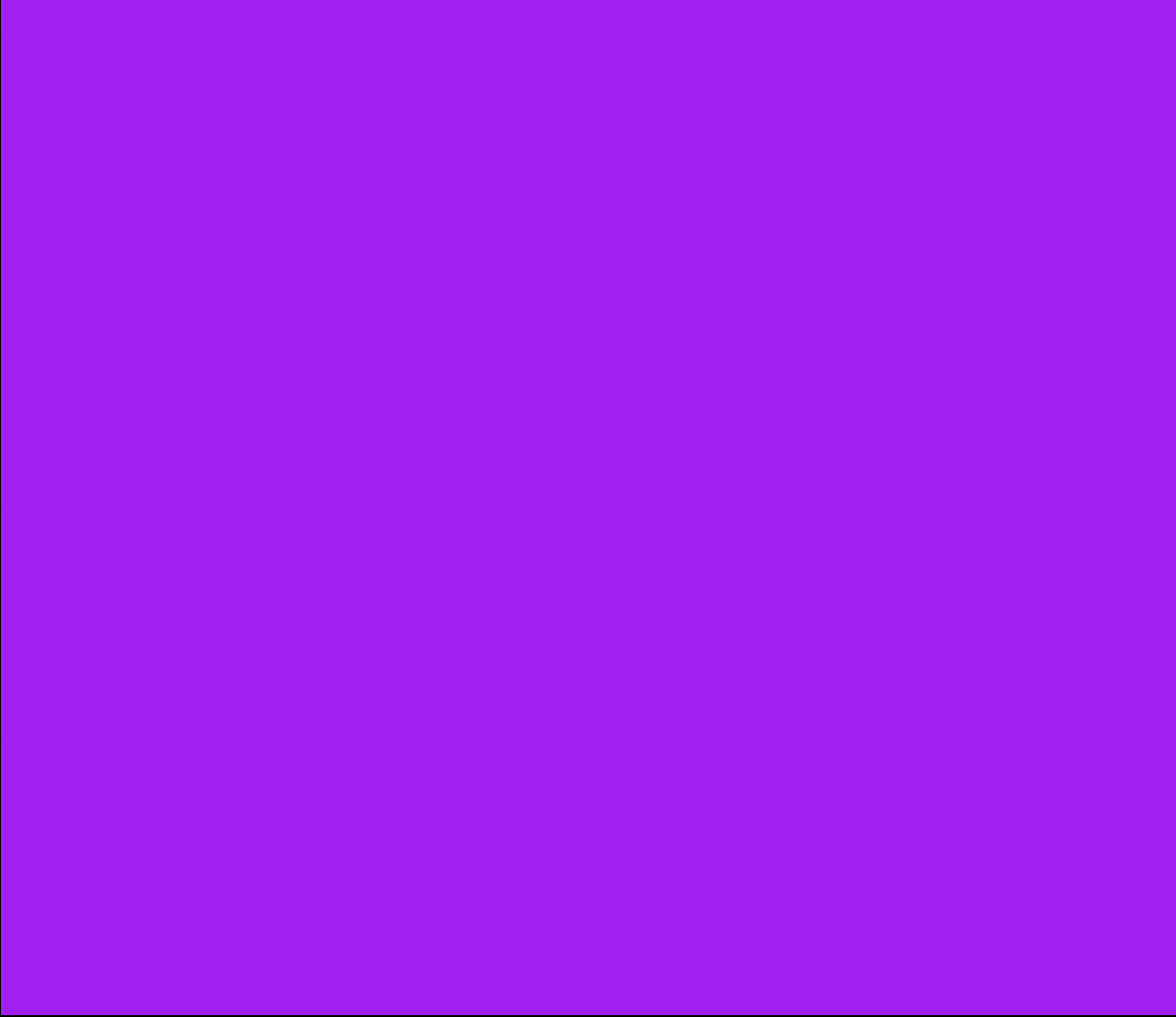} represents our three-level A-ComVar design with 7 factors and 18 runs,
\protect\includegraphics[width=.2in,height = 0.02\textheight]{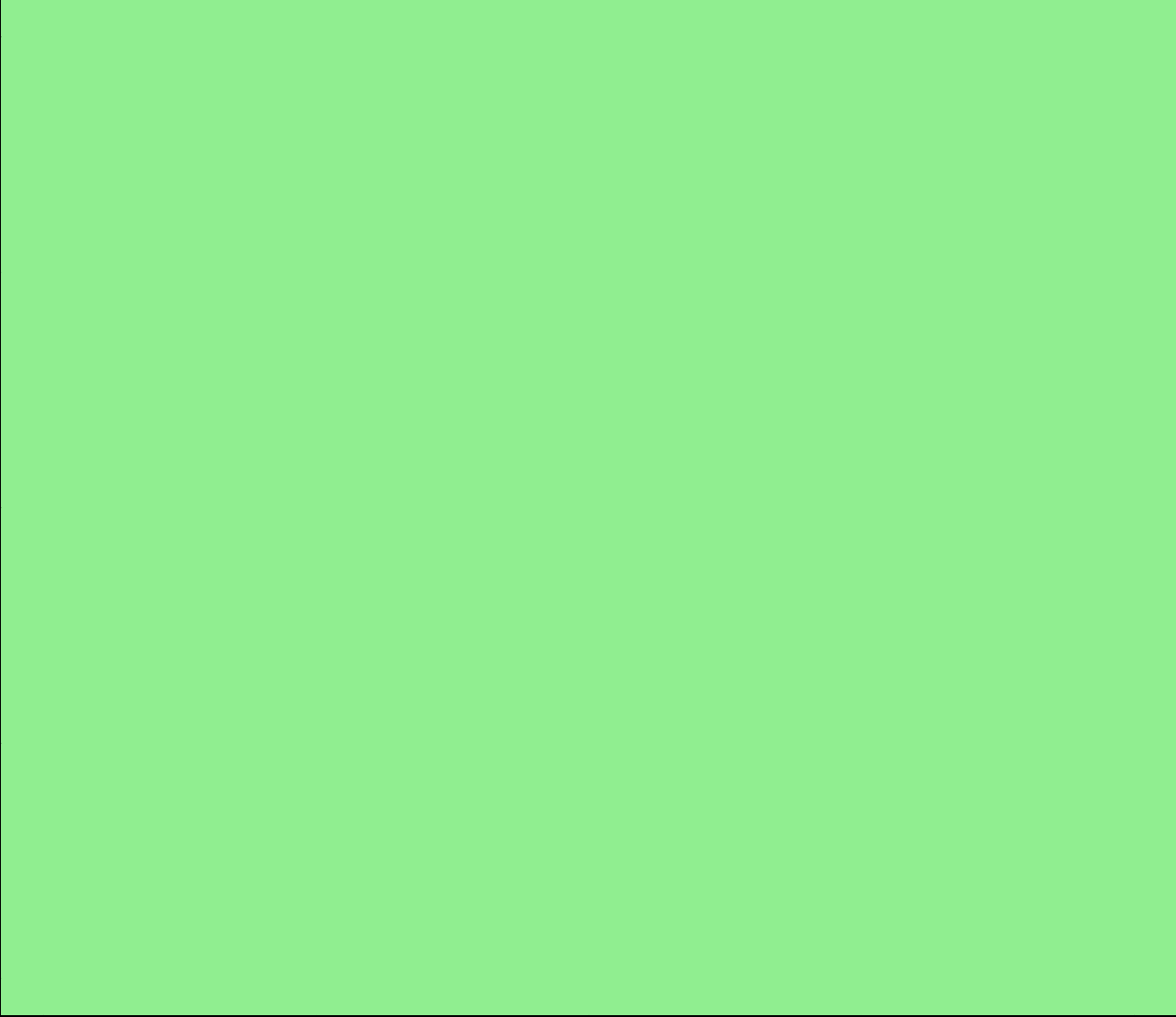} represents a orthogonal main effect plan with 7 factors and 18 runs.}
\label{fig::simresults3}
\centering
\begin{tabular}{ccc}
\multicolumn{3}{c}{3 level A-ComVar vs CCD} \\
\includegraphics[scale=.22]{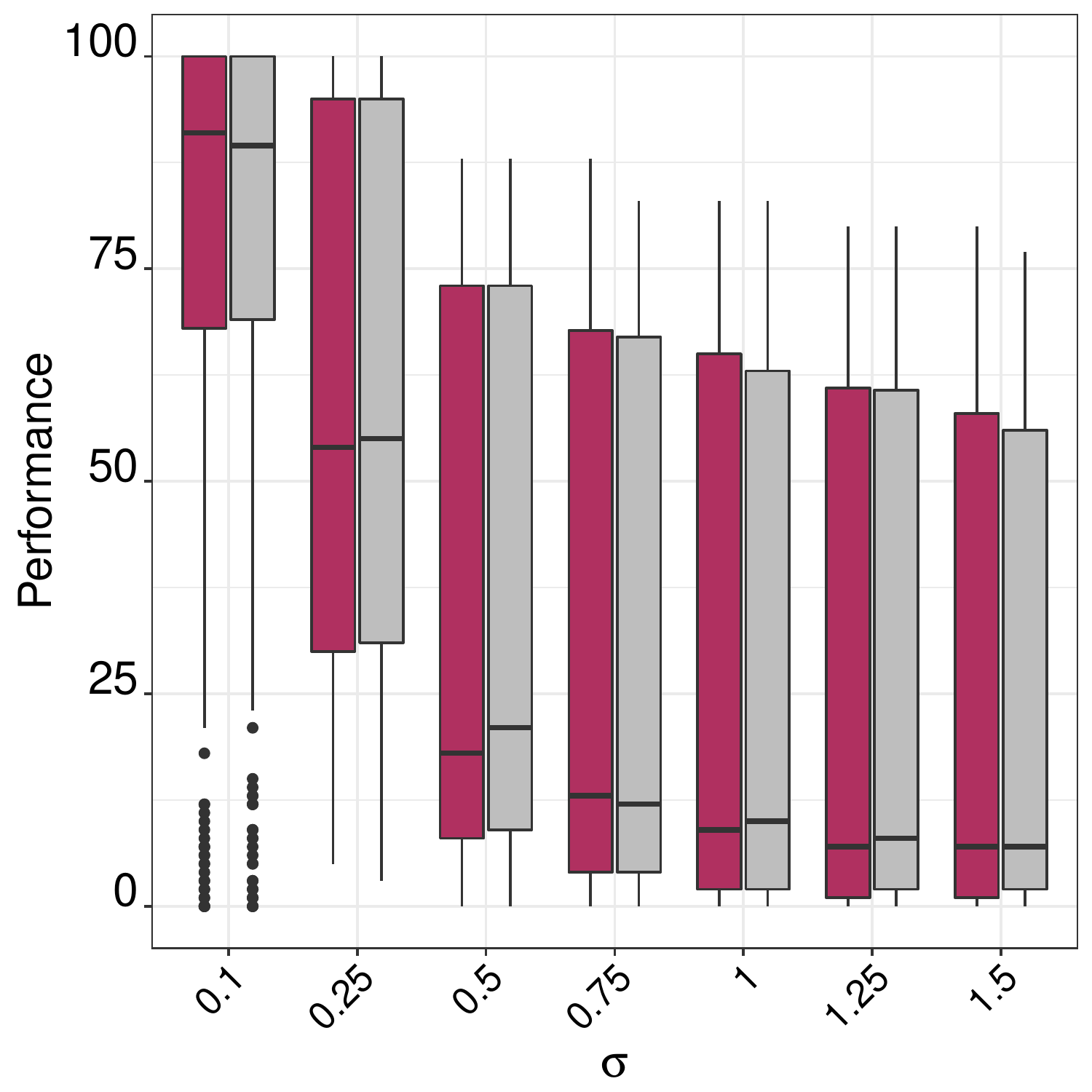} &
\includegraphics[scale=.22]{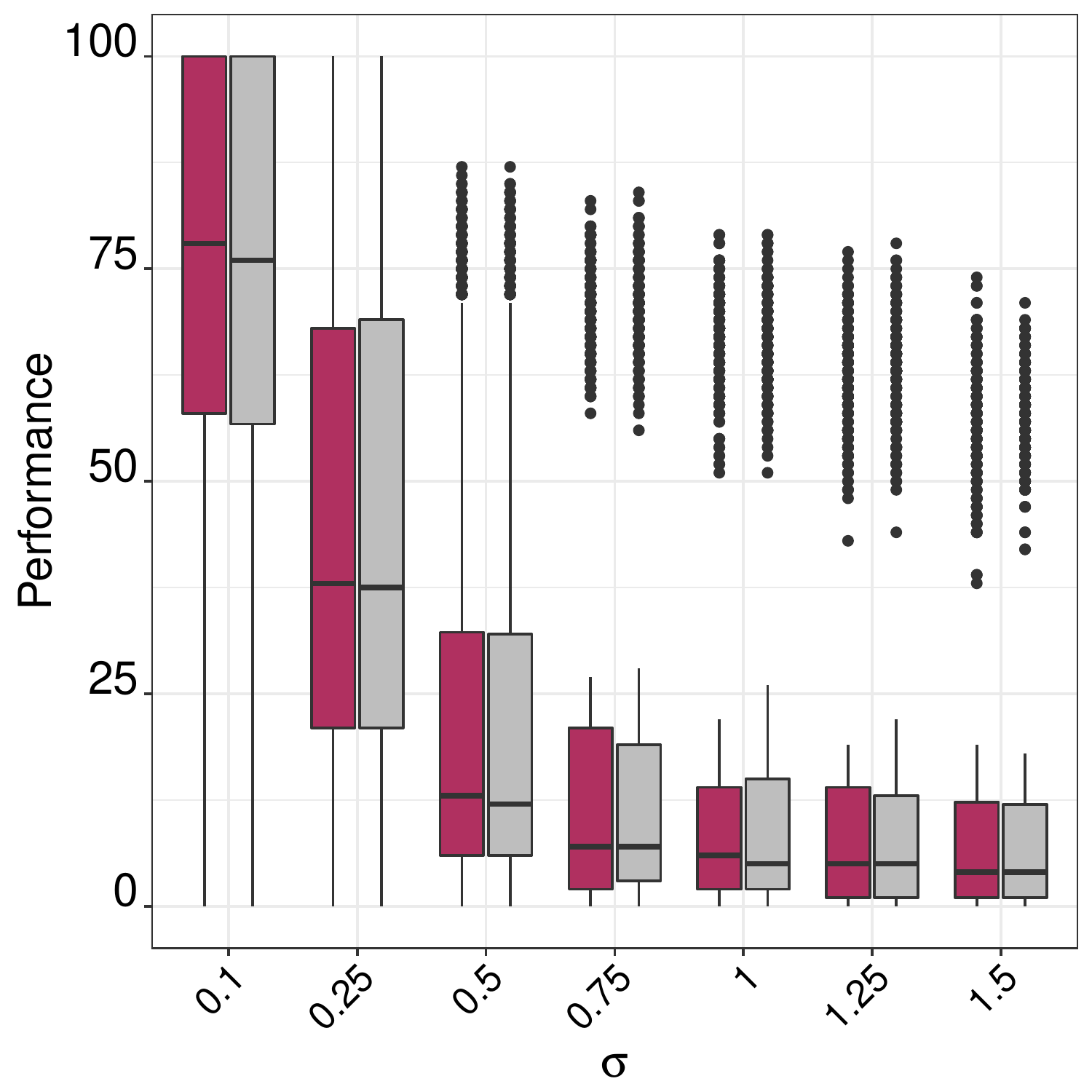} &
\includegraphics[scale=.22]{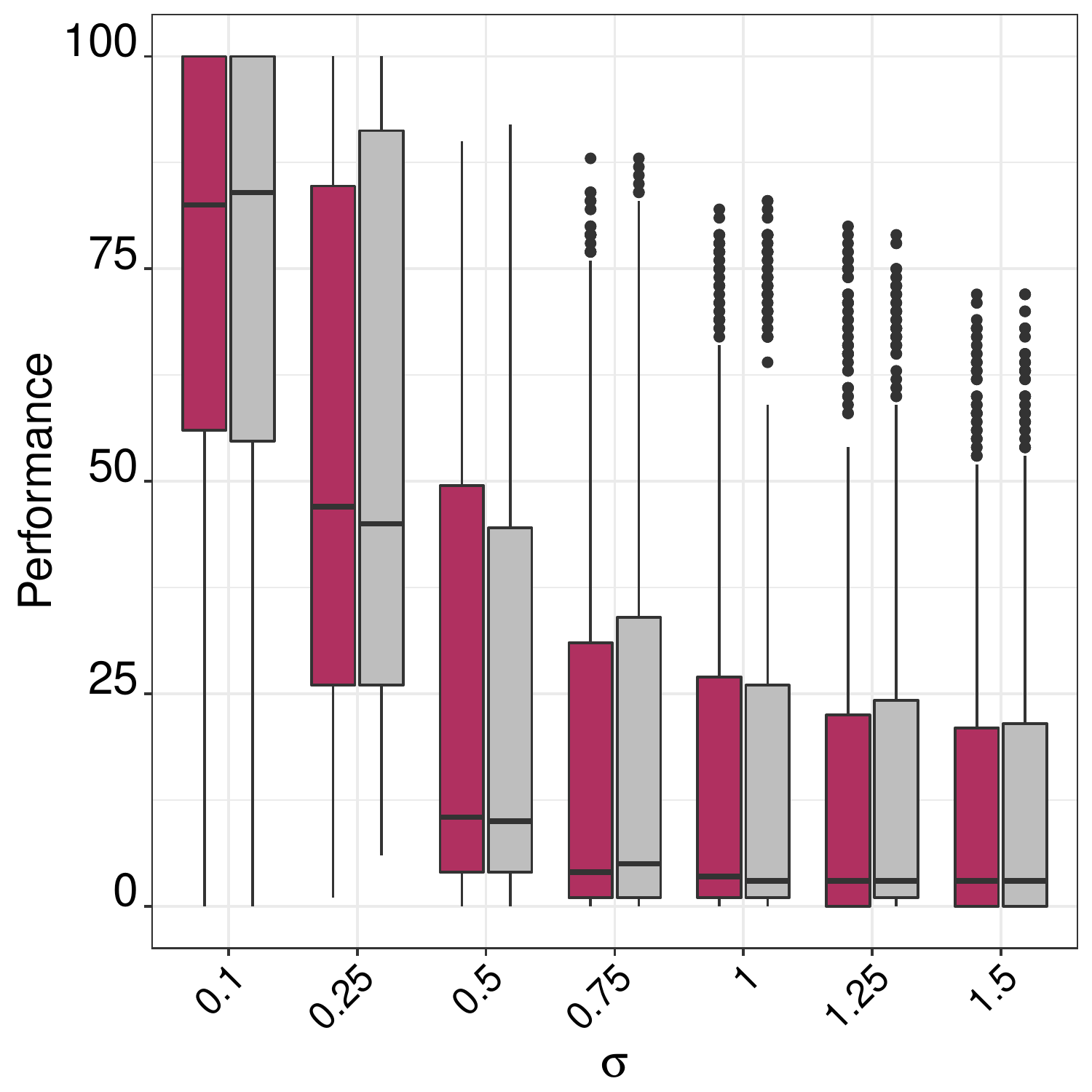} \\

\multicolumn{3}{c}{3 level A-ComVar vs OME} \\
\includegraphics[scale=.22]{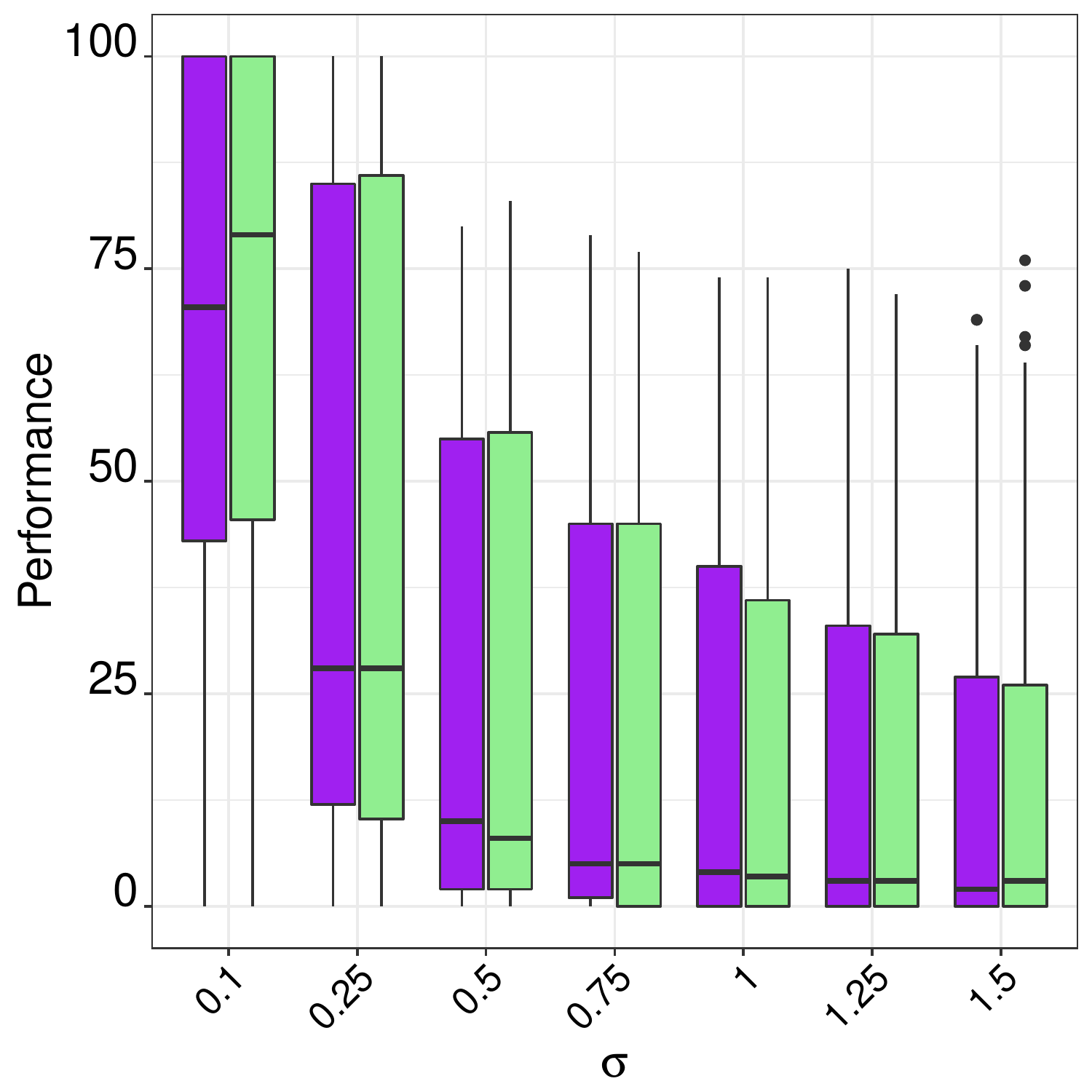} &
\includegraphics[scale=.22]{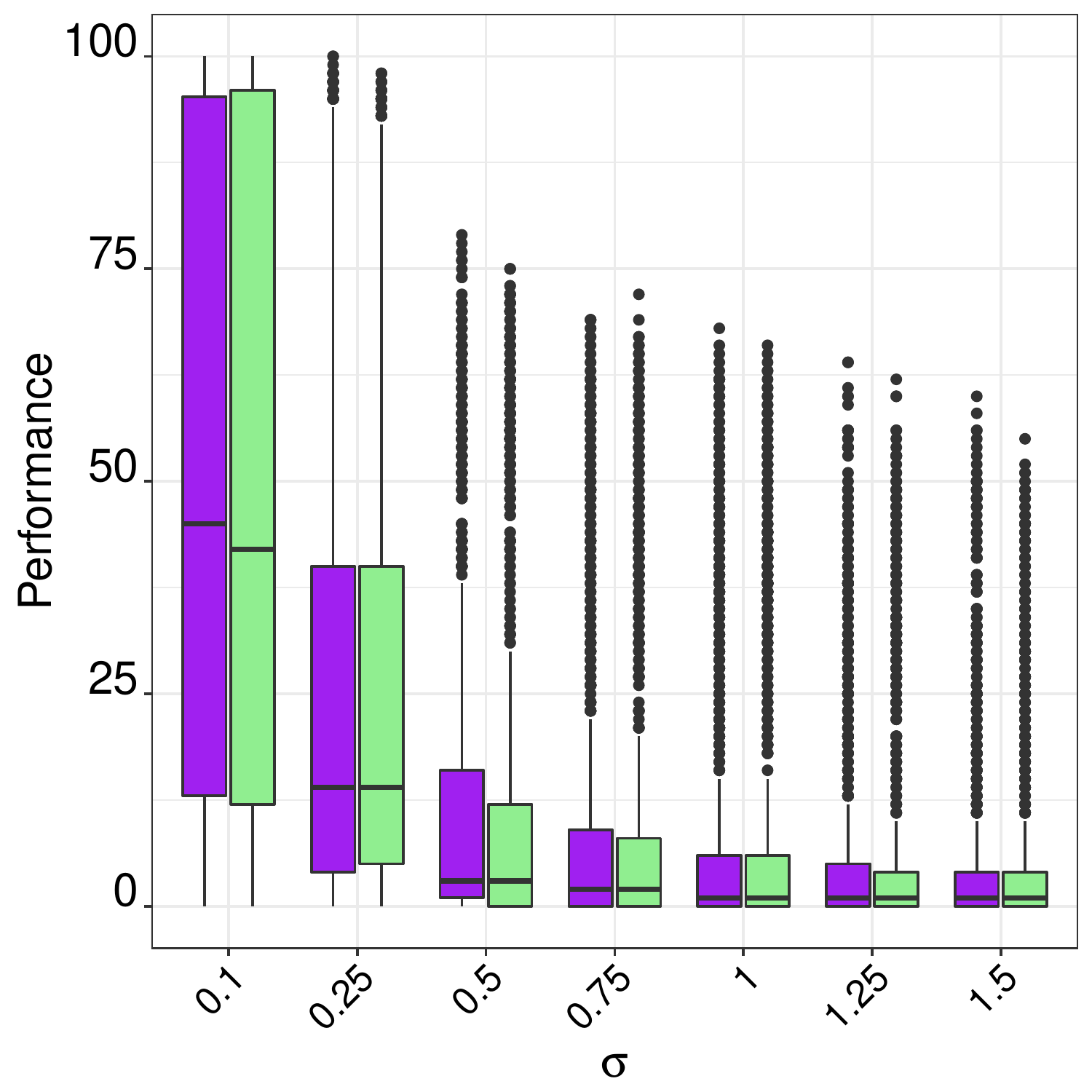} &
\includegraphics[scale=.22]{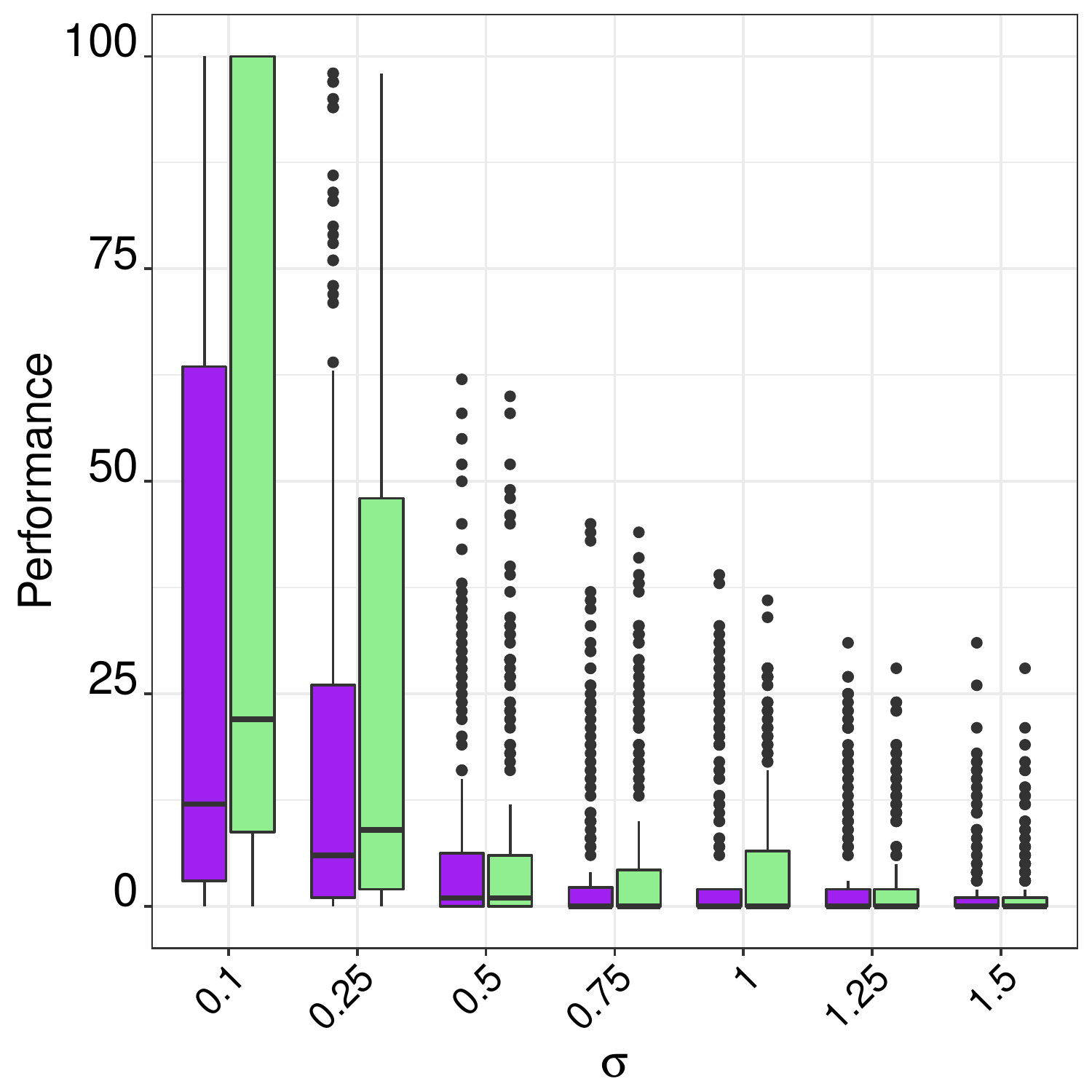} \\

\end{tabular}
\end{figure}
        
\begin{table}[ht]
\begin{minipage}[b]{0.4\hsize}
\centering
\begin{tabular}{|ccccc|} \hline
            $A$ & $B$ & $C$ & $D$ & $E$ \\ \hline
            -1 & -1 & -1 & -1 & \phone \\
            \phone & \phone & -1 & -1 & -1  \\
            \phone  & -1 & \phone & -1 & -1 \\
            \phone  & -1  & -1 & \phone  & -1  \\
            -1 & \phone  & \phone  & -1 & -1  \\
            -1 & \phone  & -1  & \phone & -1 \\
            -1  & -1 &  \phone & \phone & -1\\
            \phone   & \phone & \phone  & \phone & -1 \\
            \phone  & \phone  & \phone  & -1 & \phone \\
            \phone  & \phone & -1  & \phone  & \phone  \\
            \phone  & -1 & \phone & \phone & \phone \\
            -1  & \phone  & \phone & \phone & \phone \\ \hline
        \end{tabular}
\end{minipage}
\begin{minipage}[b]{0.5\hsize}
\centering
        \begin{tabular}{|ccccccccccc|} \hline
$A$ & $B$ & $C$ & $D$ & $E$ & $F$ & $G$ & $H$ & $I$ & $J$ & $K$ \\ \hline        
 \phone & \phone & -1 & \phone & \phone & \phone & -1 & -1 & -1 & \phone & -1 \\
\phone & -1 & \phone & \phone & \phone & -1 & -1 & -1 & \phone & -1 & \phone \\
-1 & \phone & \phone & \phone & -1 & -1 & -1 & \phone & -1 & \phone & \phone \\
\phone & \phone & \phone & -1 & -1 & -1 & \phone & -1 & \phone & \phone & -1 \\
\phone & \phone & -1 & -1 & -1 & \phone & -1 & \phone & \phone & -1 & \phone \\
\phone & -1 & -1 & -1 & \phone & -1 & \phone & \phone & -1 & \phone & \phone \\
-1 & -1 & -1 & \phone & -1 & \phone & \phone & -1 & \phone & \phone & \phone \\
-1 & -1 & \phone & -1 & \phone & \phone & -1 & \phone & \phone & \phone & -1 \\
-1 & \phone & -1 & \phone & \phone & -1 & \phone & \phone & \phone & -1 & -1 \\
\phone & -1 & \phone & \phone & -1 & \phone & \phone & \phone & -1 & -1 & -1 \\
-1 & \phone & \phone & -1 & \phone & \phone & \phone & -1 & -1 & -1 & \phone \\
-1 & -1 & -1 & -1 & -1 & -1 & -1 & -1 & -1 & -1 & -1 \\ \hline
        \end{tabular}
        \end{minipage}
        \caption{Design $\boldsymbol{d}_5^{(12)}$ with common variance for 5 factors and 12 runs (left) and Plackett-Burman Design with 11 factors and 12 runs (right).}
        \label{table:design1}
        \end{table}

\begin{table}
\centering
\makebox[0pt][c]{\parbox{1.2\textwidth}{%
    \begin{minipage}[b]{0.3\hsize}\centering
        \begin{tabular}{|ccccc|} \hline
        & & $D^1$ & & \\
        \hline
        $A$ & $B$ & $C$ & $D$ & $E$ \\ \hline
            -1 & -1 & -1 & \phone & \phone \\
            -1 & -1 & 1 & -1 & \phone  \\
            -1  & -1 & \phone & \phone & -1\\
            -1  & -1  & \phone & \phone  & \phone  \\
            -1  & \phone  & -1 & -1 & -1 \\
            -1 & \phone  & \phone  & \phone & \phone  \\
            \phone & -1  & -1 & -1  & -1 \\
            \phone   & -1 & \phone & \phone & \phone \\
            \phone  & \phone  & -1 & -1  & -1 \\
            \phone  & \phone  & -1  & -1 & \phone \\
            \phone  & \phone & -1  & \phone  & -1  \\
            \phone  & \phone & \phone & -1 & -1 \\
 \hline
        \end{tabular}
    \end{minipage}
    \begin{minipage}[b]{0.3\hsize}\centering
        \begin{tabular}{|cccc|} \hline
        & & $D^2$ &  \\
        \hline
        $A$ & $B$ & $C$ & $D$  \\ \hline
            -1 & -1 & -1 & -1 \\
            -1 & -1 & -1 & 0  \\
            -1 & -1 & 0  & 0  \\
            -1 & -1  & \phone & \phone \\
            -1 & 0  & 0  & 0 \\
            -1 & 0  & \phone  & 0  \\
            -1  & \phone & -1 & -1 \\
            -1  & \phone & -1 & 0  \\
            -1  & \phone & 0  & \phone \\
            -1  & \phone  & \phone & -1 \\
            0  & 0  & 0 & 0  \\
            0  & 0  & 0  & \phone \\ 
            0  & 0  & \phone  & \phone \\ 
            0  & \phone  & \phone  & 0 \\
            \phone  & -1  & -1  & \phone \\
            \phone  & -1  & \phone  & -1 \\
            \phone  & -1  & \phone  & 0 \\
            \phone  & 0  & 0  & \phone \\
            \phone  & \phone  & -1  & -1 \\
            \phone  & \phone  & \phone  & 0 \\
            \hline
        \end{tabular}
    \end{minipage}
    \begin{minipage}[b]{0.3\hsize}\centering
        \begin{tabular}{|ccccc|} \hline
        & & $D^3$ & & \\
        \hline
        $A$ & $B$ & $C$ & $D$ & $E$ \\ \hline
            -1 & -1 & -1 & -1 & -1 \\
            -1 & -1 & \phone & \phone & 0 \\
            -1  & \phone &-1 & -1  & -1  \\
            -1 & \phone  & -1 & \phone  & -1 \\
            -1 & \phone  & -1 & \phone  & 0  \\
            -1 & \phone  & 0  & \phone & \phone  \\
            -1 & \phone  & \phone  & \phone  & -1 \\
            0  & -1 & 0 & 0  & -1  \\
            0  & 0 & 0 & 0  & \phone  \\
            0  & 0 & 0 & \phone  & -1  \\
            \phone  & -1 & -1  & -1 & 0  \\
             \phone  & -1 & -1  & \phone & -1  \\
              \phone  & -1 & \phone  & -1 & -1  \\
            \phone  & -1 & \phone  & -1  & 0 \\
             \phone  & -1 & \phone  & 0 & 0  \\
              \phone  & -1 & \phone  & \phone & 0  \\
            \phone  & \phone  & -1 & -1 & -1  \\
            \phone  & \phone  & \phone  & \phone & 0 \\ \hline
        \end{tabular}
    \end{minipage}
}}
\caption{(1) $D^1$(left): two-level A-ComVar design for  $k=1$ with $m=5$ and $n=12$, (2) $D^2$ (middle): three-level A-ComVar design for $k=1$ with $m=4$ and $n=20$ and (3) $D^3$ (right): three-level A-ComVar design for $k=1$, $m=7$, $n=18$, used in Exzmple 3.}
\label{table::acvdesigns}
\end{table}

\begin{table}
\centering
\makebox[0pt][c]{\parbox{1.2\textwidth}{%
    \begin{minipage}[b]{0.3\hsize}\centering
        \begin{tabular}{|ccccc|} \hline
        & & $D^4$ & & \\
        \hline
        $A$ & $B$ & $C$ & $D$ & $E$ \\ \hline
        \phone	&	\phone	&	-1	&	\phone	&	\phone\\
-1	&	\phone	&	\phone	&	-1	&	\phone\\
\phone	&	-1	&	\phone	&	\phone	&	-1\\
-1	&	\phone	&	-1	&	\phone	&	\phone\\
-1	&	-1	&	\phone	&	-1	&	\phone\\
-1	&	-1	&	-1	&	\phone	&	-1\\
\phone	&	-1	&	-1	&	-1	&	\phone\\
\phone	&	\phone	&	-1	&	-1	&	-1\\
\phone	&	\phone	&	\phone	&	-1	&	-1\\
-1	&	\phone	&	\phone	&	\phone	&	-1\\
\phone	&	-1	&	\phone	&	\phone	&	\phone\\
-1	&	-1	&	-1	&	-1	&	-1\\
\hline
        \end{tabular}
    \end{minipage}
    \begin{minipage}[b]{0.3\hsize}\centering
        \begin{tabular}{|ccccc|} \hline
        & & $D^5$ & & \\
        \hline
        $A$ & $B$ & $C$ & $D$ & $E$ \\ \hline
 \phone	&	\phone	&	\phone	&	-1	&	\phone\\
-1	&	\phone	&	\phone	&	\phone	&	-1\\
-1	&	\phone	&	-1	&	\phone	&	-1\\
-1	&	-1	&	\phone	&	-1	&	\phone\\
-1	&	\phone	&	-1	&	-1	&	-1\\
-1	&	-1	&	\phone	&	\phone	&	-1\\
\phone	&	-1	&	-1	&	-1	&	\phone\\
-1	&	\phone	&	-1	&	-1	&	\phone\\
\phone	&	-1	&	\phone	&	-1	&	-1\\
\phone	&	-1	&	-1	&	\phone	&	-1\\
\phone	&	-1	&	\phone	&	\phone	&	\phone\\
\phone	&	\phone	&	-1	&	\phone	&	\phone\\
\hline
        \end{tabular}
    \end{minipage}
    \begin{minipage}[b]{0.3\hsize}\centering
        \begin{tabular}{|ccccc|} \hline
        & & $D^6$ & & \\
        \hline
        $A$ & $B$ & $C$ & $D$ & $E$ \\ \hline
\phone	&	\phone	&	\phone	&	\phone	&	-1\\
\phone	&	\phone	&	\phone	&	-1	&	\phone\\
\phone	&	\phone	&	-1	&	\phone	&	\phone\\
\phone	&	-1	&	\phone	&	\phone	&	\phone\\
-1	&	\phone	&	\phone	&	\phone	&	\phone\\
\phone	&	\phone	&	-1	&	-1	&	-1\\
\phone	&	-1	&	\phone	&	-1	&	-1\\
\phone	&	-1	&	-1	&	\phone	&	-1\\
-1	&	\phone	&	\phone	&	-1	&	-1\\
-1	&	\phone	&	-1	&	\phone	&	-1\\
-1	&	-1	&	\phone	&	\phone	&	-1\\
-1	&	-1	&	-1	&	-1	&	\phone\\
        \hline
        \end{tabular}
    \end{minipage}
}}
        \caption{(1) $D^4$(left): two-level Bayes optimal design with $m=5$ and $n=12$ \cite{bingham2007incorporating}, (2) $D^5$ (middle): two-level design with $m=5$ and $n=12$ from Li and Nachesheim (2000) and (3) $D^6$ (right): two-level design with $m=5$, $n=12$ from \cite{ghosh2006optimum}, used in Example 3.}
        \label{table::designexamples}
\end{table}

\begin{table}
\centering
\makebox[0pt][c]{\parbox{1.2\textwidth}{%
    \begin{minipage}[b]{0.5\hsize}\centering
        \begin{tabular}{|rrr|} \hline
        &  $D^7$ &  \\
        \hline
        $A$ & $B$ & $C$ \\ \hline
-1	&	-1	&	-1\\
\phone	&	-1	&	-1\\
-1	&	\phone	&	-1\\
\phone	&	\phone	&	-1\\
-1	&	-1	&	\phone\\
\phone	&	-1	&	\phone\\
-1	&	\phone	&	\phone\\
\phone	&	\phone	&	\phone\\
-1.682	&	0	&	0\\
1.682	&	0	&	0\\
0	&	-1.682	&	0\\
0	&	1.682	&	0\\
0	&	0	&	-1.682\\
0	&	0	&	1.682\\
0	&	0	&	0\\
0	&	0	&	0\\
0	&	0	&	0\\
0	&	0	&	0\\
0	&	0	&	0\\
0	&	0	&	0\\
\hline
        \end{tabular}
    \end{minipage}
    \begin{minipage}[b]{0.5\hsize}\centering
        \begin{tabular}{|rrrrrrr|} \hline
        & & & $D^8$ & & & \\
        \hline
        $A$ & $B$ & $C$ & $D$ & $E$ &$F$ &$G$\\ \hline
-1 & -1 & -1 &-1 &-1 &-1 &-1\\
0 &0 &0 &0 &0 &0 &-1\\
1 &1 &1 &1 &1 &1 &-1\\
-1 &-1 &0 &1 &0 &1 &-1\\
0 &0 &1 &-1 &1 &-1 &-1\\
1 &1 &-1 &0 &-1 &0 &-1\\
-1 &0 &-1 &1 &1 &0 &0\\
0 &1 &0 &-1 &-1 &1 &0\\
1 &-1 &1 &0 &0 &-1 &0\\
-1 &1 &1 &-1 &0 &0 &0\\
0 &-1 &-1 &0 &1 &1 &0\\
1 &0 &0 &1 &-1 &-1 &0\\
-1 &0 &1 &0 &-1 &1 &1\\
0 &1 &-1 &1 &0 &-1 &1\\
1 &-1 &0 &-1 &1 &0 &1\\
-1 &1 &0 &0 &1 &-1 &1\\
0 &-1 &1 &1 &-1 &0 &1\\
1 &0 &-1 &-1 &0 &1 &1\\
    \hline
        \end{tabular}
    \end{minipage}
}}
        \caption{(1) $D^7$(left): Central Composite Design with $m=3$ and $n=20$, (2) $D^8$ (middle): three-level orthogonal main effect plan with $m=7$ and $n=18$ used in Example 3.}
        \label{sim::3levelcompet}
\end{table}

\section{Discussion}
In this work we introduced A-ComVar Designs, an extension of common variance designs. Our proposed approach addresses the difficulties associated with finding common variance designs via exhaustive search. Through several examples, we demonstrated that the proposed algorithmic approach allows us to quickly find common variance designs that overlap with those known in the literature. Furthermore, in cases where common variance designs do not exist or cannot be found, our approach allows identification of designs with close to common variance. Comparisons to a Plackett-Burman design and several other standard optimal designs from literature demonstrated that such designs perform quite well in practice, and that in many cases these A-ComVar designs perform as well as common variance designs.
 
There are several avenues here for future work. First, we considered only the cases with two-level and three-level factors. Future work could consider finding A-ComVar designs with mixed$-$level factors. Second, we utilized a genetic algorithm to find these designs. There are numerous other optimization approaches that could be used to maximize the objective function in (\ref{obj_func}). In some cases, these other approaches may succeed in finding designs with a better ratio of minimum to maximum variance of the uncommon parameters. Third, there is another approach to finding common variance designs through hierarchical designs \citep{Chowdhury2016}. These designs are found by identifying a common variance design for a smaller number of runs and then adding runs while trying to preserve the common variance property. It is possible that a similar idea could be developed for A-ComVar designs. Finally, future work could study the types of A-ComVar designs that can be found when the number of interactions in the model increases beyond two. 

\bibliographystyle{apalike}
\bibliography{comvarbib}

\begin{thebibliography}{}

\bibitem[Bingham and Chipman, 2007]{bingham2007incorporating}
Bingham, D.~R. and Chipman, H.~A. (2007).
\newblock Incorporating prior information in optimal design for model
  selection.
\newblock {\em Technometrics}, 49(2):155--163.

\bibitem[Breiman, 1995]{breiman1995better}
Breiman, L. (1995).
\newblock Better subset regression using the nonnegative garrote.
\newblock {\em Technometrics}, 37(4):373--384.

\bibitem[Calvin, 1986]{calvin1986new}
Calvin, J.~A. (1986).
\newblock A new class of variance balanced designs.
\newblock {\em Journal of statistical planning and inference},
  14(2-3):251--254.

\bibitem[Candes et~al., 2007]{candes2007dantzig}
Candes, E., Tao, T., et~al. (2007).
\newblock The {D}antzig selector: {S}tatistical estimation when p is much
  larger than n.
\newblock {\em The annals of Statistics}, 35(6):2313--2351.

\bibitem[Cheng, 1986]{cheng1986method}
Cheng, C.-S. (1986).
\newblock A method for constructing balanced incomplete-block designs with
  nested rows and columns.
\newblock {\em Biometrika}, 73(3):695--700.

\bibitem[Chowdhury, 2016]{Chowdhury2016}
Chowdhury, S. (2016).
\newblock {\em Common Variance Fractional Factorial Designs for Model
  Comparisons}.
\newblock PhD thesis, University of California Riverside.

\bibitem[Ghosh and Chowdhury, 2017]{ghosh2017cv}
Ghosh, S. and Chowdhury, S. (2017).
\newblock {CV}, {ECV}, and robust {CV} designs for replications under a class
  of linear models in factorial experiments.
\newblock {\em Journal of Statistical Planning and Inference}, 188:1--7.

\bibitem[Ghosh and Flores, 2013]{ghosh2013common}
Ghosh, S. and Flores, A. (2013).
\newblock Common variance fractional factorial designs and their optimality to
  identify a class of models.
\newblock {\em Journal of Statistical Planning and Inference},
  143(10):1807--1815.

\bibitem[Ghosh and Tian, 2006]{ghosh2006optimum}
Ghosh, S. and Tian, Y. (2006).
\newblock Optimum two level fractional factorial plans for model identification
  and discrimination.
\newblock {\em Journal of multivariate analysis}, 97(6):1437--1450.

\bibitem[Gupta and Jones, 1983]{gupta1983equireplicate}
Gupta, S. and Jones, B. (1983).
\newblock Equireplicate balanced block designs with unequal block sizes.
\newblock {\em Biometrika}, 70(2):433--440.

\bibitem[Hedayat and Stufken, 1989]{hedayat1989relation}
Hedayat, A. and Stufken, J. (1989).
\newblock A relation between pairwise balanced and variance balanced block
  designs.
\newblock {\em Journal of the American Statistical Association},
  84(407):753--755.

\bibitem[Kane and Mandal, 2019]{kane2019}
Kane, A. and Mandal, A. (2019).
\newblock A new analysis strategy for designs with complex aliasing.
\newblock {\em The American Statistician}.

\bibitem[Khatri, 1982]{khatri1982note}
Khatri, C. (1982).
\newblock A note on variance balanced designs.
\newblock {\em Journal of Statistical Planning and inference}, 6(2):173--177.

\bibitem[Li and Nachtsheim, 2000]{li2000model}
Li, W. and Nachtsheim, C.~J. (2000).
\newblock Model-robust factorial designs.
\newblock {\em Technometrics}, 42(4):345--352.

\bibitem[Lin et~al., 2015]{lin2015using}
Lin, C.~D., Anderson-Cook, C.~M., Hamada, M.~S., Moore, L.~M., and Sitter,
  R.~R. (2015).
\newblock Using genetic algorithms to design experiments: a review.
\newblock {\em Quality and Reliability Engineering International},
  31(2):155--167.

\bibitem[Mandal et~al., 2015]{mandal2015algorithmic}
Mandal, A., Wong, W.~K., and Yu, Y. (2015).
\newblock Algorithmic searches for optimal designs.
\newblock {\em Handbook of Design and Analysis of Experiments}, pages 755--783.

\bibitem[Mukerjee and Kageyama, 1985]{mukerjee1985resolvable}
Mukerjee, R. and Kageyama, S. (1985).
\newblock On resolvable and affine resolvable variance-balanced designs.
\newblock {\em Biometrika}, 72(1):165--172.

\bibitem[Srivastava, 1976]{srivastava1976some}
Srivastava, J. (1976).
\newblock Some further theory of search linear models.
\newblock In {\em Contribution to Applied Statistics}, pages 249--256.
  Springer.

\bibitem[Srivastava and Ghosh, 1976]{srivastava1976series}
Srivastava, J. and Ghosh, S. (1976).
\newblock A series of balanced factorial designs of resolution v which allow
  search and estimation of one extra unknown effect.
\newblock {\em Sankhy{\=a}: The Indian Journal of Statistics, Series B}, pages
  280--289.

\bibitem[Srivastava and Gupta, 1979]{srivastava1979main}
Srivastava, J. and Gupta, B. (1979).
\newblock Main effect plan for 2m factorials which allow search and estimation
  of one unknown effect.
\newblock {\em Journal of Statistical Planning and Inference}, 3(3):259--265.

\bibitem[Yuan et~al., 2007]{yuan2007efficient}
Yuan, M., Joseph, V.~R., and Lin, Y. (2007).
\newblock An efficient variable selection approach for analyzing designed
  experiments.
\newblock {\em Technometrics}, 49(4):430--439.

\bibitem[Yuan et~al., 2009]{yuan2009structured}
Yuan, M., Joseph, V.~R., Zou, H., et~al. (2009).
\newblock Structured variable selection and estimation.
\newblock {\em The Annals of Applied Statistics}, 3(4):1738--1757.

\end{thebibliography}


\section*{Appendix $-$ Genetic Algorithm}

{\color{black}
In this work we used a genetic algorithm to find designs that maximize the A-ComVar objective function. This appendix provides specific details on the algorithm. Keeping with the standard genetic algorithm terminology, we use the word {\it chromosome} to describe a single candidate design. Each chromosome is comprised of the factor settings for each factor at each design point. Each of these individual factor settings is known as a {\it gene}. The {\it population} is the set of all chromosomes, i.e. all designs that we are currently considering. 

We illustrate a simple version of the genetic algorithm below. In this example we search for a 6 run A-ComVar design for an experiment with three two-level factors and one interaction. We label the factors as $A$, $B$, and $C$. For simplicity, we assume that the population size is 3, although in real applications it will generally be larger.

Since this experiment has three two-level factors, there are 8 possible design points to pick the 6 points for our design from. The 8 points are shown in Table \ref{table:designpointbank}. There are ${3}\choose{2}$ $= 3$ possible models with all main effects and one interaction. For notational simplicity we label these models by the corresponding interaction: $(AB)$, $(AC)$, and $(BC)$. Our goal is to obtain a design under which the variance of the interaction term is identical, or close to identical, under all three of these models.
	
\begin{table}[ht]
		\centering
		\begin{tabular}{ccc}
		$A$ & $B$ & $C$ \\ \hline
		-1 & -1 & -1 \\
		-1 & -1 & \phone \\
		-1 & \phone & -1 \\
		-1 & \phone & \phone \\
		\phone & -1 & -1 \\
		\phone & -1 & \phone \\
		\phone & \phone & -1 \\
		\phone & \phone & \phone
		\end{tabular}
	\caption{Set of possible design points for the Appendix example.}
	\label{table:designpointbank}
	\end{table}

\underline{\it 0. Initialization}

First, each of the three chromosomes is initialized to a random start. To obtain the random start for a specific chromosome, we simply sample six of the rows in Table \ref{table:designpointbank} without replacement. Our initialization procedure results in the following three chromosomes:

\begin{center}
\begin{tikzpicture}
\node[rectangle, draw, text centered, minimum height=4cm, minimum width=\linewidth,
    fill=white] (background){};
\matrix [below right= 0.5 and 0.5 of background.north west] (chromosome1) [matrix of nodes,
    every even row/.style = {nodes={fill=green!10}},
    every odd row/.style = {nodes={fill=green!20}},
    column 1/.style = {nodes={minimum width=1cm}},
    column 2/.style = {nodes={minimum width=1cm}},
    column 3/.style = {nodes={minimum width=1cm}},
    ]{
    -1 & -1 & -1 \\
    -1 & \phone & \phone \\
    \phone & -1 & -1 \\ 
    \phone & -1 & \phone\\
    \phone & \phone & -1 \\ 
    \phone & \phone & \phone\\
    };
\matrix [below=0.5 of background.north] (chromosome2) [matrix of nodes,
    every even row/.style = {nodes={fill=red!10}},
    every odd row/.style = {nodes={fill=red!20}},
    column 1/.style = {nodes={minimum width=1cm}},
    column 2/.style = {nodes={minimum width=1cm}},
    column 3/.style = {nodes={minimum width=1cm}},
    ]{
    -1 & -1 & -1 \\
    -1 & \phone & -1 \\
    \phone & -1 & -1 \\ 
    \phone & -1 & \phone\\
    \phone & \phone & -1 \\ 
    \phone & \phone & \phone\\
    };
\matrix [below left= 0.5 and 0.5 of background.north east] (chromosome3) [matrix of nodes,
    every even row/.style = {nodes={fill=blue!10}},
    every odd row/.style = {nodes={fill=blue!20}},
    column 1/.style = {nodes={minimum width=1cm}},
    column 2/.style = {nodes={minimum width=1cm}},
    column 3/.style = {nodes={minimum width=1cm}},
    ]{
    -1 & -1 & -1 \\
    -1 & -1 & \phone \\
    -1 & \phone & \phone \\ 
    \phone & -1 & \phone\\
    \phone & \phone & -1 \\ 
    \phone & \phone & \phone\\
    };
\node [above=0.1of background.north, text centered]{Population};
\node [above=0.01of chromosome1.north, text centered]{Chromosome 1};
\node [above=0.01of chromosome2.north, text centered]{Chromosome 2};
\node [above=0.01of chromosome3.north, text centered]{Chromosome 3};
\end{tikzpicture}
\end{center}

After initializing, we need to calculate the fitness for each of these chromosomes using the objective function in expression (2). In order to evaluate the objective function, we need to calculate $\sigma_{2i}^2$ for $i = 1, 2, 3$, which correspond to models $(AB)$, $(AC)$, and $(BC)$, respectively. Then, we take the average of these three values to be $\sigma_{2}^2$ and can evaluate the objective function. These steps are illustrated below for the first chromosome.

\begin{center}
    \begin{tikzpicture}
        \node[rectangle, draw, text centered, minimum height=14cm, minimum width=\linewidth,
        fill=white] (background){};
        \matrix [below = 0.1cm of background.north] (chromosome1) [matrix of nodes,
            every even row/.style = {nodes={fill=green!10}},
            every odd row/.style = {nodes={fill=green!20}},
            column 1/.style = {nodes={minimum width=1cm}},
            column 2/.style = {nodes={minimum width=1cm}},
            column 3/.style = {nodes={minimum width=1cm}},
            ]{
            -1 & -1 & -1 \\
            -1 & \phone & \phone \\
            \phone & -1 & -1 \\ 
            \phone & -1 & \phone\\
            \phone & \phone & -1 \\ 
            \phone & \phone & \phone\\
            };
        \matrix [below left = 1.3cm and 1.3cm of chromosome1.south west] (XAB) [matrix of nodes,
            column 1/.style = {nodes={minimum width=1cm}},
            column 2/.style = {nodes={minimum width=1cm}},
            column 3/.style = {nodes={minimum width=1cm}},
            column 4/.style = {nodes={minimum width=1cm}},
            column 5/.style = {nodes={minimum width=1cm, fill=gray!20}},
            ]{
            1 & -1 & -1 & -1 & \phone \\
            1 & -1 & \phone & \phone & -1 \\
            1 & \phone & -1 & -1 & -1\\ 
            1 & \phone & -1 & \phone & -1\\
            1 & \phone & \phone & -1 & \phone\\ 
            1 & \phone & \phone & \phone & \phone\\
        };
        \matrix [right = 0.3cm of XAB] (XAC) [matrix of nodes,
            column 1/.style = {nodes={minimum width=1cm}},
            column 2/.style = {nodes={minimum width=1cm}},
            column 3/.style = {nodes={minimum width=1cm}},
            column 4/.style = {nodes={minimum width=1cm}},
            column 5/.style = {nodes={minimum width=1cm, fill=gray!20}},
            ]{
            1 & -1 & -1 & -1 & \phone \\
            1 & -1 & \phone & \phone & -1  \\
            1 & \phone & -1 & -1 & -1\\ 
            1 & \phone & -1 & \phone & \phone\\
            1 & \phone & \phone & -1 & -1\\ 
            1 & \phone & \phone & \phone & \phone \\
        };
        \matrix [right = 0.3cm of XAC] (XBC) [matrix of nodes,
            column 1/.style = {nodes={minimum width=1cm}},
            column 2/.style = {nodes={minimum width=1cm}},
            column 3/.style = {nodes={minimum width=1cm}},
            column 4/.style = {nodes={minimum width=1cm}},
            column 5/.style = {nodes={minimum width=1cm, fill=gray!20}},
            ]{
            1 & -1 & -1 & -1 &  \phone\\
            1 & -1 & \phone & \phone & \phone  \\
            1 & \phone & -1 & -1 & \phone \\ 
            1 & \phone & -1 & \phone & -1\\
            1 & \phone & \phone & -1 & -1\\ 
            1 & \phone & \phone & \phone & \phone \\
        };
        \matrix [below =0.8cm of XAB.south] (InvFisherAB) [matrix of nodes,
            column 1/.style = {nodes={minimum width=1cm}},
            column 2/.style = {nodes={minimum width=1cm}},
            column 3/.style = {nodes={minimum width=1cm}},
            column 4/.style = {nodes={minimum width=1cm}},
            column 5/.style = {nodes={minimum width=1cm}}
            ]{
            \phantom{-}0.19 & -0.06 & \phantom{-}0.00 & \phantom{-}0.00 & \phantom{-}0.00 \\
            -0.06 & \phantom{-}0.19 & \phantom{-}0.00 & \phantom{-}0.00 & \phantom{-}0.00 \\
            \phantom{-}0.00 & \phantom{-}0.00 & \phantom{-}0.25 & -0.13 & -0.13 \\
            \phantom{-}0.00 & \phantom{-}0.00 & -0.13 & \phantom{-}0.25 & \phantom{-}0.13 \\
            \phantom{-}0.00 & \phantom{-}0.00 & -0.13 & \phantom{-}0.13 &  |[red]| \phantom{-}0.25\\
        };
        \matrix [below =0.8cm of XAC.south] (InvFisherAC) [matrix of nodes,
            column 1/.style = {nodes={minimum width=1cm}},
            column 2/.style = {nodes={minimum width=1cm}},
            column 3/.style = {nodes={minimum width=1cm}},
            column 4/.style = {nodes={minimum width=1cm}},
            column 5/.style = {nodes={minimum width=1cm}}
            ]{
            \phantom{-}0.19 & -0.06 & \phantom{-}0.00 & \phantom{-}0.00 & \phantom{-}0.00 \\
            -0.06 & 0.19 & \phantom{-}0.00 & \phantom{-}0.00 & \phantom{-}0.00 \\
            \phantom{-}0.00 & \phantom{-}0.00 & \phantom{-}0.25 & -0.13 & \phantom{-}0.13 \\
            \phantom{-}0.00 & \phantom{-}0.00 & -0.13 & \phantom{-}0.25 & -0.13 \\
            \phantom{-}0.00 & \phantom{-}0.00 & \phantom{-}0.13 & -0.13 &  |[red]| \phantom{-}0.25\\
        };
        \matrix [below =0.8cm of XBC.south] (InvFisherBC) [matrix of nodes,
            column 1/.style = {nodes={minimum width=1cm}},
            column 2/.style = {nodes={minimum width=1cm}},
            column 3/.style = {nodes={minimum width=1cm}},
            column 4/.style = {nodes={minimum width=1cm}},
            column 5/.style = {nodes={minimum width=1cm}}
            ]{
            \phantom{-}0.25 & -0.13 & \phantom{-}0.00 & \phantom{-}0.00 & -0.13 \\
            -0.13 & \phantom{-}0.25 & \phantom{-}0.00 & \phantom{-}0.00 & \phantom{-}0.13 \\
            \phantom{-}0.00 & \phantom{-}0.00 & \phantom{-}0.19 & -0.06 & \phantom{-}0.00 \\
            \phantom{-}0.00 & \phantom{-}0.00 & -0.06 & \phantom{-}0.19 & \phantom{-}0.00 \\
            -0.13 & \phantom{-}0.13 & \phantom{-}0.0 & \phantom{-}0.0 &  |[red]| \phantom{-}0.25\\
        };
        \node [below=0.25cm of InvFisherAB.south, text centered](sigmaAB){$\sigma_{21}^2 = {\color{red} 0.25}$};
        \node [below=0.25cm of InvFisherAC.south, text centered](sigmaAC){$\sigma_{22}^2 = {\color{red} 0.25}$};
        \node [below=0.25cm of InvFisherBC.south, text centered](sigmaBC){$\sigma_{23}^2 = {\color{red} 0.25}$};
        \node [below=1.2cm of InvFisherAC.south, text centered] (fitness) {$\overline{\sigma_{\beta_{2}}^2} = 0.25$, \;\;\; Fitness = 4.0};
        \draw [dashed] (-2.7,3) -- (-2.7,-5.5);
        \draw [dashed] (2.85,3) -- (2.85,-5.5);
        \draw [dashed] (-8,3) -- (8,3);
        \draw [dashed] (-8,-5.5) -- (8,-5.5);
        \node [above=0.65of XAB.north, text centered]{Model $(AB)$};
        \node [above=0.65of XAC.north, text centered]{Model $(AC)$};
        \node [above=0.65of XBC.north, text centered]{Model $(BC)$};
        \node [above=0.01of XAB.north, text centered]{$\mathbf{X}^{(1)}$};
        \node [above=0.01of XAC.north, text centered]{$\mathbf{X}^{(2)}$};
        \node [above=0.01of XBC.north, text centered]{$\mathbf{X}^{(3)}$};
        \node [above=0.01of InvFisherAB.north, text centered]{$|\mathbf{X}^{(1)^T}\mathbf{X}^{(1)}|^{-1}$};
        \node [above=0.01of InvFisherAC.north, text centered]{$|\mathbf{X}^{(2)^T}\mathbf{X}^{(2)}|^{-1}$};
        \node [above=0.01of InvFisherBC.north, text centered]{$|\mathbf{X}^{(3)^T}\mathbf{X}^{(3)}|^{-1}$};
        \node [below right=0.01 and 0.01 of background.north west, text centered]{Illustration of fitness calculation.};
    \end{tikzpicture}
\end{center}


The above procedure is repeated for each of the three chromosomes. In this case, all three designs end up having the same fitness value. We now summarize each chromosome below:

\begin{center}
\begin{tikzpicture}
\node[rectangle, draw, text centered, minimum height=4.6cm, minimum width=\linewidth,
    fill=white] (background){};
\matrix [below right= 0.5 and 0.5 of background.north west] (chromosome1) [matrix of nodes,
    every even row/.style = {nodes={fill=green!10}},
    every odd row/.style = {nodes={fill=green!20}},
    column 1/.style = {nodes={minimum width=1cm}},
    column 2/.style = {nodes={minimum width=1cm}},
    column 3/.style = {nodes={minimum width=1cm}},
    ]{
    -1 & -1 & -1 \\
    -1 & \phone & \phone \\
    \phone & -1 & -1 \\ 
    \phone & -1 & \phone\\
    \phone & \phone & -1 \\ 
    \phone & \phone & \phone\\
    };
\matrix [below=0.5 of background.north] (chromosome2) [matrix of nodes,
    every even row/.style = {nodes={fill=red!10}},
    every odd row/.style = {nodes={fill=red!20}},
    column 1/.style = {nodes={minimum width=1cm}},
    column 2/.style = {nodes={minimum width=1cm}},
    column 3/.style = {nodes={minimum width=1cm}},
    ]{
    -1 & -1 & -1 \\
    -1 & \phone & -1 \\
    \phone & -1 & -1 \\ 
    \phone & -1 & \phone\\
    \phone & \phone & -1 \\ 
    \phone & \phone & \phone\\
    };
\matrix [below left= 0.5 and 0.5 of background.north east] (chromosome3) [matrix of nodes,
    every even row/.style = {nodes={fill=blue!10}},
    every odd row/.style = {nodes={fill=blue!20}},
    column 1/.style = {nodes={minimum width=1cm}},
    column 2/.style = {nodes={minimum width=1cm}},
    column 3/.style = {nodes={minimum width=1cm}},
    ]{
    -1 & -1 & -1 \\
    -1 & -1 & \phone \\
    -1 & \phone & \phone \\ 
    \phone & -1 & \phone\\
    \phone & \phone & -1 \\ 
    \phone & \phone & \phone\\
    };
\node [below=0.1cm of chromosome1.south, text centered, fill=gray!10, draw]{Fitness: 4.0};
\node [below=0.1cm of chromosome2.south, text centered, fill=gray!10, draw]{Fitness: 4.0};
\node [below=0.1cm of chromosome3.south, text centered, fill=gray!10, draw]{Fitness: 4.0};
\node [above=0.1of background.north, text centered]{Population};
\node [above=0.01of chromosome1.north, text centered]{Chromosome 1};
\node [above=0.01of chromosome2.north, text centered]{Chromosome 2};
\node [above=0.01of chromosome3.north, text centered]{Chromosome 3};
\end{tikzpicture}
\end{center}

Now that we have completed the initialization process, we can begin the main loop over the algorithm.

\underline{\it 1. Identify worst chromosome(s)}

The first step is to identify the worst chromosomes. These are the chromosomes that will be replaced by new offspring. Since we only have three chromosomes in the population, we will only identify and replace the single worst chromosome. In the case of a tie (as we have here), the chromosome to be replaced is randomly chosen. In this case we have chosen chromosome 3 to be replaced.

\underline{\it 2. Generate replacement using crossover}

We next generate a replacement for the worst chromosome (3) using crossover from 2 randomly selecting remaining chromosomes. Since our example only has three chromosomes, we simply use the remaining chromosomes (1 and 2). In the crossover, a random cut point is selected, and the two chromosomes are combined using the values from the first chromosome for the factors to the left of the cut point, and the values from the second chromosome for the factors to the right of the cut point. This process is illustrated below:

\begin{center}
\begin{tikzpicture}
\node[rectangle, draw, text centered, minimum height=4.5cm, minimum width=\linewidth,
    fill=white] (background){};
\matrix [below right= 0.6 and 0.5 of background.north west] (chromosome1) [matrix of nodes,
    every even row/.style = {nodes={fill=green!10}},
    every odd row/.style = {nodes={fill=green!20}},
    column 1/.style = {nodes={minimum width=1cm}},
    column 2/.style = {nodes={minimum width=1cm}},
    column 3/.style = {nodes={minimum width=1cm}},
    ]{
    -1 & -1 & -1 \\
    -1 & \phone & \phone \\
    \phone & -1 & -1 \\ 
    \phone & -1 & \phone\\
    \phone & \phone & -1 \\ 
    \phone & \phone & \phone\\
    };
\matrix [right=1cm of chromosome1.east] (chromosome2) [matrix of nodes,
    every even row/.style = {nodes={fill=red!10}},
    every odd row/.style = {nodes={fill=red!20}},
    column 1/.style = {nodes={minimum width=1cm}},
    column 2/.style = {nodes={minimum width=1cm}},
    column 3/.style = {nodes={minimum width=1cm}},
    ]{
    -1 & -1 & -1 \\
    -1 & \phone & -1 \\
    \phone & -1 & -1 \\ 
    \phone & -1 & \phone\\
    \phone & \phone & -1 \\ 
    \phone & \phone & \phone\\
    };
\matrix [below left= 0.6 and 0.5 of background.north east] (offspring) [matrix of nodes,
    column 1/.style = {nodes={minimum width=1cm, fill=green!20}},
    column 2/.style = {nodes={minimum width=1cm, fill=green!20}},
    column 3/.style = {nodes={minimum width=1cm, fill=red!10}},
    ]{
    -1 & -1 & -1 \\
    -1 & -1 & \phone \\
    -1 & \phone & \phone \\ 
    \phone & -1 & \phone\\
    \phone & \phone & -1 \\ 
    \phone & \phone & \phone\\
    };
\draw [->, ultra thick] ( [xshift=1cm] chromosome2.east) -- ([xshift=-0.5cm]offspring.west);
\node [above=0.1of background.north, text centered]{Crossover};
\draw [dashed, red] (-5.65,1.5) -- (-5.65,-1.8);
\draw [dashed, red] (-1.3,1.5) -- (-1.3,-1.8);
\node [above=0.01of chromosome1.north, text centered]{Chromosome 1};
\node [above=0.01of chromosome2.north, text centered]{Chromosome 2};
\node [above=0.01of offspring.north, text centered]{Offspring};
\node [below right=0.1 and 0.01 of chromosome1.south, text centered]{cut point};
\node [below right=0.1 and 0.01 of chromosome2.south, text centered]{cut point};
\end{tikzpicture}
\end{center}

Note that it is possible to consider other ways of producing offspring via crossover. For example, the cut point could be different for each support point, or they could be ``horizontal" instead of ``vertical," choosing certain rows from the first chromosome and the remaining rows from the second.

\underline{\it 3. Mutation}

In addition to crossover, more novelty can be introduced to the solution by randomly changing, or mutating, some of factor settings. For our purpose, the probability of each factor setting (gene) mutating is identical.

\underline{\it 4. Replacement and Fitness Evaluation}

Following Steps 3 and 4, we are now ready to replace the old chromosome with the offspring. In this step, the worst chromosome(s) is replaced by the offspring created in Steps 2$-$3. The fitness of this new chromosome is evaluated and stored. 

}

\section*{Appendix $-$ Tables for Example 3}

Tables \ref{table:sim1}$-$\ref{table:sim3} below present detailed results for each of the comparisons in Example 3.

\begin{table}
		\centering
		\resizebox{\textwidth}{!}{
		\begin{tabular}{ccc|cc|cc|cc|cc|cc|cc|cc}
			\hline
		\multirow{2}{*}{} & \multirow{2}{*}{Model} & \multirow{2}{*}{Size} & \multicolumn{2}{c}{$\sigma=0.1$} & \multicolumn{2}{c}{$\sigma=0.25$} & \multicolumn{2}{c}{$\sigma=0.5$} & \multicolumn{2}{c}{$\sigma=0.75$} & \multicolumn{2}{c}{$\sigma=1$} & \multicolumn{2}{c}{$\sigma=1.25$} & \multicolumn{2}{c}{$\sigma=1.5$}  \\
			\cline{4-17}
			& & & CV & PB & CV & PB & CV & PB & CV & PB & CV & PB & CV & PB & CV & PB\\
			\hline
 1 & F1 & b & 100.00 & 100.00 & 95.38 & 95.14 & 74.86 & 75.88 & 68.08 & 70.70 & 63.68 & 66.18 & 59.32 & 61.46 & 55.40 & 55.54 \\ 
  2 & F1 & s & 78.14 & 78.84 & 34.82 & 33.72 & 10.22 & 9.76 & 5.92 & 6.16 & 4.22 & 4.00 & 3.58 & 3.08 & 3.00 & 2.92 \\ 
  3 & F1+F2 & b+b & 100.00 & 100.00 & 96.04 & 95.66 & 68.18 & 69.28 & 54.46 & 58.34 & 47.70 & 49.34 & 39.98 & 40.18 & 32.96 & 32.90 \\ 
  4 & F1+F2 & b+s & 56.90 & 55.54 & 34.20 & 33.08 & 8.20 & 7.82 & 4.26 & 4.28 & 3.10 & 3.10 & 2.32 & 2.44 & 2.38 & 1.88 \\ 
  5 & F1+F2 & s+s & 77.32 & 73.44 & 14.66 & 12.32 & 1.58 & 1.40 & 0.46 & 0.62 & 0.28 & 0.26 & 0.26 & 0.14 & 0.14 & 0.16 \\ 
  6 & F1+F1F2 & b+b & 100.00 & 100.00 & 98.12 & 97.08 & 77.10 & 76.62 & 66.78 & 66.54 & 57.10 & 57.88 & 51.32 & 49.20 & 43.66 & 41.00 \\ 
  7 & F1+F1F2 & b+s & 56.70 & 56.58 & 38.52 & 37.56 & 10.94 & 11.16 & 6.42 & 6.10 & 4.76 & 4.44 & 3.74 & 3.22 & 3.40 & 3.14 \\ 
  8 & F1+F1F2 & s+s & 84.54 & 79.68 & 29.98 & 25.62 & 7.30 & 5.74 & 4.02 & 3.02 & 2.00 & 2.00 & 1.72 & 1.30 & 1.26 & 1.54 \\ 
  9 & F1+F2+F1F2 & b+b+b & 100.00 & 100.00 & 98.50 & 97.20 & 67.74 & 68.34 & 48.92 & 51.74 & 42.68 & 39.90 & 32.12 & 32.04 & 28.04 & 25.68 \\ 
  10 & F1+F2+F1F2 & b+b+s & 37.82 & 36.88 & 45.96 & 40.48 & 13.38 & 9.98 & 7.42 & 5.14 & 6.02 & 4.06 & 4.58 & 3.26 & 3.54 & 2.58 \\ 
  11 & F1+F2+F1F2 & b+s+b & 35.52 & 38.36 & 40.50 & 37.10 & 14.14 & 10.94 & 6.70 & 5.08 & 4.86 & 4.08 & 4.54 & 3.64 & 3.90 & 3.04 \\ 
  12 & F1+F2+F1F2 & b+s+s & 50.68 & 49.40 & 28.02 & 22.66 & 4.56 & 2.64 & 1.20 & 0.88 & 0.62 & 0.48 & 0.56 & 0.62 & 0.46 & 0.44 \\ 
  13 & F1+F2+F1F2 & s+s+s & 81.68 & 75.38 & 17.04 & 11.72 & 2.38 & 1.18 & 0.86 & 0.62 & 0.30 & 0.30 & 0.30 & 0.22 & 0.16 & 0.22 \\ 
  14 & F1+F2+F3 & b+b+b & 100.00 & 100.00 & 97.52 & 96.08 & 61.90 & 54.18 & 42.96 & 36.66 & 31.64 & 25.60 & 23.54 & 20.44 & 18.66 & 16.32 \\ 
  15 & F1+F2+F3 & b+b+s & 38.06 & 37.36 & 36.96 & 34.76 & 10.14 & 8.28 & 4.42 & 4.12 & 3.06 & 3.40 & 2.94 & 2.54 & 2.18 & 1.98 \\ 
  16 & F1+F2+F3 & b+s+s & 37.22 & 36.44 & 19.02 & 17.84 & 2.02 & 1.68 & 0.70 & 0.68 & 0.46 & 0.36 & 0.22 & 0.22 & 0.20 & 0.18 \\ 
  17 & F1+F2+F3 & s+s+s & 77.30 & 69.38 & 8.94 & 6.54 & 0.46 & 0.38 & 0.06 & 0.14 & 0.10 & 0.02 & 0.02 & 0.00 & 0.06 & 0.04 \\ 
  18 & F1+F2+F1F3 & b+b+b & 100.00 & 96.30 & 98.66 & 91.14 & 69.58 & 60.16 & 50.40 & 44.04 & 41.16 & 31.04 & 31.16 & 22.62 & 25.30 & 18.46 \\ 
  19 & F1+F2+F1F3 & b+b+s & 41.62 & 30.92 & 36.56 & 24.74 & 15.18 & 9.52 & 7.20 & 4.82 & 4.12 & 4.02 & 3.74 & 2.92 & 3.10 & 2.52 \\ 
  20 & F1+F2+F1F3 & b+s+b & 35.54 & 28.18 & 37.66 & 30.10 & 12.14 & 8.70 & 5.54 & 4.08 & 4.64 & 3.06 & 3.28 & 2.68 & 2.40 & 2.40 \\ 
  21 & F1+F2+F1F3 & b+s+s & 41.84 & 31.94 & 20.48 & 12.30 & 3.32 & 2.20 & 1.28 & 0.52 & 0.70 & 0.68 & 0.42 & 0.34 & 0.46 & 0.20 \\ 
  22 & F1+F2+F1F3 & s+b+s & 40.74 & 36.20 & 23.18 & 12.86 & 3.80 & 2.64 & 1.44 & 0.76 & 0.56 & 0.36 & 0.26 & 0.26 & 0.16 & 0.28 \\ 
  23 & F1+F2+F1F3 & s+s+s & 81.24 & 61.44 & 12.70 & 7.64 & 1.02 & 0.42 & 0.26 & 0.16 & 0.06 & 0.06 & 0.06 & 0.14 & 0.06 & 0.02 \\ 
  24 & F1+F2+F3+F1F3 & b+b+b+b & 100.00 & 86.52 & 98.80 & 87.86 & 63.30 & 51.68 & 36.18 & 28.08 & 23.32 & 19.72 & 15.42 & 11.78 & 13.30 & 10.08 \\ 
  25 & F1+F2+F3+F1F3 & b+b+s+s & 23.88 & 15.08 & 24.82 & 13.82 & 5.22 & 1.86 & 1.42 & 0.72 & 0.96 & 0.36 & 0.56 & 0.28 & 0.38 & 0.24 \\ 
  26 & F1+F2+F3+F1F3 & b+s+s+b & 27.62 & 24.40 & 17.58 & 13.38 & 3.90 & 1.42 & 1.66 & 0.80 & 1.00 & 0.42 & 0.84 & 0.28 & 0.60 & 0.34 \\ 
  27 & F1+F2+F3+F1F3 & s+s+s+s & 84.62 & 48.26 & 8.96 & 3.40 & 0.38 & 0.18 & 0.08 & 0.06 & 0.04 & 0.02 & 0.02 & 0.04 & 0.00 & 0.00 \\ 
  28 & F1+F2+F3+F1F3+F2F3 & b+b+b+b+b & 95.16 & 45.48 & 95.38 & 34.42 & 70.66 & 31.50 & 39.42 & 12.04 & 27.16 & 8.22 & 19.22 & 8.20 & 13.30 & 4.52 \\ 
  29 & F1+F2+F3+F1F3+F2F3 & b+b+s+s+s & 12.74 & 2.58 & 6.96 & 2.76 & 1.66 & 0.56 & 0.52 & 0.18 & 0.34 & 0.16 & 0.14 & 0.14 & 0.14 & 0.10 \\ 
  30 & F1+F2+F3+F1F3+F2F3 & b+s+s+b+b & 12.28 & 6.44 & 10.80 & 4.72 & 5.86 & 1.92 & 3.56 & 1.04 & 1.70 & 0.48 & 1.22 & 0.50 & 0.74 & 0.40 \\ 
  31 & F1+F2+F3+F1F3+F2F3 & s+s+s+s+s & 57.96 & 24.54 & 4.56 & 1.54 & 0.16 & 0.12 & 0.06 & 0.02 & 0.00 & 0.00 & 0.02 & 0.00 & 0.02 & 0.00 \\ 
  32 & F1+F2+F3+F4+F5+F1F2 & b+b+b+b+b+b & 97.76 & 19.12 & 95.10 & 15.06 & 79.38 & 12.74 & 49.36 & 6.86 & 32.74 & 3.40 & 18.88 & 1.82 & 13.22 & 1.42 \\ 
  33 & F1+F2+F3+F4+F5+F1F2 & b+b+s+s+s+s & 0.66 & 0.54 & 0.30 & 0.20 & 0.10 & 0.04 & 0.12 & 0.02 & 0.06 & 0.04 & 0.10 & 0.02 & 0.02 & 0.02 \\ 
  34 & F1+F2+F3+F4+F5+F1F2 & b+s+s+b+b+b & 4.48 & 0.36 & 6.84 & 2.66 & 4.88 & 2.00 & 2.64 & 1.16 & 1.88 & 0.56 & 1.40 & 0.46 & 0.76 & 0.26 \\ 
  35 & F1+F2+F3+F4+F5+F1F2 & s+s+s+s+s+s & 23.78 & 2.60 & 1.52 & 0.38 & 0.08 & 0.00 & 0.04 & 0.00 & 0.02 & 0.00 & 0.00 & 0.00 & 0.00 & 0.00 \\
			\hline
		\end{tabular}
		}
	\caption{Average Percentage of Correctly Identified Models for the common variance design $D^1$ and the Plackett-Burman design.}
    \label{table:sim1}
	\end{table}

\begin{table}
		\centering
		\resizebox{\textwidth}{!}{
		\begin{tabular}{ccc|cc|cc|cc|cc|cc|cc|cc}
		\hline
		\multirow{2}{*}{} & \multirow{2}{*}{Model} & \multirow{2}{*}{Size} & \multicolumn{2}{c}{$\sigma=0.1$} & \multicolumn{2}{c}{$\sigma=0.25$} & \multicolumn{2}{c}{$\sigma=0.5$} & \multicolumn{2}{c}{$\sigma=0.75$} & \multicolumn{2}{c}{$\sigma=1$} & \multicolumn{2}{c}{$\sigma=1.25$} & \multicolumn{2}{c}{$\sigma=1.5$}  \\
			\cline{4-17}
			& & & A-ComVar & PB & A-ComVar & PB & A-ComVar & PB & A-ComVar & PB & A-ComVar & PB & A-ComVar & PB & A-ComVar & PB\\
			\hline
1 & F1 & b & 100.00 & 100.00 & 94.48 & 95.16 & 74.48 & 77.18 & 68.42 & 71.04 & 61.98 & 65.46 & 56.88 & 61.56 & 53.58 & 57.36 \\ 
  2 & F1 & s & 75.58 & 79.76 & 33.24 & 34.52 & 10.00 & 9.74 & 6.88 & 5.60 & 4.70 & 4.16 & 3.20 & 3.14 & 3.36 & 3.02 \\ 
  3 & F1+F2 & b+b & 100.00 & 100.00 & 91.44 & 95.88 & 65.14 & 69.96 & 54.94 & 59.42 & 48.96 & 48.48 & 42.64 & 40.48 & 33.78 & 32.78 \\ 
  4 & F1+F2 & b+s & 44.64 & 49.28 & 33.22 & 35.96 & 10.56 & 8.92 & 5.80 & 4.04 & 4.04 & 2.86 & 3.44 & 2.36 & 2.76 & 1.68 \\ 
  5 & F1+F2 & s+s & 63.08 & 71.40 & 15.64 & 14.44 & 2.06 & 1.74 & 0.70 & 0.50 & 0.48 & 0.28 & 0.22 & 0.16 & 0.26 & 0.12 \\ 
  6 & F1+F1F2 & b+b & 100.00 & 100.00 & 96.44 & 97.40 & 71.22 & 77.90 & 61.44 & 68.20 & 52.14 & 56.54 & 44.82 & 49.74 & 38.26 & 44.20 \\ 
  7 & F1+F1F2 & b+s & 57.04 & 63.24 & 31.96 & 38.66 & 7.98 & 10.74 & 4.22 & 5.80 & 2.90 & 3.90 & 2.00 & 3.52 & 2.38 & 2.48 \\ 
  8 & F1+F1F2 & s+s & 71.52 & 78.20 & 20.52 & 24.04 & 4.88 & 6.20 & 2.06 & 2.54 & 1.70 & 1.58 & 1.44 & 1.30 & 1.10 & 1.36 \\ 
  9 & F1+F2+F1F2 & b+b+b & 100.00 & 100.00 & 95.82 & 98.08 & 60.24 & 69.26 & 45.58 & 54.36 & 38.70 & 42.74 & 31.16 & 30.80 & 26.42 & 26.68 \\ 
  10 & F1+F2+F1F2 & b+b+s & 49.40 & 50.78 & 35.46 & 41.84 & 9.06 & 11.16 & 4.00 & 6.04 & 2.86 & 3.36 & 2.26 & 2.44 & 1.64 & 2.28 \\ 
  11 & F1+F2+F1F2 & b+s+b & 45.72 & 43.06 & 37.64 & 34.38 & 13.16 & 9.72 & 7.52 & 5.66 & 5.74 & 4.16 & 4.48 & 2.70 & 4.66 & 3.58 \\ 
  12 & F1+F2+F1F2 & b+s+s & 43.42 & 46.72 & 17.60 & 20.92 & 2.78 & 2.80 & 1.02 & 0.78 & 0.62 & 0.64 & 0.40 & 0.30 & 0.42 & 0.22 \\ 
  13 & F1+F2+F1F2 & s+s+s & 66.24 & 75.08 & 11.96 & 11.94 & 1.40 & 1.46 & 0.58 & 0.64 & 0.48 & 0.40 & 0.22 & 0.24 & 0.10 & 0.14 \\ 
  14 & F1+F2+F3 & b+b+b & 100.00 & 100.00 & 91.70 & 96.78 & 43.58 & 51.96 & 30.92 & 33.12 & 27.02 & 26.10 & 27.02 & 21.24 & 23.70 & 17.02 \\ 
  15 & F1+F2+F3 & b+b+s & 47.14 & 42.76 & 34.52 & 33.60 & 12.16 & 9.24 & 5.60 & 4.48 & 3.60 & 2.62 & 3.24 & 2.26 & 2.70 & 1.88 \\ 
  16 & F1+F2+F3 & b+s+s & 22.68 & 32.52 & 18.18 & 20.60 & 2.04 & 1.90 & 0.88 & 0.44 & 0.64 & 0.36 & 0.66 & 0.12 & 0.46 & 0.24 \\ 
  17 & F1+F2+F3 & s+s+s & 65.00 & 70.60 & 6.92 & 6.38 & 0.64 & 0.32 & 0.12 & 0.06 & 0.10 & 0.02 & 0.04 & 0.00 & 0.14 & 0.02 \\ 
  18 & F1+F2+F1F3 & b+b+b & 100.00 & 97.28 & 94.60 & 92.96 & 57.02 & 60.60 & 41.00 & 44.78 & 36.78 & 32.48 & 30.02 & 25.36 & 25.12 & 20.24 \\ 
  19 & F1+F2+F1F3 & b+b+s & 47.68 & 40.20 & 30.34 & 28.82 & 9.52 & 8.76 & 3.76 & 4.48 & 3.10 & 3.58 & 2.34 & 3.04 & 1.62 & 1.98 \\ 
  20 & F1+F2+F1F3 & b+s+b & 42.98 & 33.18 & 45.38 & 33.68 & 13.98 & 8.80 & 6.70 & 4.26 & 4.68 & 3.54 & 3.90 & 3.16 & 3.28 & 2.22 \\ 
  21 & F1+F2+F1F3 & b+s+s & 39.96 & 31.36 & 19.72 & 16.24 & 2.30 & 1.96 & 0.94 & 0.76 & 0.48 & 0.40 & 0.56 & 0.22 & 0.20 & 0.16 \\ 
  22 & F1+F2+F1F3 & s+b+s & 32.62 & 32.32 & 16.50 & 13.42 & 2.72 & 2.42 & 0.84 & 0.76 & 0.58 & 0.52 & 0.48 & 0.38 & 0.26 & 0.30 \\ 
  23 & F1+F2+F1F3 & s+s+s & 67.20 & 58.36 & 7.38 & 6.30 & 0.82 & 0.44 & 0.10 & 0.20 & 0.06 & 0.04 & 0.06 & 0.04 & 0.08 & 0.08 \\ 
  24 & F1+F2+F3+F1F3 & b+b+b+b & 99.90 & 84.42 & 95.72 & 78.94 & 55.94 & 44.94 & 32.72 & 25.64 & 25.32 & 19.80 & 21.10 & 13.88 & 15.00 & 8.70 \\ 
  25 & F1+F2+F3+F1F3 & b+b+s+s & 23.04 & 15.04 & 15.74 & 8.82 & 3.34 & 2.38 & 0.98 & 0.70 & 0.64 & 0.54 & 0.38 & 0.46 & 0.38 & 0.36 \\ 
  26 & F1+F2+F3+F1F3 & b+s+s+b & 14.88 & 13.84 & 16.38 & 10.16 & 3.76 & 1.80 & 2.18 & 0.88 & 1.34 & 0.34 & 0.66 & 0.40 & 0.42 & 0.20 \\ 
  27 & F1+F2+F3+F1F3 & s+s+s+s & 66.74 & 55.54 & 4.48 & 3.84 & 0.20 & 0.16 & 0.04 & 0.04 & 0.00 & 0.00 & 0.04 & 0.00 & 0.00 & 0.00 \\ 
  28 & F1+F2+F3+F1F3+F2F3 & b+b+b+b+b & 80.00 & 29.72 & 60.46 & 49.98 & 41.84 & 33.08 & 21.62 & 17.04 & 15.24 & 8.34 & 7.14 & 5.20 & 6.54 & 5.90 \\ 
  29 & F1+F2+F3+F1F3+F2F3 & b+b+s+s+s & 1.94 & 2.26 & 1.08 & 1.88 & 0.22 & 0.68 & 0.14 & 0.22 & 0.12 & 0.10 & 0.10 & 0.06 & 0.10 & 0.08 \\ 
  30 & F1+F2+F3+F1F3+F2F3 & b+s+s+b+b & 7.10 & 3.88 & 4.90 & 6.14 & 1.70 & 1.86 & 1.48 & 1.48 & 0.62 & 0.62 & 0.70 & 0.26 & 0.70 & 0.26 \\ 
  31 & F1+F2+F3+F1F3+F2F3 & s+s+s+s+s & 28.96 & 22.60 & 3.58 & 2.00 & 0.06 & 0.06 & 0.06 & 0.04 & 0.00 & 0.00 & 0.00 & 0.00 & 0.00 & 0.00 \\ 
  32 & F1+F2+F3+F4+F5+F1F2 & b+b+b+b+b+b & 90.96 & 18.12 & 80.90 & 16.62 & 60.14 & 12.80 & 33.42 & 5.92 & 19.70 & 2.38 & 11.04 & 2.14 & 7.30 & 1.76 \\ 
  33 & F1+F2+F3+F4+F5+F1F2 & b+b+s+s+s+s & 7.78 & 0.42 & 1.76 & 0.52 & 0.26 & 0.12 & 0.30 & 0.02 & 0.06 & 0.02 & 0.16 & 0.04 & 0.06 & 0.00 \\ 
  34 & F1+F2+F3+F4+F5+F1F2 & b+s+s+b+b+b & 9.98 & 3.02 & 9.50 & 1.38 & 5.66 & 1.60 & 3.38 & 1.14 & 2.06 & 0.40 & 1.64 & 0.26 & 0.86 & 0.16 \\ 
  35 & F1+F2+F3+F4+F5+F1F2 & s+s+s+s+s+s & 30.52 & 3.00 & 1.60 & 0.30 & 0.04 & 0.02 & 0.02 & 0.00 & 0.02 & 0.00 & 0.08 & 0.00 & 0.00 & 0.00 \\ 
			\hline
		\end{tabular}
		}
	\caption{Average Percentage of Correctly Identified Models for A-ComVar design $D^2$ and the Plackett-Burman design.}
    \label{table:sim2}
	\end{table}
	
\begin{table}
		\centering
		\resizebox{\textwidth}{!}{
		\begin{tabular}{ccc|cc|cc|cc|cc|cc|cc|cc}
			\hline
			\multirow{2}{*}{} & \multirow{2}{*}{Model} & \multirow{2}{*}{Size} & \multicolumn{2}{c}{$\sigma=0.1$} & \multicolumn{2}{c}{$\sigma=0.25$} & \multicolumn{2}{c}{$\sigma=0.5$} & \multicolumn{2}{c}{$\sigma=0.75$} & \multicolumn{2}{c}{$\sigma=1$} & \multicolumn{2}{c}{$\sigma=1.25$} & \multicolumn{2}{c}{$\sigma=1.5$}  \\
			\cline{4-17}
			& & & A-ComVar & Bayes & A-ComVar & Bayes & A-ComVar & Bayes & A-ComVar & Bayes & A-ComVar & Bayes & A-ComVar & Bayes & A-ComVar & Bayes\\
			\hline
1 & F1 & b & 100.00 & 100.00 & 95.68 & 95.98 & 76.08 & 77.82 & 66.98 & 69.48 & 61.70 & 64.56 & 57.26 & 58.62 & 54.26 & 55.62 \\ 
  2 & F1 & s & 75.26 & 80.12 & 35.42 & 34.28 & 12.28 & 11.68 & 6.74 & 5.52 & 4.38 & 4.22 & 3.86 & 3.62 & 3.24 & 3.06 \\ 
  3 & F1+F2 & b+b & 100.00 & 100.00 & 93.92 & 95.82 & 63.48 & 69.86 & 53.68 & 59.24 & 48.52 & 49.12 & 40.42 & 39.14 & 34.44 & 31.08 \\ 
  4 & F1+F2 & b+s & 53.24 & 58.86 & 29.78 & 31.78 & 10.02 & 8.10 & 5.92 & 4.94 & 4.36 & 2.82 & 3.32 & 1.94 & 3.18 & 1.68 \\ 
  5 & F1+F2 & s+s & 63.96 & 71.90 & 13.62 & 12.04 & 1.84 & 1.22 & 0.48 & 0.50 & 0.32 & 0.28 & 0.30 & 0.10 & 0.34 & 0.12 \\ 
  6 & F1+F1F2 & b+b & 100.00 & 100.00 & 96.12 & 96.82 & 71.04 & 76.08 & 61.32 & 66.18 & 53.06 & 57.74 & 43.76 & 48.94 & 38.30 & 42.60 \\ 
  7 & F1+F1F2 & b+s & 55.92 & 62.68 & 26.06 & 33.22 & 8.22 & 9.48 & 4.10 & 5.70 & 2.64 & 3.60 & 2.28 & 3.20 & 1.74 & 2.98 \\ 
  8 & F1+F1F2 & s+s & 75.86 & 81.54 & 20.96 & 24.34 & 4.34 & 5.32 & 2.44 & 2.66 & 1.18 & 2.28 & 1.16 & 1.54 & 0.94 & 1.14 \\ 
  9 & F1+F2+F1F2 & b+b+b & 100.00 & 100.00 & 95.14 & 97.58 & 63.36 & 70.28 & 43.76 & 53.06 & 37.56 & 40.76 & 31.18 & 31.48 & 27.42 & 25.70 \\ 
  10 & F1+F2+F1F2 & b+b+s & 36.04 & 41.98 & 37.62 & 43.92 & 9.66 & 10.90 & 4.04 & 5.42 & 2.28 & 3.64 & 2.30 & 3.06 & 1.66 & 2.60 \\ 
  11 & F1+F2+F1F2 & b+s+b & 39.22 & 38.96 & 46.70 & 42.58 & 14.30 & 10.82 & 7.24 & 5.84 & 5.24 & 3.96 & 4.60 & 3.78 & 4.46 & 3.60 \\ 
  12 & F1+F2+F1F2 & b+s+s & 41.06 & 45.52 & 16.94 & 20.60 & 2.54 & 2.68 & 0.70 & 0.76 & 0.68 & 0.54 & 0.40 & 0.18 & 0.34 & 0.42 \\ 
  13 & F1+F2+F1F2 & s+s+s & 68.24 & 75.40 & 11.50 & 11.72 & 2.26 & 1.82 & 0.60 & 0.60 & 0.48 & 0.26 & 0.22 & 0.16 & 0.24 & 0.10 \\ 
  14 & F1+F2+F3 & b+b+b & 100.00 & 100.00 & 92.24 & 96.56 & 44.58 & 54.62 & 30.22 & 36.40 & 28.46 & 26.80 & 28.64 & 20.00 & 23.12 & 15.90 \\ 
  15 & F1+F2+F3 & b+b+s & 46.38 & 42.38 & 37.58 & 37.56 & 11.52 & 9.22 & 6.70 & 5.32 & 3.94 & 2.76 & 3.52 & 2.14 & 2.72 & 1.78 \\ 
  16 & F1+F2+F3 & b+s+s & 34.76 & 50.40 & 10.26 & 14.26 & 2.56 & 1.52 & 1.22 & 0.62 & 0.64 & 0.32 & 0.42 & 0.20 & 0.44 & 0.12 \\ 
  17 & F1+F2+F3 & s+s+s & 61.98 & 67.32 & 6.04 & 5.06 & 0.70 & 0.38 & 0.22 & 0.10 & 0.10 & 0.08 & 0.02 & 0.04 & 0.08 & 0.00 \\ 
  18 & F1+F2+F1F3 & b+b+b & 100.00 & 97.48 & 94.26 & 89.32 & 58.04 & 59.28 & 44.68 & 37.54 & 36.10 & 28.78 & 32.24 & 23.28 & 23.82 & 18.28 \\ 
  19 & F1+F2+F1F3 & b+b+s & 44.02 & 36.26 & 31.46 & 26.06 & 7.90 & 9.12 & 4.76 & 4.34 & 3.32 & 3.80 & 1.78 & 2.18 & 1.62 & 2.30 \\ 
  20 & F1+F2+F1F3 & b+s+b & 46.56 & 38.66 & 44.34 & 33.52 & 11.74 & 7.60 & 7.78 & 4.38 & 6.02 & 3.12 & 4.46 & 2.64 & 3.68 & 2.10 \\ 
  21 & F1+F2+F1F3 & b+s+s & 33.38 & 26.12 & 17.52 & 11.26 & 2.92 & 1.94 & 1.00 & 0.62 & 0.60 & 0.34 & 0.42 & 0.26 & 0.30 & 0.18 \\ 
  22 & F1+F2+F1F3 & s+b+s & 36.84 & 33.70 & 16.72 & 15.46 & 2.92 & 2.32 & 0.90 & 0.68 & 0.46 & 0.64 & 0.70 & 0.30 & 0.28 & 0.28 \\ 
  23 & F1+F2+F1F3 & s+s+s & 68.38 & 61.24 & 6.14 & 5.84 & 0.62 & 0.26 & 0.18 & 0.06 & 0.06 & 0.02 & 0.10 & 0.08 & 0.00 & 0.02 \\ 
  24 & F1+F2+F3+F1F3 & b+b+b+b & 99.94 & 90.34 & 95.58 & 83.50 & 54.14 & 47.02 & 31.40 & 27.72 & 23.24 & 17.26 & 19.54 & 13.10 & 16.02 & 8.48 \\ 
  25 & F1+F2+F3+F1F3 & b+b+s+s & 24.28 & 18.78 & 16.38 & 7.92 & 3.42 & 2.04 & 1.36 & 0.86 & 0.52 & 0.44 & 0.44 & 0.48 & 0.22 & 0.30 \\ 
  26 & F1+F2+F3+F1F3 & b+s+s+b & 21.52 & 15.70 & 14.48 & 7.86 & 3.98 & 1.26 & 2.10 & 0.68 & 1.10 & 0.26 & 0.66 & 0.34 & 0.66 & 0.36 \\ 
  27 & F1+F2+F3+F1F3 & s+s+s+s & 67.98 & 56.30 & 3.80 & 2.56 & 0.18 & 0.18 & 0.10 & 0.04 & 0.04 & 0.00 & 0.00 & 0.00 & 0.04 & 0.00 \\ 
  28 & F1+F2+F3+F1F3+F2F3 & b+b+b+b+b & 76.06 & 35.42 & 64.30 & 32.20 & 37.64 & 23.04 & 16.26 & 14.34 & 12.66 & 10.46 & 9.02 & 7.24 & 5.04 & 4.06 \\ 
  29 & F1+F2+F3+F1F3+F2F3 & b+b+s+s+s & 2.22 & 2.12 & 1.48 & 1.02 & 0.20 & 0.28 & 0.24 & 0.20 & 0.20 & 0.04 & 0.10 & 0.14 & 0.04 & 0.06 \\ 
  30 & F1+F2+F3+F1F3+F2F3 & b+s+s+b+b & 6.16 & 4.62 & 6.04 & 8.60 & 3.24 & 2.24 & 1.32 & 1.26 & 0.80 & 0.64 & 0.78 & 0.30 & 0.60 & 0.28 \\ 
  31 & F1+F2+F3+F1F3+F2F3 & s+s+s+s+s & 34.24 & 30.44 & 3.46 & 1.18 & 0.20 & 0.10 & 0.04 & 0.00 & 0.02 & 0.00 & 0.02 & 0.00 & 0.00 & 0.02 \\ 
  32 & F1+F2+F3+F4+F5+F1F2 & b+b+b+b+b+b & 83.12 & 23.72 & 83.90 & 26.24 & 62.28 & 15.98 & 35.18 & 6.58 & 19.10 & 4.40 & 11.66 & 2.34 & 6.70 & 1.44 \\ 
  33 & F1+F2+F3+F4+F5+F1F2 & b+b+s+s+s+s & 4.30 & 0.74 & 1.48 & 0.10 & 0.36 & 0.02 & 0.10 & 0.02 & 0.10 & 0.00 & 0.08 & 0.00 & 0.04 & 0.02 \\ 
  34 & F1+F2+F3+F4+F5+F1F2 & b+s+s+b+b+b & 11.14 & 2.42 & 11.58 & 2.96 & 7.30 & 1.28 & 3.08 & 0.68 & 2.00 & 0.48 & 1.64 & 0.32 & 1.14 & 0.26 \\ 
  35 & F1+F2+F3+F4+F5+F1F2 & s+s+s+s+s+s & 35.38 & 2.30 & 1.90 & 0.36 & 0.14 & 0.02 & 0.06 & 0.00 & 0.00 & 0.00 & 0.00 & 0.00 & 0.00 & 0.00 \\ 
			\hline
		\end{tabular}
		}
	\caption{Average Percentage of Correctly Identified Models for A-ComVar design $D^3$ and Bayes Optimal design $D^4$ from \cite{bingham2007incorporating}.}
    \label{table:sim3}
	\end{table}

\begin{table}
		\centering
		\resizebox{\textwidth}{!}{
		\begin{tabular}{ccc|cc|cc|cc|cc|cc|cc|cc}
			\hline
			\multirow{2}{*}{} & \multirow{2}{*}{Model} & \multirow{2}{*}{Size} & \multicolumn{2}{c}{$\sigma=0.1$} & \multicolumn{2}{c}{$\sigma=0.25$} & \multicolumn{2}{c}{$\sigma=0.5$} & \multicolumn{2}{c}{$\sigma=0.75$} & \multicolumn{2}{c}{$\sigma=1$} & \multicolumn{2}{c}{$\sigma=1.25$} & \multicolumn{2}{c}{$\sigma=1.5$}  \\
			\cline{4-17}
			& & & A-ComVar & Nachtsheim & A-ComVar & Nachtsheim & A-ComVar & Nachtsheim & A-ComVar & Nachtsheim & A-ComVar & Nachtsheim & A-ComVar & Nachtsheim & A-ComVar & Nachtsheim\\
			\hline
1 & F1 & b & 100.00 & 100.00 & 94.60 & 94.70 & 75.24 & 76.30 & 67.78 & 69.22 & 62.76 & 65.16 & 57.78 & 60.58 & 53.98 & 55.42 \\ 
  2 & F1 & s & 75.40 & 79.34 & 36.32 & 36.74 & 12.72 & 11.86 & 6.30 & 6.30 & 4.86 & 4.06 & 3.52 & 3.08 & 3.40 & 3.68 \\ 
  3 & F1+F2 & b+b & 100.00 & 99.04 & 91.34 & 95.26 & 64.94 & 70.58 & 53.42 & 60.66 & 50.90 & 50.44 & 45.04 & 40.56 & 36.64 & 34.40 \\ 
  4 & F1+F2 & b+s & 48.16 & 51.68 & 33.42 & 32.64 & 8.82 & 6.26 & 4.96 & 3.62 & 3.44 & 2.42 & 3.12 & 2.48 & 2.54 & 2.20 \\ 
  5 & F1+F2 & s+s & 62.60 & 65.90 & 11.72 & 12.30 & 2.24 & 1.64 & 0.78 & 0.60 & 0.50 & 0.24 & 0.34 & 0.20 & 0.28 & 0.18 \\ 
  6 & F1+F1F2 & b+b & 100.00 & 100.00 & 95.32 & 96.88 & 70.74 & 73.46 & 61.02 & 65.22 & 53.02 & 54.06 & 44.12 & 48.44 & 39.74 & 40.98 \\ 
  7 & F1+F1F2 & b+s & 65.12 & 69.86 & 32.06 & 36.06 & 8.32 & 10.62 & 3.88 & 5.30 & 2.72 & 3.80 & 2.42 & 2.92 & 2.08 & 2.40 \\ 
  8 & F1+F1F2 & s+s & 74.26 & 79.96 & 19.72 & 22.58 & 5.14 & 6.34 & 2.32 & 2.90 & 1.90 & 2.24 & 1.36 & 1.82 & 0.90 & 1.14 \\ 
  9 & F1+F2+F1F2 & b+b+b & 100.00 & 100.00 & 94.22 & 97.36 & 61.38 & 67.32 & 46.70 & 52.20 & 36.96 & 42.46 & 31.56 & 34.12 & 27.88 & 27.24 \\ 
  10 & F1+F2+F1F2 & b+b+s & 48.84 & 40.06 & 30.52 & 28.32 & 9.34 & 8.90 & 4.32 & 5.50 & 3.34 & 3.92 & 2.34 & 2.82 & 1.54 & 2.12 \\ 
  11 & F1+F2+F1F2 & b+s+b & 46.60 & 42.90 & 41.00 & 37.82 & 14.04 & 11.34 & 7.48 & 5.16 & 5.52 & 4.26 & 5.32 & 3.44 & 4.60 & 4.00 \\ 
  12 & F1+F2+F1F2 & b+s+s & 38.72 & 38.36 & 15.28 & 16.58 & 2.76 & 2.52 & 0.86 & 0.78 & 0.62 & 0.58 & 0.42 & 0.52 & 0.38 & 0.16 \\ 
  13 & F1+F2+F1F2 & s+s+s & 71.42 & 71.20 & 11.54 & 11.78 & 1.90 & 1.60 & 0.58 & 0.36 & 0.46 & 0.38 & 0.28 & 0.20 & 0.26 & 0.12 \\ 
  14 & F1+F2+F3 & b+b+b & 100.00 & 97.34 & 93.98 & 88.64 & 46.60 & 60.04 & 27.30 & 44.80 & 28.24 & 37.44 & 25.16 & 31.02 & 21.92 & 20.80 \\ 
  15 & F1+F2+F3 & b+b+s & 38.78 & 25.56 & 32.72 & 20.66 & 10.80 & 5.84 & 5.82 & 3.92 & 4.74 & 2.88 & 3.14 & 2.72 & 3.26 & 1.64 \\ 
  16 & F1+F2+F3 & b+s+s & 28.92 & 25.14 & 13.46 & 14.18 & 2.00 & 1.26 & 1.20 & 0.64 & 0.56 & 0.28 & 0.60 & 0.44 & 0.42 & 0.04 \\ 
  17 & F1+F2+F3 & s+s+s & 62.20 & 59.68 & 6.42 & 6.78 & 1.02 & 0.40 & 0.28 & 0.00 & 0.08 & 0.04 & 0.06 & 0.02 & 0.06 & 0.02 \\ 
  18 & F1+F2+F1F3 & b+b+b & 100.00 & 98.30 & 94.74 & 96.94 & 57.28 & 68.26 & 43.18 & 51.42 & 35.26 & 36.08 & 28.88 & 33.54 & 23.30 & 23.38 \\ 
  19 & F1+F2+F1F3 & b+b+s & 37.68 & 29.58 & 29.96 & 25.76 & 9.28 & 10.04 & 4.14 & 5.32 & 3.14 & 3.68 & 1.90 & 2.98 & 1.52 & 2.70 \\ 
  20 & F1+F2+F1F3 & b+s+b & 48.30 & 41.20 & 39.40 & 34.16 & 13.64 & 10.40 & 6.48 & 5.16 & 5.16 & 4.38 & 4.18 & 3.52 & 3.30 & 2.94 \\ 
  21 & F1+F2+F1F3 & b+s+s & 32.10 & 35.14 & 17.12 & 20.26 & 3.12 & 2.76 & 1.10 & 0.90 & 0.64 & 0.46 & 0.50 & 0.38 & 0.32 & 0.36 \\ 
  22 & F1+F2+F1F3 & s+b+s & 39.32 & 36.94 & 14.06 & 15.54 & 2.80 & 2.68 & 0.84 & 0.78 & 0.46 & 0.60 & 0.42 & 0.32 & 0.34 & 0.30 \\ 
  23 & F1+F2+F1F3 & s+s+s & 64.94 & 67.38 & 7.30 & 7.84 & 0.70 & 0.70 & 0.08 & 0.14 & 0.06 & 0.12 & 0.14 & 0.10 & 0.08 & 0.06 \\ 
  24 & F1+F2+F3+F1F3 & b+b+b+b & 100.00 & 98.96 & 96.54 & 93.96 & 52.84 & 65.94 & 32.32 & 38.26 & 24.60 & 28.18 & 20.28 & 21.50 & 15.26 & 13.22 \\ 
  25 & F1+F2+F3+F1F3 & b+b+s+s & 30.10 & 11.60 & 15.76 & 7.86 & 2.66 & 1.94 & 1.12 & 0.78 & 0.80 & 0.54 & 0.36 & 0.26 & 0.28 & 0.24 \\ 
  26 & F1+F2+F3+F1F3 & b+s+s+b & 22.56 & 18.12 & 16.50 & 8.98 & 4.54 & 2.08 & 1.62 & 0.80 & 1.38 & 0.66 & 0.78 & 0.46 & 0.50 & 0.26 \\ 
  27 & F1+F2+F3+F1F3 & s+s+s+s & 63.66 & 53.70 & 5.86 & 4.24 & 0.28 & 0.18 & 0.02 & 0.02 & 0.04 & 0.04 & 0.04 & 0.00 & 0.02 & 0.02 \\ 
  28 & F1+F2+F3+F1F3+F2F3 & b+b+b+b+b & 81.88 & 58.50 & 63.34 & 62.88 & 32.62 & 35.16 & 22.28 & 22.72 & 10.56 & 13.14 & 9.58 & 9.90 & 5.90 & 6.70 \\ 
  29 & F1+F2+F3+F1F3+F2F3 & b+b+s+s+s & 1.00 & 2.80 & 2.08 & 2.74 & 0.34 & 0.32 & 0.18 & 0.18 & 0.16 & 0.06 & 0.04 & 0.10 & 0.04 & 0.08 \\ 
  30 & F1+F2+F3+F1F3+F2F3 & b+s+s+b+b & 3.38 & 5.88 & 4.98 & 4.74 & 2.40 & 2.16 & 1.38 & 0.84 & 0.78 & 0.78 & 0.56 & 0.42 & 0.68 & 0.36 \\ 
  31 & F1+F2+F3+F1F3+F2F3 & s+s+s+s+s & 32.76 & 27.20 & 2.46 & 2.80 & 0.08 & 0.12 & 0.04 & 0.00 & 0.02 & 0.04 & 0.00 & 0.04 & 0.00 & 0.00 \\ 
  32 & F1+F2+F3+F4+F5+F1F2 & b+b+b+b+b+b & 85.74 & 4.78 & 87.16 & 4.30 & 62.10 & 2.66 & 33.26 & 2.06 & 18.52 & 2.02 & 11.68 & 0.62 & 7.54 & 0.90 \\ 
  33 & F1+F2+F3+F4+F5+F1F2 & b+b+s+s+s+s & 3.48 & 0.06 & 1.40 & 0.08 & 0.30 & 0.02 & 0.12 & 0.00 & 0.10 & 0.00 & 0.14 & 0.00 & 0.16 & 0.02 \\ 
  34 & F1+F2+F3+F4+F5+F1F2 & b+s+s+b+b+b & 9.72 & 0.16 & 9.46 & 0.58 & 6.14 & 0.50 & 3.82 & 0.48 & 2.08 & 0.58 & 1.16 & 0.20 & 0.94 & 0.26 \\ 
  35 & F1+F2+F3+F4+F5+F1F2 & s+s+s+s+s+s & 34.28 & 1.24 & 1.86 & 0.02 & 0.10 & 0.00 & 0.04 & 0.02 & 0.06 & 0.00 & 0.02 & 0.00 & 0.02 & 0.00 \\ 
			\hline
		\end{tabular}
		}
	\caption{Average Percentage of Correctly Identified Models for A-ComVar design $D^3$ and design $D^5$ from \cite{li2000model}.}
    \label{table:sim3}
	\end{table}

\begin{table}
		\centering
		\resizebox{\textwidth}{!}{
		\begin{tabular}{ccc|cc|cc|cc|cc|cc|cc|cc}
			\hline
			\multirow{2}{*}{} & \multirow{2}{*}{Model} & \multirow{2}{*}{Size} & \multicolumn{2}{c}{$\sigma=0.1$} & \multicolumn{2}{c}{$\sigma=0.25$} & \multicolumn{2}{c}{$\sigma=0.5$} & \multicolumn{2}{c}{$\sigma=0.75$} & \multicolumn{2}{c}{$\sigma=1$} & \multicolumn{2}{c}{$\sigma=1.25$} & \multicolumn{2}{c}{$\sigma=1.5$}  \\
			\cline{4-17}
			& & & A-ComVar & Tian & A-ComVar & Tian & A-ComVar & Tian & A-ComVar & Tian & A-ComVar & Tian & A-ComVar & Tian & A-ComVar & Tian\\
			\hline
 1 & F1 & b & 100.00 & 100.00 & 96.12 & 95.78 & 75.22 & 75.58 & 67.58 & 67.82 & 63.18 & 63.72 & 57.32 & 58.14 & 53.04 & 53.10 \\ 
  2 & F1 & s & 77.68 & 80.32 & 29.36 & 30.58 & 10.28 & 9.72 & 5.72 & 5.12 & 5.30 & 4.16 & 4.08 & 3.42 & 3.54 & 3.18 \\ 
  3 & F1+F2 & b+b & 100.00 & 100.00 & 93.00 & 96.26 & 61.86 & 68.26 & 54.50 & 56.08 & 49.66 & 48.18 & 42.86 & 40.28 & 35.66 & 32.78 \\ 
  4 & F1+F2 & b+s & 42.60 & 49.34 & 36.26 & 40.94 & 10.08 & 8.88 & 5.18 & 3.98 & 4.18 & 3.16 & 3.22 & 2.48 & 3.02 & 1.90 \\ 
  5 & F1+F2 & s+s & 63.60 & 74.36 & 14.58 & 15.12 & 1.82 & 1.78 & 0.78 & 0.38 & 0.40 & 0.42 & 0.36 & 0.22 & 0.26 & 0.14 \\ 
  6 & F1+F1F2 & b+b & 100.00 & 100.00 & 95.14 & 97.50 & 71.52 & 76.86 & 60.10 & 66.28 & 50.82 & 56.94 & 45.86 & 50.12 & 38.24 & 45.48 \\ 
  7 & F1+F1F2 & b+s & 56.04 & 62.22 & 30.78 & 41.08 & 8.16 & 11.02 & 4.06 & 5.78 & 2.60 & 4.44 & 2.08 & 3.58 & 1.98 & 3.02 \\ 
  8 & F1+F1F2 & s+s & 73.00 & 83.06 & 19.90 & 29.64 & 4.74 & 6.84 & 2.24 & 3.34 & 1.76 & 3.08 & 1.68 & 1.56 & 0.86 & 1.32 \\ 
  9 & F1+F2+F1F2 & b+b+b & 100.00 & 100.00 & 95.66 & 98.88 & 59.06 & 66.88 & 45.52 & 50.78 & 36.82 & 39.50 & 32.00 & 34.64 & 27.22 & 29.40 \\ 
  10 & F1+F2+F1F2 & b+b+s & 44.38 & 46.18 & 35.06 & 44.38 & 8.26 & 13.66 & 4.26 & 6.38 & 3.18 & 5.18 & 2.34 & 3.82 & 1.82 & 2.98 \\ 
  11 & F1+F2+F1F2 & b+s+b & 52.88 & 45.86 & 39.60 & 37.40 & 14.76 & 14.44 & 8.52 & 7.08 & 5.34 & 4.48 & 5.24 & 3.96 & 4.38 & 4.58 \\ 
  12 & F1+F2+F1F2 & b+s+s & 39.40 & 48.62 & 20.78 & 31.20 & 2.02 & 4.30 & 0.74 & 0.90 & 0.62 & 0.60 & 0.40 & 0.44 & 0.24 & 0.52 \\ 
  13 & F1+F2+F1F2 & s+s+s & 68.96 & 83.90 & 13.40 & 18.72 & 1.32 & 1.98 & 0.64 & 0.70 & 0.52 & 0.28 & 0.22 & 0.26 & 0.38 & 0.28 \\ 
  14 & F1+F2+F3 & b+b+b & 100.00 & 100.00 & 92.84 & 97.36 & 42.16 & 61.20 & 34.08 & 40.74 & 29.50 & 32.66 & 27.52 & 26.44 & 24.20 & 20.06 \\ 
  15 & F1+F2+F3 & b+b+s & 47.30 & 47.36 & 37.42 & 37.18 & 9.54 & 9.38 & 5.26 & 4.74 & 4.30 & 3.20 & 3.60 & 2.80 & 2.70 & 2.06 \\ 
  16 & F1+F2+F3 & b+s+s & 28.52 & 40.26 & 14.46 & 17.48 & 2.98 & 2.88 & 1.02 & 0.74 & 0.76 & 0.48 & 0.48 & 0.46 & 0.32 & 0.18 \\ 
  17 & F1+F2+F3 & s+s+s & 64.14 & 77.44 & 6.50 & 7.88 & 0.46 & 0.54 & 0.30 & 0.12 & 0.06 & 0.06 & 0.10 & 0.00 & 0.00 & 0.02 \\ 
  18 & F1+F2+F1F3 & b+b+b & 100.00 & 100.00 & 95.64 & 98.30 & 57.84 & 68.56 & 44.06 & 49.12 & 36.04 & 38.68 & 28.94 & 29.62 & 23.22 & 25.40 \\ 
  19 & F1+F2+F1F3 & b+b+s & 48.34 & 44.18 & 30.96 & 40.40 & 9.50 & 13.30 & 4.18 & 6.64 & 2.84 & 3.80 & 2.14 & 3.92 & 2.02 & 2.62 \\ 
  20 & F1+F2+F1F3 & b+s+b & 42.44 & 34.94 & 40.88 & 39.40 & 13.62 & 14.16 & 7.30 & 6.06 & 5.34 & 4.38 & 4.10 & 3.48 & 3.02 & 2.66 \\ 
  21 & F1+F2+F1F3 & b+s+s & 44.42 & 44.30 & 17.78 & 24.50 & 2.70 & 3.28 & 1.10 & 1.70 & 0.54 & 0.70 & 0.40 & 0.24 & 0.40 & 0.22 \\ 
  22 & F1+F2+F1F3 & s+b+s & 28.98 & 34.10 & 14.74 & 24.62 & 2.62 & 4.02 & 0.94 & 1.72 & 0.56 & 0.80 & 0.36 & 0.56 & 0.42 & 0.36 \\ 
  23 & F1+F2+F1F3 & s+s+s & 65.72 & 82.84 & 6.82 & 11.16 & 0.70 & 1.10 & 0.14 & 0.18 & 0.20 & 0.12 & 0.06 & 0.02 & 0.06 & 0.06 \\ 
  24 & F1+F2+F3+F1F3 & b+b+b+b & 99.92 & 100.00 & 96.14 & 99.24 & 54.44 & 60.80 & 33.04 & 33.54 & 24.12 & 22.66 & 20.86 & 16.66 & 15.48 & 14.12 \\ 
  25 & F1+F2+F3+F1F3 & b+b+s+s & 27.32 & 22.44 & 15.40 & 23.02 & 2.40 & 4.22 & 1.16 & 1.58 & 0.80 & 0.86 & 0.54 & 0.56 & 0.30 & 0.34 \\ 
  26 & F1+F2+F3+F1F3 & b+s+s+b & 20.20 & 19.94 & 16.72 & 18.52 & 3.56 & 3.88 & 1.80 & 1.50 & 0.90 & 0.86 & 0.62 & 0.46 & 1.02 & 0.42 \\ 
  27 & F1+F2+F3+F1F3 & s+s+s+s & 64.62 & 82.02 & 3.94 & 8.26 & 0.28 & 0.44 & 0.06 & 0.08 & 0.02 & 0.02 & 0.00 & 0.00 & 0.00 & 0.02 \\ 
  28 & F1+F2+F3+F1F3+F2F3 & b+b+b+b+b & 80.30 & 91.90 & 62.82 & 95.18 & 34.38 & 68.36 & 20.78 & 39.74 & 12.32 & 25.46 & 9.08 & 18.46 & 5.04 & 14.28 \\ 
  29 & F1+F2+F3+F1F3+F2F3 & b+b+s+s+s & 3.32 & 20.20 & 1.54 & 7.36 & 0.28 & 2.00 & 0.22 & 0.58 & 0.12 & 0.32 & 0.14 & 0.24 & 0.12 & 0.20 \\ 
  30 & F1+F2+F3+F1F3+F2F3 & b+s+s+b+b & 5.06 & 7.12 & 4.46 & 8.96 & 2.70 & 5.90 & 1.66 & 3.04 & 1.02 & 1.92 & 0.78 & 1.12 & 0.40 & 0.88 \\ 
  31 & F1+F2+F3+F1F3+F2F3 & s+s+s+s+s & 24.42 & 58.00 & 1.40 & 5.72 & 0.16 & 0.28 & 0.08 & 0.04 & 0.00 & 0.02 & 0.02 & 0.02 & 0.00 & 0.00 \\ 
  32 & F1+F2+F3+F4+F5+F1F2 & b+b+b+b+b+b & 87.12 & 97.18 & 83.34 & 96.94 & 61.18 & 77.40 & 32.78 & 48.54 & 18.52 & 30.96 & 11.18 & 19.32 & 7.76 & 13.26 \\ 
  33 & F1+F2+F3+F4+F5+F1F2 & b+b+s+s+s+s & 4.42 & 0.56 & 1.52 & 0.36 & 0.30 & 0.22 & 0.10 & 0.06 & 0.04 & 0.06 & 0.12 & 0.14 & 0.06 & 0.04 \\ 
  34 & F1+F2+F3+F4+F5+F1F2 & b+s+s+b+b+b & 12.96 & 7.40 & 11.02 & 8.74 & 5.46 & 5.32 & 3.34 & 2.60 & 2.04 & 1.88 & 1.30 & 1.30 & 1.14 & 0.70 \\ 
  35 & F1+F2+F3+F4+F5+F1F2 & s+s+s+s+s+s & 34.18 & 22.26 & 1.56 & 1.44 & 0.06 & 0.02 & 0.02 & 0.00 & 0.02 & 0.00 & 0.00 & 0.00 & 0.04 & 0.00 \\ 
			\hline
		\end{tabular}
		}
	\caption{Average Percentage of Correctly Identified Models for A-ComVar design $D^3$ and design $D^6$ from \cite{ghosh2006optimum}.}
    \label{table:sim3}
	\end{table}

\begin{table}
		\centering
		\resizebox{\textwidth}{!}{
		\begin{tabular}{ccc|cc|cc|cc|cc|cc|cc|cc}
			\hline
			\multirow{2}{*}{} & \multirow{2}{*}{Model} & \multirow{2}{*}{Size} & \multicolumn{2}{c}{$\sigma=0.1$} & \multicolumn{2}{c}{$\sigma=0.25$} & \multicolumn{2}{c}{$\sigma=0.5$} & \multicolumn{2}{c}{$\sigma=0.75$} & \multicolumn{2}{c}{$\sigma=1$} & \multicolumn{2}{c}{$\sigma=1.25$} & \multicolumn{2}{c}{$\sigma=1.5$}  \\
			\cline{4-17}
			& & & A-ComVar & CCD & A-ComVar & CCD & A-ComVar & CCD & A-ComVar & CCD & A-ComVar & CCD & A-ComVar & CCD & A-ComVar & CCD\\
			\hline
1 & F1 & b & 100.00 & 100.00 & 95.90 & 96.16 & 78.24 & 79.10 & 75.30 & 74.48 & 75.12 & 70.80 & 70.20 & 69.98 & 69.30 & 66.98 \\ 
  2 & F1 & s & 72.86 & 73.04 & 52.26 & 49.62 & 21.30 & 22.00 & 13.28 & 12.70 & 9.48 & 10.36 & 7.78 & 8.30 & 7.00 & 6.56 \\ 
  3 & F1+F2 & b+b & 100.00 & 100.00 & 95.34 & 96.22 & 75.70 & 74.12 & 69.62 & 69.00 & 65.10 & 64.24 & 62.76 & 61.84 & 60.10 & 57.46 \\ 
  4 & F1+F2 & b+s & 63.76 & 62.90 & 39.32 & 41.38 & 15.78 & 15.68 & 11.22 & 11.84 & 7.58 & 8.30 & 6.22 & 6.94 & 6.28 & 6.32 \\ 
  5 & F1+F2 & s+s & 67.04 & 66.74 & 35.06 & 34.42 & 7.40 & 6.76 & 2.56 & 3.02 & 1.28 & 1.42 & 0.94 & 0.86 & 0.86 & 1.10 \\ 
  6 & F1+F1F2 & b+b & 100.00 & 100.00 & 92.94 & 93.80 & 75.22 & 75.40 & 71.56 & 72.14 & 69.10 & 66.98 & 65.44 & 64.74 & 60.48 & 57.62 \\ 
  7 & F1+F1F2 & b+s & 58.70 & 53.14 & 34.38 & 27.88 & 15.36 & 13.62 & 9.98 & 8.96 & 7.64 & 7.20 & 7.52 & 6.08 & 5.44 & 5.80 \\ 
  8 & F1+F1F2 & s+s & 69.70 & 68.42 & 36.84 & 34.18 & 13.30 & 12.18 & 6.82 & 7.06 & 5.30 & 5.14 & 4.50 & 4.60 & 3.80 & 3.52 \\ 
  9 & F1+F2+F1F2 & b+b+b & 100.00 & 100.00 & 93.58 & 93.26 & 72.32 & 73.82 & 68.24 & 68.10 & 63.78 & 63.74 & 59.22 & 60.28 & 54.58 & 55.94 \\ 
  10 & F1+F2+F1F2 & b+b+s & 42.50 & 33.62 & 31.88 & 29.36 & 15.86 & 14.16 & 12.38 & 10.18 & 9.06 & 9.68 & 8.22 & 8.90 & 8.36 & 8.56 \\ 
  11 & F1+F2+F1F2 & b+s+b & 75.32 & 78.08 & 40.28 & 43.80 & 24.34 & 24.90 & 13.90 & 14.76 & 10.36 & 10.92 & 10.16 & 9.84 & 8.36 & 8.74 \\ 
  12 & F1+F2+F1F2 & b+s+s & 69.64 & 68.36 & 27.22 & 26.36 & 7.08 & 6.66 & 3.06 & 3.08 & 2.20 & 2.14 & 1.48 & 1.50 & 1.38 & 1.52 \\ 
  13 & F1+F2+F1F2 & s+s+s & 65.38 & 66.00 & 29.28 & 28.02 & 3.58 & 3.82 & 1.02 & 1.12 & 0.62 & 0.58 & 0.46 & 0.44 & 0.30 & 0.42 \\ 
  14 & F1+F2+F3 & b+b+b & 100.00 & 100.00 & 96.96 & 96.22 & 76.50 & 75.64 & 70.44 & 69.36 & 67.64 & 65.98 & 63.10 & 62.58 & 59.72 & 58.88 \\ 
  15 & F1+F2+F3 & b+b+s & 68.86 & 73.80 & 38.26 & 39.32 & 23.06 & 24.48 & 14.78 & 14.52 & 11.38 & 12.02 & 9.22 & 9.50 & 8.82 & 8.98 \\ 
  16 & F1+F2+F3 & b+s+s & 76.42 & 77.70 & 38.28 & 40.14 & 10.42 & 10.50 & 4.04 & 3.98 & 2.12 & 2.36 & 1.60 & 1.84 & 1.50 & 1.92 \\ 
  17 & F1+F2+F3 & s+s+s & 68.62 & 70.40 & 28.04 & 28.90 & 5.02 & 5.20 & 1.46 & 1.50 & 0.54 & 1.00 & 0.24 & 0.44 & 0.24 & 0.52 \\ 
  18 & F1+F2+F1F3 & b+b+b & 100.00 & 100.00 & 93.90 & 94.06 & 72.88 & 71.66 & 67.86 & 67.26 & 64.92 & 63.86 & 59.32 & 60.94 & 56.94 & 56.64 \\ 
  19 & F1+F2+F1F3 & b+b+s & 46.02 & 40.10 & 30.52 & 28.40 & 16.06 & 14.98 & 11.34 & 11.56 & 9.54 & 9.76 & 8.40 & 9.00 & 8.22 & 8.14 \\ 
  20 & F1+F2+F1F3 & b+s+b & 65.34 & 70.84 & 40.60 & 42.72 & 22.42 & 22.34 & 13.06 & 13.84 & 10.70 & 12.18 & 10.08 & 8.82 & 8.60 & 8.82 \\ 
  21 & F1+F2+F1F3 & b+s+s & 70.14 & 68.44 & 28.44 & 28.72 & 7.06 & 7.02 & 3.14 & 2.90 & 2.22 & 1.76 & 1.60 & 1.68 & 1.40 & 1.32 \\ 
  22 & F1+F2+F1F3 & s+b+s & 62.18 & 57.40 & 27.12 & 26.96 & 7.88 & 7.94 & 3.48 & 3.48 & 2.56 & 2.20 & 1.36 & 1.78 & 1.32 & 1.42 \\ 
  23 & F1+F2+F1F3 & s+s+s & 66.44 & 65.94 & 23.78 & 25.24 & 3.98 & 4.30 & 1.04 & 1.24 & 0.68 & 0.66 & 0.36 & 0.26 & 0.24 & 0.32 \\ 
  24 & F1+F2+F3+F1F3 & b+b+b+b & 100.00 & 100.00 & 95.98 & 96.62 & 77.84 & 77.98 & 72.38 & 73.38 & 67.74 & 69.08 & 62.90 & 65.64 & 55.50 & 59.94 \\ 
  25 & F1+F2+F3+F1F3 & b+b+s+s & 47.24 & 46.70 & 36.76 & 36.94 & 9.56 & 9.80 & 4.38 & 4.38 & 3.10 & 3.20 & 2.22 & 2.72 & 2.18 & 2.22 \\ 
  26 & F1+F2+F3+F1F3 & b+s+s+b & 75.60 & 77.60 & 45.90 & 48.00 & 13.40 & 13.92 & 5.22 & 6.20 & 3.90 & 3.64 & 2.80 & 3.24 & 1.66 & 2.26 \\ 
  27 & F1+F2+F3+F1F3 & s+s+s+s & 71.72 & 73.32 & 25.94 & 26.32 & 3.12 & 3.22 & 0.64 & 0.84 & 0.36 & 0.46 & 0.14 & 0.34 & 0.18 & 0.10 \\ 
  28 & F1+F2+F3+F1F3+F2F3 & b+b+b+b+b & 100.00 & 100.00 & 95.42 & 96.90 & 84.44 & 84.62 & 77.00 & 79.24 & 72.62 & 73.66 & 68.44 & 68.68 & 60.28 & 60.16 \\ 
  29 & F1+F2+F3+F1F3+F2F3 & b+b+s+s+s & 49.14 & 45.36 & 35.18 & 32.44 & 8.02 & 6.82 & 2.32 & 2.50 & 1.72 & 1.14 & 1.44 & 1.04 & 1.12 & 1.30 \\ 
  30 & F1+F2+F3+F1F3+F2F3 & b+s+s+b+b & 61.28 & 63.64 & 54.34 & 59.20 & 16.20 & 17.52 & 9.88 & 12.24 & 6.92 & 7.12 & 5.74 & 6.70 & 5.42 & 6.34 \\ 
  31 & F1+F2+F3+F1F3+F2F3 & s+s+s+s+s & 80.08 & 83.16 & 26.12 & 27.80 & 2.68 & 3.02 & 0.56 & 0.88 & 0.26 & 0.48 & 0.18 & 0.38 & 0.20 & 0.14 \\ 
			\hline
		\end{tabular}
		}
	    \caption{Average Percentage of Correctly Identified Models for $3-$level A-ComVar design $D^2$ and CCD  $D^7$.}
    \label{table:sim3}
	\end{table}

\begin{table}
		\centering
		\resizebox{\textwidth}{!}{
		\begin{tabular}{ccc|cc|cc|cc|cc|cc|cc|cc}
			\hline
			\multirow{3}{*}{} & \multirow{3}{*}{Model} & \multirow{3}{*}{Size} & \multicolumn{2}{c}{$\sigma=0.1$} & \multicolumn{2}{c}{$\sigma=0.25$} & \multicolumn{2}{c}{$\sigma=0.5$} & \multicolumn{2}{c}{$\sigma=0.75$} & \multicolumn{2}{c}{$\sigma=1$} & \multicolumn{2}{c}{$\sigma=1.25$} & \multicolumn{2}{c}{$\sigma=1.5$}  \\
			\cline{4-17}
			& & & A-ComVar & OME & A-ComVar & OME & A-ComVar & OME & A-ComVar & OME & A-ComVar & OME & A-ComVar & OME & A-ComVar & OME\\
			\hline
 1 & F1 & b & 100.00 & 100.00 & 89.86 & 89.82 & 71.44 & 71.24 & 67.02 & 68.26 & 63.22 & 65.52 & 61.12 & 60.74 & 57.60 & 57.10 \\ 
  2 & F1 & s & 62.82 & 64.26 & 30.72 & 30.64 & 12.28 & 12.32 & 7.14 & 6.36 & 4.76 & 4.74 & 3.80 & 3.66 & 3.04 & 3.04 \\ 
  3 & F1+F2 & b+b & 99.98 & 100.00 & 89.82 & 90.00 & 63.38 & 61.96 & 53.76 & 55.32 & 49.76 & 49.46 & 44.20 & 43.40 & 35.60 & 36.88 \\ 
  4 & F1+F2 & b+s & 52.38 & 64.90 & 23.38 & 24.40 & 9.16 & 7.74 & 5.36 & 4.16 & 4.16 & 3.48 & 3.18 & 2.56 & 2.56 & 2.50 \\ 
  5 & F1+F2 & s+s & 48.70 & 47.64 & 14.42 & 12.40 & 2.84 & 2.16 & 0.80 & 0.44 & 0.44 & 0.36 & 0.18 & 0.22 & 0.20 & 0.14 \\ 
  6 & F1+F1F2 & b+b & 99.76 & 100.00 & 84.96 & 86.94 & 62.02 & 67.10 & 56.18 & 59.44 & 51.72 & 53.94 & 43.68 & 47.20 & 39.34 & 41.68 \\ 
  7 & F1+F1F2 & b+s & 52.28 & 45.64 & 19.04 & 18.38 & 7.78 & 6.40 & 4.50 & 4.50 & 4.00 & 3.38 & 3.54 & 3.54 & 3.08 & 2.68 \\ 
  8 & F1+F1F2 & s+s & 49.68 & 50.30 & 20.96 & 18.02 & 5.90 & 4.36 & 2.32 & 2.12 & 1.76 & 1.64 & 1.50 & 1.70 & 1.50 & 1.34 \\ 
  9 & F1+F2+F1F2 & b+b+b & 99.88 & 100.00 & 81.68 & 86.50 & 51.34 & 56.92 & 41.60 & 45.56 & 33.76 & 38.32 & 27.18 & 32.84 & 21.80 & 26.22 \\ 
  10 & F1+F2+F1F2 & b+b+s & 30.06 & 29.14 & 14.78 & 16.44 & 7.86 & 6.76 & 3.24 & 4.42 & 3.36 & 3.90 & 2.54 & 2.96 & 2.30 & 3.06 \\ 
  11 & F1+F2+F1F2 & b+s+b & 58.80 & 69.10 & 20.84 & 27.14 & 10.82 & 9.10 & 4.62 & 5.36 & 4.28 & 4.40 & 2.74 & 2.94 & 2.80 & 3.04 \\ 
  12 & F1+F2+F1F2 & b+s+s & 38.66 & 47.06 & 12.56 & 13.54 & 1.88 & 1.66 & 0.94 & 0.68 & 0.52 & 0.38 & 0.32 & 0.28 & 0.28 & 0.30 \\ 
  13 & F1+F2+F1F2 & s+s+s & 32.78 & 36.64 & 7.90 & 7.58 & 1.22 & 1.00 & 0.26 & 0.24 & 0.20 & 0.14 & 0.22 & 0.10 & 0.12 & 0.12 \\ 
  14 & F1+F2+F3 & b+b+b & 100.00 & 100.00 & 87.54 & 86.80 & 53.26 & 52.66 & 42.66 & 41.44 & 37.66 & 33.06 & 30.50 & 29.08 & 25.82 & 23.42 \\ 
  15 & F1+F2+F3 & b+b+s & 53.94 & 61.40 & 23.22 & 24.84 & 7.80 & 6.56 & 5.04 & 4.36 & 3.64 & 3.50 & 3.50 & 2.64 & 2.86 & 2.62 \\ 
  16 & F1+F2+F3 & b+s+s & 56.34 & 63.86 & 12.86 & 11.34 & 2.48 & 1.62 & 0.62 & 0.52 & 0.56 & 0.30 & 0.32 & 0.32 & 0.24 & 0.18 \\ 
  17 & F1+F2+F3 & s+s+s & 35.14 & 34.28 & 9.64 & 6.76 & 0.80 & 0.52 & 0.12 & 0.06 & 0.04 & 0.04 & 0.10 & 0.04 & 0.00 & 0.10 \\ 
  18 & F1+F2+F1F3 & b+b+b & 99.40 & 100.00 & 80.92 & 84.14 & 51.74 & 56.48 & 41.22 & 42.32 & 36.52 & 37.28 & 31.50 & 29.46 & 23.14 & 21.62 \\ 
  19 & F1+F2+F1F3 & b+b+s & 39.90 & 30.16 & 20.04 & 15.46 & 7.18 & 6.16 & 5.34 & 4.80 & 3.28 & 3.00 & 3.12 & 2.46 & 2.74 & 2.76 \\ 
  20 & F1+F2+F1F3 & b+s+b & 59.36 & 57.88 & 22.30 & 24.94 & 9.52 & 7.50 & 5.88 & 4.48 & 3.82 & 3.60 & 3.30 & 2.54 & 2.60 & 2.70 \\ 
  21 & F1+F2+F1F3 & b+s+s & 45.74 & 34.28 & 12.84 & 12.10 & 2.18 & 1.80 & 0.68 & 0.50 & 0.50 & 0.32 & 0.44 & 0.34 & 0.44 & 0.28 \\ 
  22 & F1+F2+F1F3 & s+b+s & 47.24 & 50.74 & 9.94 & 9.16 & 1.88 & 1.42 & 0.74 & 0.52 & 0.46 & 0.54 & 0.36 & 0.50 & 0.34 & 0.40 \\ 
  23 & F1+F2+F1F3 & s+s+s & 34.90 & 33.98 & 6.66 & 5.66 & 0.82 & 0.48 & 0.10 & 0.10 & 0.08 & 0.04 & 0.06 & 0.00 & 0.00 & 0.00 \\ 
  24 & F1+F2+F3+F1F3 & b+b+b+b & 99.56 & 100.00 & 77.32 & 80.80 & 38.48 & 42.02 & 30.90 & 30.64 & 24.20 & 22.76 & 17.12 & 17.20 & 14.46 & 13.46 \\ 
  25 & F1+F2+F3+F1F3 & b+b+s+s & 31.46 & 24.52 & 9.30 & 10.48 & 1.82 & 1.38 & 0.78 & 0.64 & 0.60 & 0.46 & 0.24 & 0.14 & 0.22 & 0.20 \\ 
  26 & F1+F2+F3+F1F3 & b+s+s+b & 38.18 & 46.34 & 14.02 & 15.10 & 2.06 & 1.78 & 0.76 & 0.84 & 0.68 & 0.56 & 0.16 & 0.26 & 0.24 & 0.32 \\ 
  27 & F1+F2+F3+F1F3 & s+s+s+s & 21.30 & 23.86 & 4.42 & 3.38 & 0.18 & 0.16 & 0.08 & 0.04 & 0.04 & 0.00 & 0.00 & 0.02 & 0.00 & 0.02 \\ 
  28 & F1+F2+F3+F1F3+F2F3 & b+b+b+b+b & 88.32 & 99.96 & 58.76 & 72.88 & 27.84 & 28.82 & 18.96 & 24.74 & 19.10 & 19.10 & 14.02 & 11.86 & 7.52 & 8.14 \\ 
  29 & F1+F2+F3+F1F3+F2F3 & b+b+s+s+s & 10.96 & 18.20 & 5.48 & 6.60 & 0.60 & 0.28 & 0.12 & 0.10 & 0.06 & 0.06 & 0.04 & 0.04 & 0.08 & 0.08 \\ 
  30 & F1+F2+F3+F1F3+F2F3 & b+s+s+b+b & 16.98 & 33.28 & 11.02 & 16.10 & 2.18 & 2.32 & 1.00 & 0.74 & 0.36 & 0.64 & 0.46 & 0.52 & 0.24 & 0.16 \\ 
  31 & F1+F2+F3+F1F3+F2F3 & s+s+s+s+s & 13.18 & 16.50 & 1.36 & 0.98 & 0.14 & 0.06 & 0.00 & 0.00 & 0.00 & 0.00 & 0.00 & 0.00 & 0.00 & 0.00 \\ 
  32 & F1+F2+F3+F4+F5+F1F2 & b+b+b+b+b+b & 97.18 & 83.78 & 71.82 & 68.82 & 27.60 & 20.50 & 20.46 & 12.78 & 12.80 & 7.52 & 12.40 & 5.32 & 9.70 & 3.04 \\ 
  33 & F1+F2+F3+F4+F5+F1F2 & b+b+s+s+s+s & 10.08 & 5.94 & 5.38 & 1.90 & 0.32 & 0.08 & 0.10 & 0.04 & 0.08 & 0.00 & 0.00 & 0.00 & 0.02 & 0.00 \\ 
  34 & F1+F2+F3+F4+F5+F1F2 & b+s+s+b+b+b & 13.44 & 19.14 & 12.42 & 14.42 & 2.40 & 1.62 & 0.96 & 0.82 & 0.64 & 0.48 & 0.48 & 0.22 & 0.56 & 0.16 \\ 
  35 & F1+F2+F3+F4+F5+F1F2 & s+s+s+s+s+s & 12.42 & 8.58 & 1.50 & 0.24 & 0.08 & 0.04 & 0.00 & 0.00 & 0.00 & 0.00 & 0.00 & 0.00 & 0.00 & 0.00 \\ 
			\hline
		\end{tabular}
		}
	\caption{Average Percentage of Correctly Identified Models for $3-$level A-ComVar design $D^3$ and OME $D^8$.}
    \label{table:sim3}
	\end{table}

\end{document}